\documentclass[prd,aps,twocolumn,a4paper,showkeys,nofootinbib]{revtex4}
\usepackage{amsmath}
\usepackage{amsfonts}
\usepackage{amssymb}	
\usepackage{graphicx}
\usepackage{bm}
\usepackage{color}
\usepackage{ulem}
\usepackage{commath}
\allowdisplaybreaks

\def\e{{\rm e}}

\def\GMc2{G M_{\odot} c^{-2}}

\def\lm{{\ell m}}

\def\lm{{\ell m}}

\def\de{\partial}
\def\lm{{\ell m}}

\def\ii{{\rm i}}

\def\F{{\cal F}}

\newcommand\be{\begin{equation}}
\newcommand\ee{\end{equation}}

\def\TEOBResumS{\texttt{TEOBResumS}}

\begin{document}
\title{Effective one body multipolar waveform model for spin-aligned,
quasi-circular, eccentric, hyperbolic black hole binaries.}
\author{Alessandro \surname{Nagar}${}^{2,3}$}
\author{Alice \surname{Bonino}${}^{1}$}
\author{Piero \surname{Rettegno}${}^{1,2}$}
\affiliation{${}^{1}$ Dipartimento di Fisica, Universit\`a di Torino, via P. Giuria 1, 10125 Torino, Italy}
\affiliation{${}^2$INFN Sezione di Torino, Via P. Giuria 1, 10125 Torino, Italy}
\affiliation{${}^3$Institut des Hautes Etudes Scientifiques, 91440 Bures-sur-Yvette, France}

\begin{abstract}
Building upon recent work, we present an improved effective-one-body (EOB) model for 
spin-aligned, coalescing, black hole binaries with generic orbital configurations, i.e. quasi-circular, 
eccentric or hyperbolic orbits. The model, called {\tt TEOBResumSGeneral}, 
relies on the idea of incorporating general Newtonian prefactors, 
instead of the usual quasi-circular ones, in both radiation reaction and waveform. 
The major 
advance with respect to previous work is that the quasi-circular limit of the model is now correctly 
informed by numerical relativity (NR) quasi-circular simulation. This provides  EOB/NR unfaithfulness 
for the dominant quadrupolar waveform, calculated with Advanced LIGO noise, at most of the 
order of $1\%$ over a meaningful portion of the quasi-circular NR simulations computed by the 
Simulating eXtreme Spacetime (SXS) collaboration. In the presence of eccentricity, the model 
is similarly NR-faithful, $\lesssim 1\%$, all over the 28 public SXS 
NR datasets, with initial eccentricity up to $\simeq 0.2$ , mass ratio up to $q=3$ and dimensionless 
spin magnitudes as large as $+0.7$. Higher multipoles, up to $\ell=5$ are also reliably modeled 
through the eccentric inspiral, plunge, merger and ringdown. For hyperbolic-like configurations, 
we also show that the EOB computed scattering angle is in excellent agreement with all currently
available NR results.
   \end{abstract}
   
\date{\today}

\maketitle

\section{Introduction}
\label{sec:intro}
A recent work~\cite{Chiaramello:2020ehz} introduced an effective one body (EOB) 
waveform model for spin-aligned, eccentric, black hole binaries. This waveform model 
is not limited to stable configurations, but can also generate waveforms for hyperbolic 
encounters and dynamical captures from binary black holes (BBHs) coalescences~\cite{Nagar:2020xsk}. 
The pivotal technical aspect behind this waveform model is the possibility of  
accurately generalizing the EOB resummed quasi-circular radiation reaction 
(and waveform) to generic orbits by simply considering generic (i.e., non quasi-circular) 
Newtonian prefactors in these functions. 
In order to assess the quality of the model, Ref.~\cite{Chiaramello:2020ehz}
compared the so constructed EOB waveforms to 22 eccentric numerical
relativity (NR) public waveforms from the SXS 
catalog~\cite{Chu:2009md,Lovelace:2010ne,Lovelace:2011nu,Buchman:2012dw,
Hemberger:2013hsa,Scheel:2014ina,Blackman:2015pia,
Lovelace:2014twa,Mroue:2013xna,Kumar:2015tha,Chu:2015kft,
Boyle:2019kee,SXS:catalog}. 
The comparison was performed by computing the EOB/NR unfaithfulness (or mismatch)
using the Advanced LIGO power spectral density. For the model of 
Ref.~\cite{Chiaramello:2020ehz} this led to unfaithfulnesses that 
reached up to $3\%$. 
Although this result could be considered satisfactory at the time, the 
model of Ref.~\cite{Chiaramello:2020ehz} was not especially optimized 
and can be improved along several directions. In particular, it relied on the EOB 
conservative dynamics of Refs.~\cite{Nagar:2019wds,Nagar:2020pcj}, 
that was NR-informed using the {\it standard quasi-circular} EOB radiation reaction.
The model of Refs.~\cite{Nagar:2019wds,Nagar:2020pcj}, called  {\tt TEOBiResumS\_SM},
incorporates higher-order modes and is the most advanced and accurate version
of the \TEOBResumS{} model~\cite{Nagar:2018zoe}. To simplify the nomenclature, 
from now on we address as \TEOBResumS{} the model of Ref.~\cite{Nagar:2020pcj}.
The purpose of this paper is to correct the inconsistency of Refs.~\cite{Nagar:2019wds,Nagar:2020xsk},  
by determining new NR-informed EOB flexibility functions $(a_6^c,c_3)$~\cite{Nagar:2019wds,Nagar:2020pcj}, 
consistent with the general, non quasi-circular, radiation reaction and waveform. 
We will see that this modification is sufficient to lower the EOB/NR unfaithfulness in 
the eccentric sector at approximately the $1\%$ level. To distinguish it from the
quasi-circular model \TEOBResumS{}, and since it can deal also with general, spin-aligned, 
configurations, like hyperbolic scattering or capture, for convenience we will address 
it as {\tt TEOBResumSGeneral}.
The paper is organized as follows. In Sec.~\ref{sec:dynamics} we recall the structure
of the EOB dynamics and waveform, provide the new expressions of  $(a_6^c,c_3)$ 
and illustrate the related new EOB/NR waveform performance 
for quasi-circular configurations. The eccentric case is discussed in Sec.~\ref{sec:ecc},
while Sec.~\ref{sec:scattering} provides a new EOB/NR comparison of the scattering angle.
Our findings are summarized in Sec.~\ref{sec:conclusions}.  Throughout this paper we
mostly use geometric units with $G=c=1$.

\section{Quasi-circular configurations}
\label{sec:dynamics}

\subsection{Effective one body dynamics}
\begin{figure}[t]
\center
\includegraphics[width=0.45\textwidth]{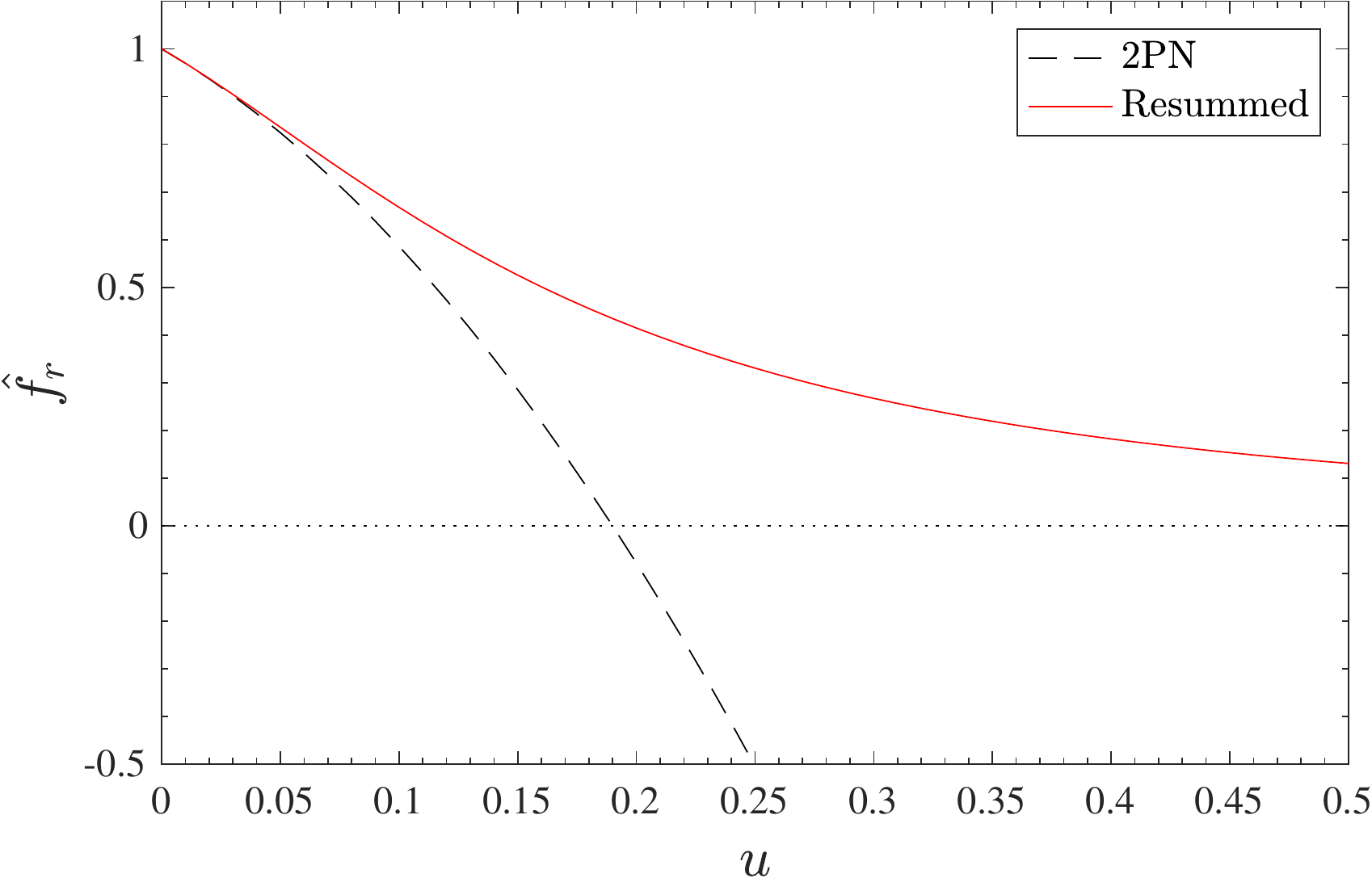}
\caption{\label{fig:Fr} Comparing the PN-expanded $\hat{f}_r^{\rm 2PN}$, Eq.~\eqref{eq:fr}, 
with its Pad\'e resummed counterpart for $q=1$. The resummation avoids the infinitely growing, 
unphysical,  strong-field behavior of $\hat{f}_r^{\rm 2PN}$.}
\end{figure}

The structure of the dynamics of the EOB eccentric model is essentially 
the same discussed in Sec.~II of Ref.~\cite{Chiaramello:2020ehz} and thus
we limit ourselves to report here the few details that we have improved on.
We adopt the usual notation within the EOB formalism.
We use mass-reduced phase-space variables $(r,\varphi,p_\varphi,p_{r_*})$,  related to the physical 
ones by $r=R/M$ (relative separation), $p_{r_*}=P_{R_*}/\mu$ (radial momentum), 
$p_\varphi=P_\varphi/(\mu M)$ (angular momentum) and $t=T/M$ (time),
where $\mu\equiv m_1 m_2/M$ and $M\equiv m_1+m_2$. The radial momentum is 
$p_{r_*}\equiv (A/B)^{1/2}p_r$, where $A$ and $B$ are the EOB potentials 
(with included spin-spin interactions~\cite{Damour:2014sva}). 
The EOB Hamiltonian is $\hat{H}_{\rm EOB}\equiv H_{\rm EOB}/\mu=\nu^{-1}\sqrt{1+2\nu(\hat{H}_{\rm eff}-1)}$, 
with $\nu\equiv \mu/M$ and $\hat{H}_{\rm eff}=\tilde{G}p_\varphi + \hat{H}^{\rm orb}_{\rm eff}$, 
where $\tilde{G}p_\varphi$ incorporates odd-in-spin (spin-orbit) effects while 
$\hat{H}^{\rm orb}_{\rm eff}$ incorporates even-in-spin effects~\cite{Nagar:2018zoe}. 
We denote dimensionless spin variables as $\chi_{i}\equiv S_i/m_i^2$, 
and adopt $\hat{\F}_{\varphi,r}\equiv \F_{\varphi,r}/\mu$ as $\mu$-rescaled radiation 
reaction forces.
The novelties here mainly pertain the analytical expressions  $(\hat{\F}_\varphi, \hat{\F}_r)$ 
and thus impact the related Hamilton's equations, that we rewrite for completeness:
\begin{align}
\label{eq:pphi}
\dot{p}_{\varphi} &=\hat{\F}_\varphi \ ,\\
\label{eq:pr}
\dot{p}_{r_*}       &=\sqrt{\dfrac{A}{B}}\left(-\de_r \hat{H}_{\rm EOB} + \hat{\F}_r\right).
\end{align}
As mentioned in the introduction, the dynamics also depends on two effective EOB 
flexibility functions that
are informed by NR simulations; i.e., the 4.5PN spin-orbit effective function 
$c_3(\nu,\chi_1,\chi_2)$, that enters $\tilde{G}$, and the effective 
5PN function $a_6^c(\nu)$, that  enters the Pad\'e resummed radial potential $A(r)$. 
Reference~\cite{Chiaramello:2020ehz} used a generic, 2PN accurate, expression 
of $\hat{\cal F}_r$, resummed by taking its inverse. Instead, $(a_6^c,c_3)$ 
were the same of the standard quasi-circular model of Ref.~\cite{Nagar:2020xsk}. 
Such a choice, done for simplicity at the time, is an evident source of systematic
errors in the waveform model. It was already pointed out in Ref.~\cite{Chiaramello:2020ehz} 
that it would be necessary to use quasi-circular NR simulations 
to NR-inform two {\it new} functions $(a_6^c,c_3)$ to be
compatible with the general expressions for $(\hat{\F}_\varphi,\hat{\F}_r)$ 
and with the corresponding waveform. 
This is our scope here, but in doing so we also modify {\it both}  the analytical expressions of 
$(\hat{\F}_\varphi,\hat{\F}_r)$ with respect  to  Ref.~\cite{Chiaramello:2020ehz}. 
For what concerns $\hat{\F}_\varphi$, the simplifying choice of~\cite{Chiaramello:2020ehz} 
of incorporating noncircular corrections with $\ell=m=2$ overall factor also 
introduces some systematics, since the subdominant flux modes are multiplied 
by an incorrect noncircular Newtonian factor. Here we go beyond the previous, 
simplified, approach and we incorporate the noncircular correction, Eq.~(7) of~\cite{Chiaramello:2020ehz}, 
as a multiplicative factor entering {\it only} the $\ell=m=2$ flux 
contribution\footnote{Evidently this is also an approximation, since for uniformity the
same approach should be applied also to the subdominant modes. Since, as we will see, 
this approximation already delivers good results, we postpone such refinement to future work.}.
In addition, while performing the EOB/NR comparison with quasi-circular data aimed
at determining a consistent expression of $a_6^c(\nu)$, we realized that the general, 
2PN resummed, expression for $\hat{\F}_r$ used in ~\cite{Chiaramello:2020ehz} 
reduces the flexibility of the quasi-circular limit of the model. This prevents us from
determining $a_6^c(\nu)$ so to yield an EOB/NR phasing agreement at the level 
of \TEOBResumS{} (see Appendix~\ref{sec:Fr_derivation}). This problem 
turns out to be alleviated by using a resummed version of the 2PN-accurate 
{\it quasi-circular} reduction of $\hat{\F}_r$, starting from Eq.~(5.16) of Ref.~\cite{Bini:2012ji}.
The nonresummed version reads
\begin{align}
\label{eq:FRcirc}
\hat{\F}_r^{\rm 2PN} &=\dfrac{32}{3}\nu p_r u^4\bigg[1 -\left(\dfrac{1133}{280} + \dfrac{118}{35}\nu\right) u \nonumber\\
        &+\dfrac{1}{15120}\left(-175549 + 322623\nu + 70794\nu^2\right) u^2 \bigg].
\end{align}
We then proceed by replacing $p_r = p_{r_*}\sqrt{B/A}$, expanding at 2PN order and getting
\begin{align}
\label{eq:Fr_PN}
\hat{\F}^{\rm 2PN}_r =\dfrac{32}{3} \nu p_{r_*} u^4\hat{f}_r^{\rm 2PN} \ ,
\end{align}
where
\begin{align}
\label{eq:fr}
\hat{f}_r^{\rm 2PN}  &= 1 -\left(\dfrac{573}{280} + \dfrac{118}{35}\nu\right) u \nonumber\\
        &+\left(-\dfrac{33919}{2160} + \dfrac{6493}{560}\nu + \dfrac{1311}{280}\nu^2\right) u^2 .
\end{align}
The PN-expanded $\hat{f}_r^{\rm 2PN}$ function becomes negative before, and up to, merger.
This means that its effect progressively becomes unphysical (see Fig.~\ref{fig:Fr}). 
To overcome this difficulty, $\hat{f}_r^{\rm 2PN}$ is resummed using a $P^0_{2}$ 
Pad\'e approximant\footnote{The natural
$P(1,1)$ approximant is unusable as it develops a spurious pole.}, not differently
from what is usually done for the gyro-gravitomagnetic functions in the spin-sector of the
{\tt TEOBResumS} model~\cite{Damour:2014sva}. In practice, our Eq.~\eqref{eq:Fr_PN}
is replaced by
\be
\label{eq:Fr_circ_resum}
\hat{\F}_r = \dfrac{32}{3}\nu p_{r_*} u^4P^0_2\left[\hat{f}_r^{\rm 2PN}(u)\right],
\ee
that is then taken as default radial force in Eq.~\eqref{eq:pr}.

%======================
% Fr comparison: PN vs resum
%=======================
%=============

\subsection{Waveform}
\label{sec:wave}
To include the effect brought by the generic dynamics in the waveform we essentially
follow the approach of Ref.~\cite{Chiaramello:2020ehz}, that is replacing the quasi-circular
Newtonian prefactor of each multipole with its generic counterpart. We report here more
technical details with respect to what briefly sketched in Ref.~\cite{Chiaramello:2020ehz}.
Let us first recall the basic notation and conventions. The strain waveform is decomposed
in multipoles $h_\lm$ as
\be
h_+-\ii h_\times=D_L^{-1}\sum_\lm h_\lm {}_{-2}Y_{\lm}, 
\ee
where $D_L$ is the luminosity distance and ${}_{-2}Y_{\lm}$ are the $s=-2$ spin-weighted
spherical harmonics. Following our usual practice, we will perform EOB/NR comparisons 
using the Regge-Wheeler-Zerilli normalized variable $\Psi_{\lm}=h_{\lm}/\sqrt{(\ell+2)(\ell+1)\ell(\ell-1)}$.
Within the EOB formalism, each multipole is factorized as~\cite{Damour:2008gu}
\be
h_{\lm}=h_\lm^{(N,\epsilon)}\hat{h}_\lm\hat{h}^{\rm NQC}_\lm,
\ee
where $h_\lm^{(N,\epsilon)}$ is the Newtonian (leading-order) prefactor, $\epsilon$ the parity of $\ell+m$, 
$\hat{h}_\lm$ is the relativistic correction that includes higher PN terms in resummed form 
and $\hat{h}_\lm^{\rm NQC}$ is the NR-informed next-to-quasi-circular factor. 
To go beyond the quasi-circular behavior, the usual quasi-circular Newtonian 
prefactor $h_\lm^{(N,\epsilon)}$ is here replaced by the general expression 
obtained computing the time-derivatives of the Newtonian mass and 
current multipoles. 
We have $h_\lm^{(N,0)}\propto e^{\ii m\varphi} I^{(\ell)}_\lm$
and  $h_\lm^{(N,1)}\propto e^{\ii m \varphi} S^{(\ell)}_\lm$,
where the superscript $(\ell)$ indicates the $\ell$-th time-derivative.
$I_\lm\equiv r^\ell e^{-\ii m\varphi}$ and
$S_\lm\equiv r^{\ell +1}\Omega e^{-\ii m\varphi}$ 
are the Newtonian mass and current multipoles,
where $\Omega\equiv\dot{\varphi}$ is the orbital frequency.
In practice, there are additional choices that can be made, multipole by multipole,
and that are better explained with explicit examples.
Let us focus first on the $\ell=m=2$ mode. The Newtonian prefactor 
is written as
\be
\label{eq:h22Newt}
h_{22}^{(N,0)}= -8\sqrt{\dfrac{\pi}{5}}\nu (r_\omega \Omega)^2\left(1+{\cal S}(t)\hat{h}_{22}^{\rm nc}\right)e^{-2{\rm i}\varphi},
\ee
where ${\cal S}(t)$ is a certain sigmoid function to be discussed below and the noncircular 
factor $\hat{h}_{22}^{\rm nc}$ reads
\be
\label{eq:h22_nc}
 \hat{h}_{22}^{\rm nc}=-\dfrac{1}{2}\left(\dfrac{\dot{r}^2}{(r\Omega)^2} + \dfrac{\ddot{r}}{r\Omega^2}\right) 
        + {\rm i}\left(\dfrac{2\dot{r}}{r\Omega} + \dfrac{\dot{\Omega}}{2\Omega^2}\right),
\ee
with an amplitude and a phase correction. Note that  we factored out the leading order term
$(r \Omega)^2$, where the modified EOB radius $r_\omega$ is replacing $r$ only in the 
quasi-circular prefactor of Eq.~\eqref{eq:h22Newt}, so to be consistent with the standard 
EOB prescription for quasi-circular configurations~\cite{Damour:2006tr,Damour:2007yf}.
The scope of the sigmoid function
\be
{\cal S}(t) = \dfrac{1}{1+e^{\alpha(t-t_0)}},
\ee
is to progressively switch off $\hat{h}^{\rm nc}_{22}$ around a given time $t_0$, sufficiently close 
to merger, as the system circularizes during late inspiral and plunge. 
The main reason for doing so is that $\hat{h}_{22}^{\rm nc}$ looks unable to 
correctly match the quasi-circular behavior of NR simulations in the late 
inspiral to plunge phase, because of the continuous growth of $(\dot{r},\dot{\Omega})$. 
Pragmatically\footnote{One could have chosen different, though Newtonian-like consistent, 
forms for $\hat{h}^{\rm nc}_{22}$, e.g. by replacing $\dot{r}$ with $p_{r_*}$ as it is done to 
obtain the analytical structure of $\hat{h}^{\rm NQC}_\lm$~\cite{Nagar:2019wds}.
We did not explore this option since the application of the straight Newtonian expression 
already looked sufficiently accurate for our purposes.}, we decided to smoothly 
switch off $h_{22}^{\rm nc}$,  so to recover the usual robustness properties 
of the simpler quasi-circular EOB waveform.
In the meanwhile, as already pointed out in Ref.~\cite{Chiaramello:2020ehz}, 
the next-to-quasi-circular (NQC) waveform factor $\hat{h}^{\rm NQC}_{22}$ is similarly switched 
on close to merger, to suitably modify the waveform with the NR-informed NQC parameters
and provide the usual good EOB/NR match for quasi-circular binaries.
For what concerns this paper, the parameters $(\alpha,t_0)$ are chosen somehow 
arbitrarily: $\alpha=0.02$ and $t_0\equiv t_{\Omega_{\rm peak}^{\rm orb}}-100$, 
where $t_{\Omega_{\rm peak}^{\rm orb}}$ is the peak time of the {\it pure orbital}
frequency, i.e. the orbital frequency with the spin-orbit contribution subtracted~\cite{Damour:2014sva}.
For what concerns the subdominant multipoles\footnote{We recall that we do not taken into account $m=0$ modes.}, 
up to $\ell=m=5$, we adopt the following expression for each Newtonian prefactor
\be
h_{\ell m}^{(N,\epsilon)}= c_{\ell m}(\nu) \left((r_\omega \Omega)^\ell + {\cal S}(t) h_\lm^{\rm nc}\right)e^{-{\rm i}m\varphi} \ ,
\ee
where $c_{\ell m}(\nu)$ schematically indicate the well known Newtonian numerical coefficients, 
analogous\footnote{See e.g. Eqs.~(3.21)-(3.30) of~\cite{Nagar:2019wds} for their explicit values.} 
of the $-8\sqrt{\pi/5} \, \nu$ in Eq.~\eqref{eq:h22Newt}, and $h_\lm^{\rm nc}$
indicates the remaining, leading order, noncircular corrections, notably not divided 
by $(r\Omega)^\ell$ as in Eq.~\eqref{eq:h22Newt}. We found that this nonfactorized
expression is generally more robust and accurate than the factorized one in all corners 
of the parameter space. By contrast, they are substantially equivalent
for the $(2,2)$ mode.
Finally, to achieve a good compromise between accuracy and robustness, we had
to slightly modify the prescription for $\hat{h}_\lm^{\rm NQC}$ presented in 
Ref.~\cite{Nagar:2019wds}, to which we refer the reader for all precise 
technical details concerning the  determination procedure
of NQC corrections. Just  to clarify the logic, we only recall here that the 
NQC correction factor reads
\be
\hat{h}^{\rm NQC}_{\lm} = (1 + a_1^\lm n_1+ a_2^\lm n_2)e^{{\rm i}(b_1^\lm n_3^\lm + b_2^\lm n_4^\lm)},
\ee
where $n_i^\lm$, with $i=1\dots 4$, are functions depending on the radial 
momentum, while $(a^\lm_i,b_i^\lm)$, with $i=1,2$, are NR-informed numerical 
coefficients~\footnote{The procedure implemented for their determination is
 detailed in Sec.~IIID of Ref.~\cite{Nagar:2019wds}.}. 
 For the $\ell=m=2$ mode, the $n_i^{22}$ functions are given by Eqs.~(3.32)-(3.35)
 of~\cite{Nagar:2019wds}. For higher modes, we implement the following modifications
 with respect to the prescriptions of Ref.~\cite{Nagar:2019wds}, Eqs.~(3.38)-(3.45) in
 order to improve their robustness all over the parameter space.
 In particular, for $\ell=m=3$ and $\ell=m=4$, we have that $n_4^\lm = n_3^\lm\Omega^{2/3}$ 
 when the spins are negative, while $n_4^\lm=n_3^\lm (r\Omega)^2$ for {\it positive} spins.
In addition, we also use $n_2^{32}=n_1^{32}\Omega^{2/3}$; $n_2^{43}=n_1^{\rm 43}\Omega^{2/3}$,
$n_2^{42}=n_2^{42}\Omega^{2/3}$ and $n_4^{42}=n_3^{42}(r\Omega)^2$  instead of 
Eq.(3.42)-(3.45) of~\cite{Nagar:2019wds}. This is meaningful seen the effective nature of
the NQC correction factor: its choice has to compensate/improve the bare, purely analytical,
part of the waveform whose (resummed) PN accuracy depends on the multipole~\cite{Nagar:2020pcj}.

\subsection{Determining the $a_6^c$ and $c_3$ functions.}
\label{sec:a6_c3_new}
To determine new expressions for $(a_6^c,c_3)$   we follow our 
usual procedure, discussed extensively e.g. in Ref.~\cite{Nagar:2018zoe,Nagar:2019wds,Nagar:2020pcj}.
The approach simply relies on monitoring the time-evolution of the EOB/NR phase difference 
between waveforms aligned in the early inspiral while varying $(a_6^c,c_3)$.
We follow a two-step procedure,  by first informing the nonspinning sector of the 
model and then the spinning one. To start with, we consider 9 nonspinning simulations 
(see Table~\ref{tab:a6c}), and determine for each one the value of $a_6^c$ 
that makes the EOB/NR phase difference at NR merger of the order of 
(or smaller than) the nominal NR uncertainty, which is estimated taking the 
phase difference between the highest and second highest numerical 
resolution (see Table~V of Ref.~\cite{Nagar:2017jdw} for this calculation).
%===============
% Bare values of a6c
%==============
\begin{table}[t]
 \caption{\label{tab:a6c} Informing the nonspinning sector of the model. From left to right the columns report: the dataset number, 
 the SXS identification number; the mass ratio $q$; the symmetric mass ratio $\nu$; the first guess value of $a_6^c$ and the fitted
 value from Eq.~\eqref{eq:a6c_fit}.}
   \begin{center}
     \begin{ruledtabular}
\begin{tabular}{c c c c c c} 
$\#$ & SXS & $q$ & $\nu$ &  $a_6^c$  & $a_6^c(\nu)$ \\
  \hline
  \hline
%  1 & SXS:BBH:0180  & 1      & 0.25  & 250 &  243.73\\
%  2 & SXS:BBH:0007  & 1.5  &  0.24 &$160$ & 171.30\\
%  3 & SXS:BBH:0184 & 2     &  $0.\bar{2}$ &$90$ &  86.73\\
%  4 &  SXS:BBH:0259 & 2.5  &  0.204 & $40$ & 38.34 \\
%  5 &  SXS:BBH:0294 & 3.5  &  0.173  &$6$ &  2.18\\
%  6 &  SXS:BBH:0295 & 4.5  &  0.149 &$-7$ & $-6.42$\\
%  7 &  SXS:BBH:0056 & 5     &  0.139 & $-8$&  $-7.7$\\
%  8 & SXS:BBH:0063  & 8     &  0.0988 &$-10$ & $-7.42$ \\ 
%  9 & SXS:BBH:0303  & 10   & 0.0826& $-10.15$ & $-6.39$
  1 & SXS:BBH:0180  & 1      & 0.25  & 280 &  281.62\\
  2 & SXS:BBH:0007  & 1.5  &  0.24 &$200$ & 198.63\\
  3 & SXS:BBH:0184 & 2     &  $0.\bar{2}$ &$110$ &  102.75\\
  4 &  SXS:BBH:0259 & 2.5  &  0.204 & $36$ & 48.61 \\
  5 &  SXS:BBH:0294 & 3.5  &  0.173  &$14$ &  8.56\\
  6 &  SXS:BBH:0295 & 4.5  &  0.149 &$0$ & $-1.18$\\
  7 &  SXS:BBH:0056 & 5     &  0.139 & $-1$&  $-2.78$\\
  8 & SXS:BBH:0063  & 8     &  0.0988 &$-5$ & $-3.61$ \\ 
  9 & SXS:BBH:0303  & 10   & 0.0826& $-5.1$ & $-3.1$
  \end{tabular}
 \end{ruledtabular}
 \end{center}
 \end{table}
 %==========================
% Table of spin-dependent datasets
%==========================
\begin{table}[t]
   \caption{\label{tab:c3}Informing the spinning sector of the model. From left to right the columns report:
   the dataset number, the SXS identification number; the mass ratio and the individual dimensionless 
   spins $(q,\chi_1,\chi_2)$; the first-guess values of $c_3$ used to inform the global interpolating fit 
   given in Eq.~\eqref{eq:c3fit}, and the corresponding $c_3^{\rm fit}$ values.}
   \begin{center}
 \begin{ruledtabular}
   \begin{tabular}{lllll}
     $\#$ & ID & $(q,\chi_1,\chi_2)$ & $c_3^{\rm first\;guess}$ & $c_3^{\rm fit}$\\
     \hline
    1 & SXS:BBH:0156 &$(1,-0.95,-0.95)$     & 89 & 88.822 \\
    2 & SXS:BBH:0159 &$(1,-0.90,-0.90)$     & 86.5 & 86.538 \\
    3 & SXS:BBH:0154 &$(1,-0.80,-0.80)$     & 81  & 81.508 \\
    4 & SXS:BBH:0215 &$(1,-0.60,-0.60)$     & 70.5 & 70.144 \\
    5 & SXS:BBH:0150 &$(1,+0.20,+0.20)$   & 26.5 & 26.677 \\
    6 & SXS:BBH:0228 &$(1,+0.60,+0.60)$   & 16.0 & 15.765 \\
    7 & SXS:BBH:0230 &$(1,+0.80,+0.80)$   & 13.0 & 12.920 \\
    8 & SXS:BBH:0153 &$(1,+0.85,+0.85)$   & 12.0 & 12.278 \\
    9 & SXS:BBH:0160 &$(1,+0.90,+0.90)$   & 11.5 & 11.595\\
    10 & SXS:BBH:0157 &$(1,+0.95,+0.95)$ & 11.0 & 10.827\\
    11 & SXS:BBH:0004 &$(1,-0.50,0)$         & 54.5 & 46.723 \\
    12 & SXS:BBH:0231 &$(1,+0.90,0)$        & 24.0 &  23.008 \\
    13 & SXS:BBH:0232 &$(1,+0.90,+0.50)$ & 15.8 & 16.082 \\
    14 & SXS:BBH:0005 &$(1,+0.50,0)$        & 34.3 & 27.136 \\
    15 & SXS:BBH:0016 &$(1.5,-0.50,0)$     & 57.0 & 49.654 \\
    16 & SXS:BBH:0016 &$(1.5,+0.95,+0.95)$ & 13.0 & 11.720 \\
    17 & SXS:BBH:0255 &$(2,+0.60,0)$          & 29.0 & 23.147\\
    18 & SXS:BBH:0256 &$(2,+0.60,+0.60)$   & 20.8 & 17.37 \\
    19 & SXS:BBH:0257 &$(2,+0.85,+0.85)$   & 14.7 & 14.56 \\ 
    20 & SXS:BBH:0036 &$(3,-0.50,0)$           &  60.0 & 53.095\\
    21 & SXS:BBH:0267 &$(3,-0.50,-0.50)$     & 69.5 & 60.37\\ 
    22 & SXS:BBH:0174 &$(3,+0.50,0)$           &  30.0 & 24.210\\
    23 & SXS:BBH:0291 &$(3,+0.60,+0.60)$        &  23.4 & 19.635 \\
    24 & SXS:BBH:0293 &$(3,+0.85,+0.85)$        & 16.2 & 17.759 \\
    25 & SXS:BBH:1434 &$(4.368,+0.80,+0.80)$ & 20.3 & 20.715 \\
    26 & SXS:BBH:0060 &$(5,-0.50,0)$                & 62.0 & 55.385\\
    27 & SXS:BBH:0110 &$(5,+0.50,0)$               & 31.0 & 24.488\\
    28 & SXS:BBH:1375 &$(8,-0.90,0)$               & 64.0 & 71.91 \\
    29 & SXS:BBH:0064 &$(8,-0.50,0)$               & 57.0 & 55.385 \\
    30 & SXS:BBH:0065 &$(8,+0.50,0)$              & 28.5 & 24.306\\
    31 & BAM                  &$(8,+0.80,0)$              & 24.5 & 22.605\\
    32 & BAM                 &$(8,+0.85,+0.85)$         & 14.0 & 22.199
 \end{tabular}
 \end{ruledtabular}
 \end{center}
 \end{table}
%=============
% q=3 comparison
%=============
\begin{figure*}[t]
\center
\includegraphics[width=0.45\textwidth]{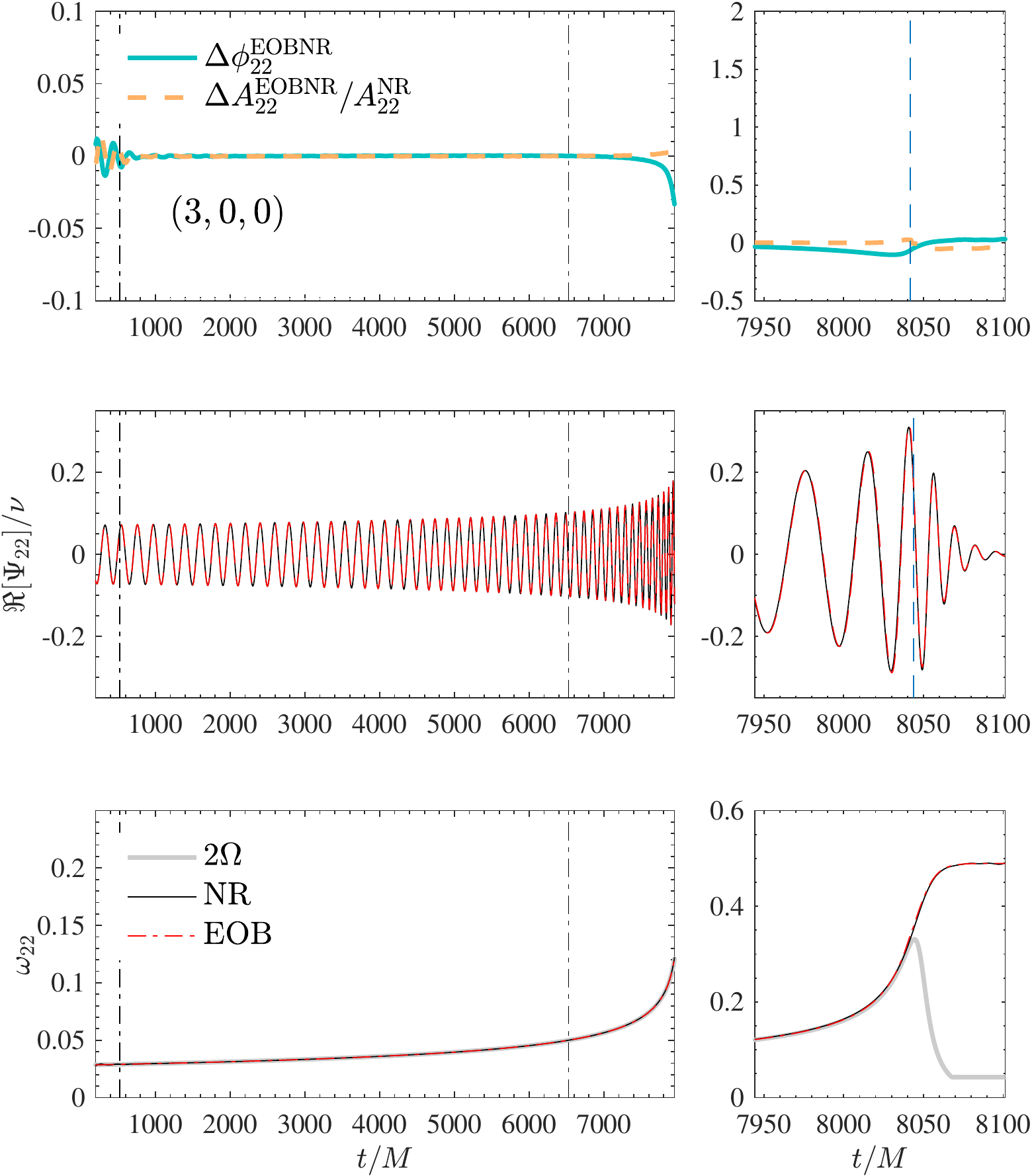}\qquad
\includegraphics[width=0.45\textwidth]{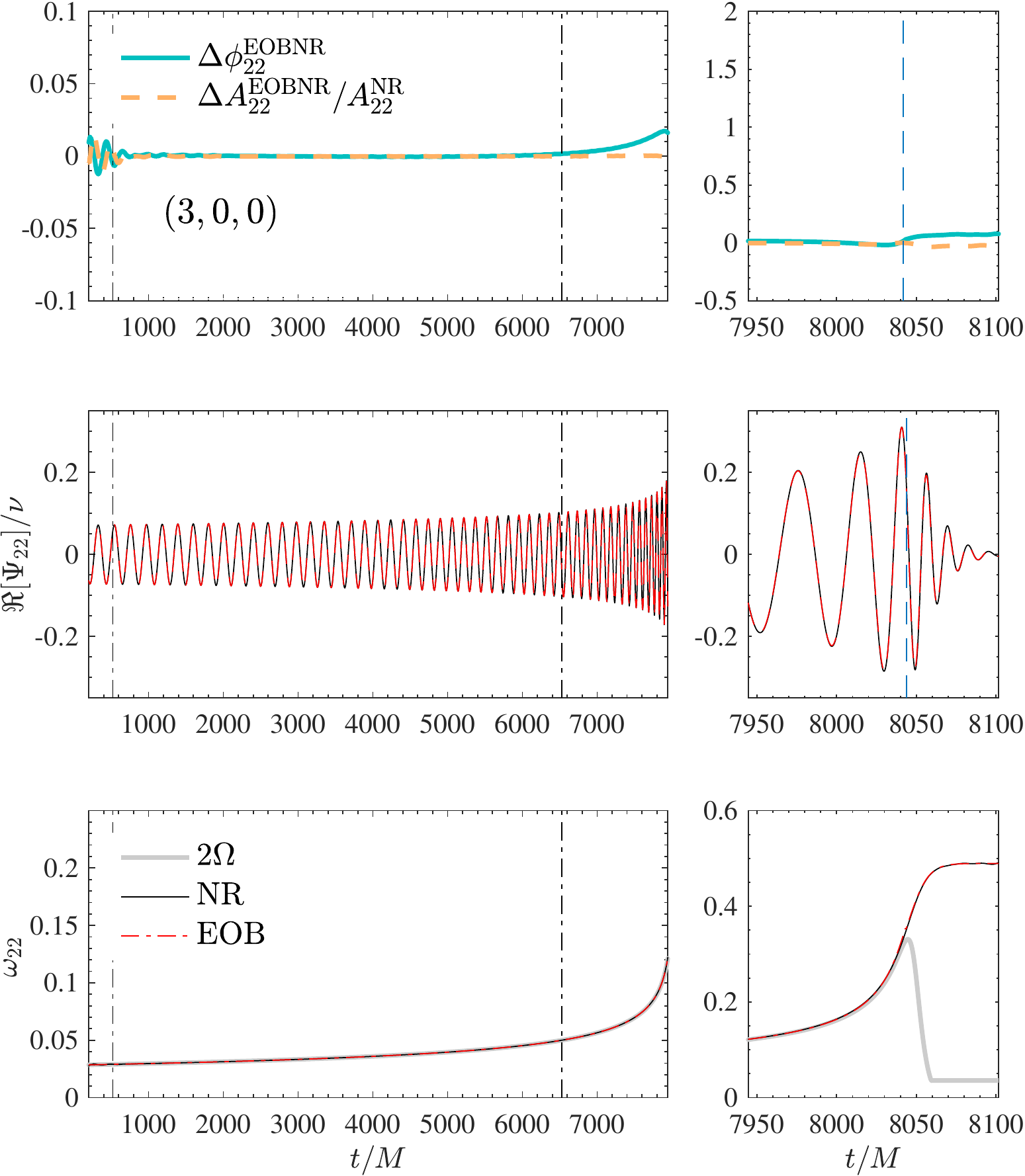}
\caption{\label{fig:q3} EOB/NR comparison with the 27-orbits long $q=3$, quasi-circular, nonspinning dataset SXS:BBH:1221. 
Left panel: phasing comparison with the {\tt TEOBResumSGeneral} model discussed in this paper, endowed with the
generic waveform and radiation reaction. Right panel: phasing comparison with the quasi-circular {\tt TEOBResumS} model.
Top panel: (relative) amplitude and phase difference (in radians). Middle panel: real part of the waveform. Bottom panel: gravitational frequencies. 
For convenience, also twice the EOB orbital frequency 2$\Omega$ is shown on the plot. 
The dash-dotted vertical lines indicate the alignment frequency region, while the dashed one the merger time.}
\end{figure*}
%=============
%================
% quasi-circular results
%================
\begin{figure*}[t]
\center
\includegraphics[width=0.45\textwidth]{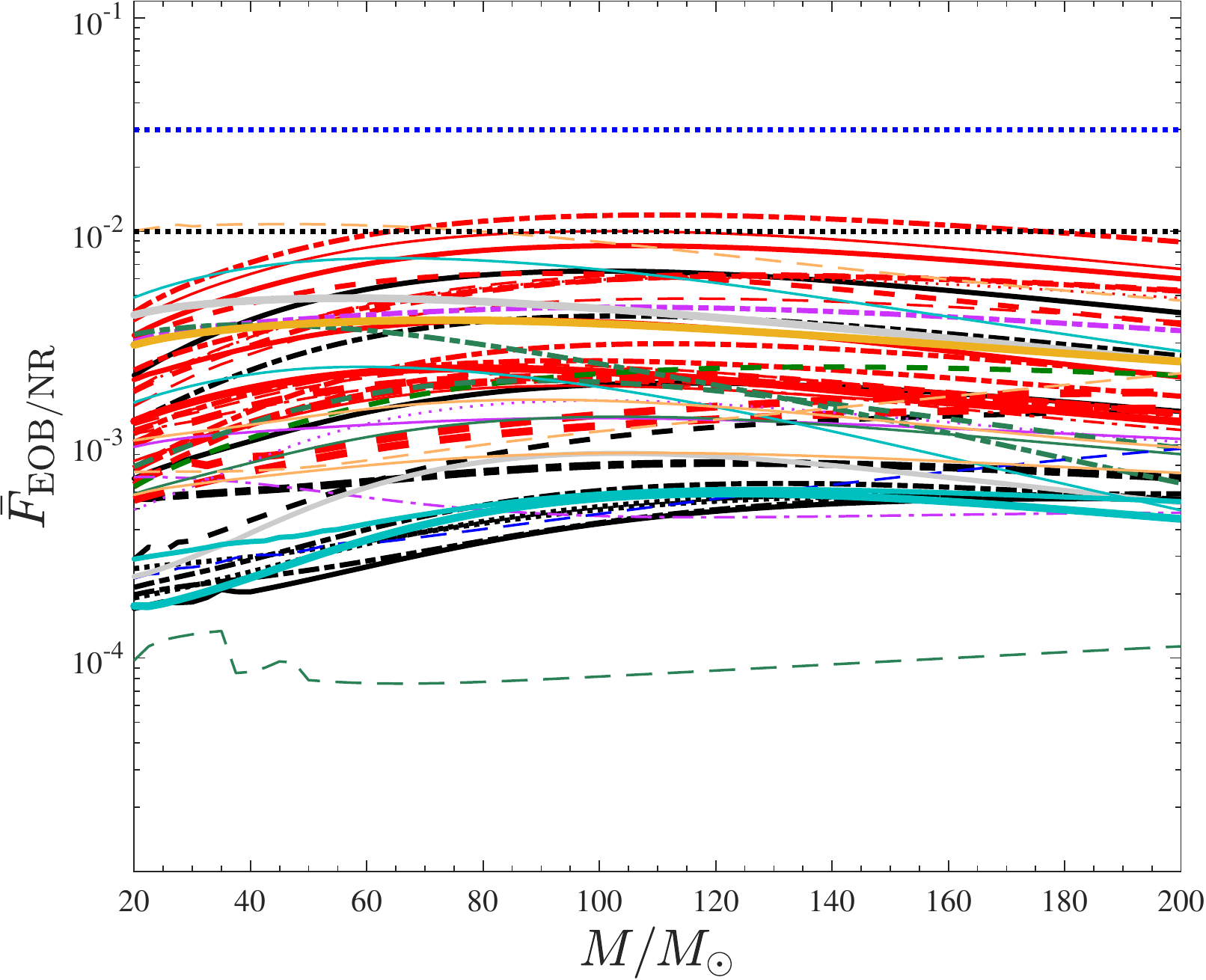}
\qquad
\includegraphics[width=0.45\textwidth]{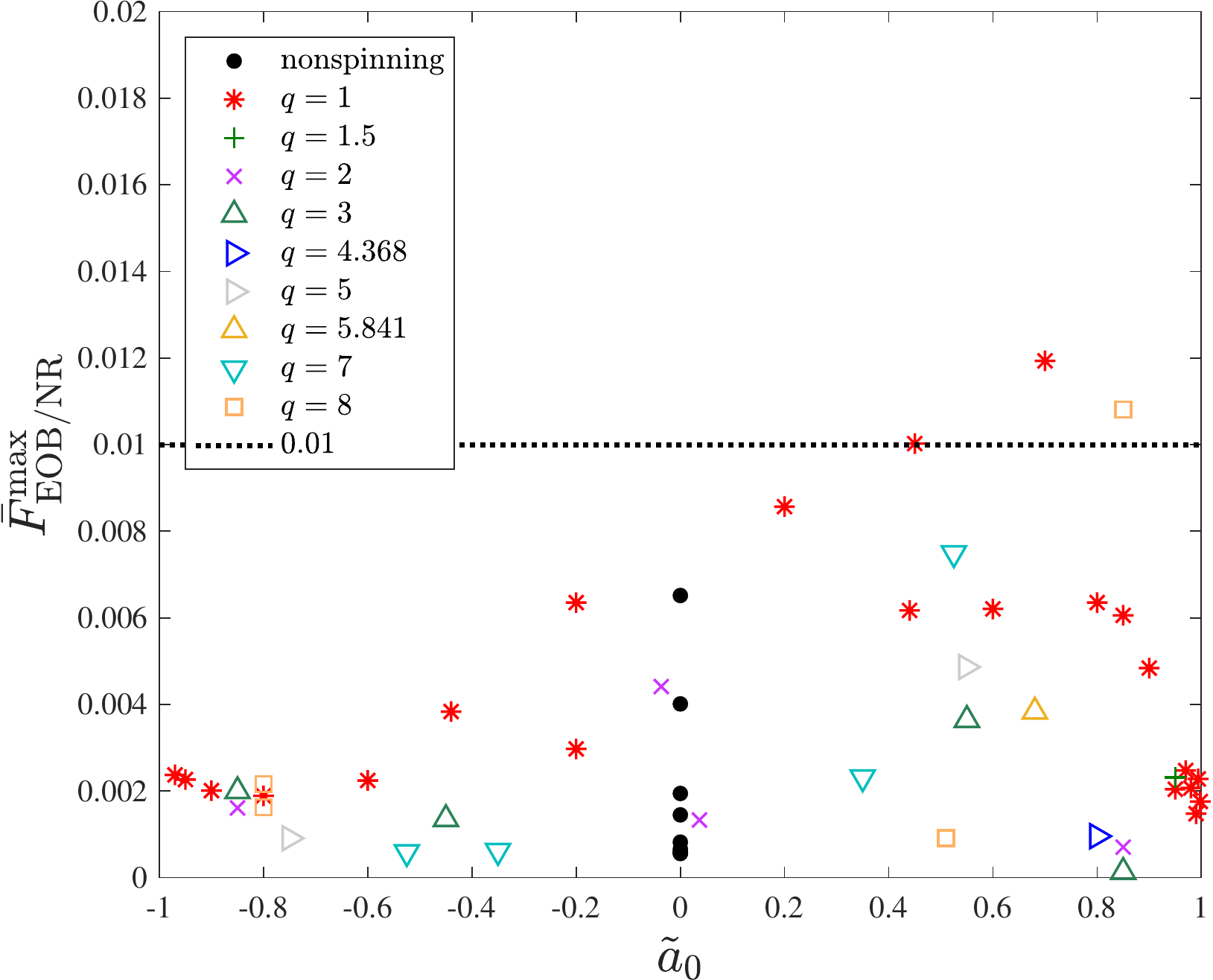}
\caption{\label{fig:barF_eobnr_circ} Quasi-circular configurations. Left panel: EOB/NR unfaithfulness for the $\ell=m=2$ mode.
A subset of this data is used to inform the $(a_6^c,c_3)$ functions as explained in the text. The phasing performance
is acceptably at most of the order of $0.01$, though it is less good than the standard quasi-circular \TEOBResumS{} model~\cite{Nagar:2020xsk}. 
The maximal values $\bar{F}_{\rm EOB/NR}^{\rm max}\equiv \max_M[\bar{F}_{\rm EOB/NR}(M)]$
are shown in the right panel (see also Tables~\ref{tab:equal_SXS}-\ref{tab:unequal_SXS}).
Note the two outliers slightly above the $0.01$ threshold.}
\end{figure*}
Following the standard procedure within the \TEOBResumS{} 
waveform paradigm~\cite{Nagar:2017jdw,Nagar:2018zoe,Nagar:2019wds,Rettegno:2019tzh,Nagar:2020pcj},
these few points are determined, by hand and without any  
automatized procedure, by simply inspecting the phase differences on 
time-domain phasing plots and checking that the EOB/NR phase difference
at merger is within the corresponding NR uncertainty\footnote{This is estimated taking the
difference between the highest and second highest resolution available for each dataset}. 
Due to the robustness of the theoretical  framework, these points are easily fitted to get
\begin{equation}
\label{eq:a6c_fit}
a_6^c(\nu) =\left(-0.50395 - 4.8547\nu+52.96\nu^2\right)e^{20.7013\nu} \ .
\end{equation}
Although this result is reliable and accurate for our purposes, we note that
the functional form is {\it not} quasi-linear as for \TEOBResumS{} case~\cite{Nagar:2020pcj}.
The exponential is needed to accomodate the rather large values of $a_6^c$ found
as $\nu\to 0.25$ and it is related to the fact that $\bar{\F}_r\neq 0$. 
To get a quantitative idea of the EOB/NR agreement yielded by our new
analytical choices, Fig.~\ref{fig:q3} reports two EOB/NR phasing comparisons 
corresponding to the $q=3$ case (EOB/NR comparison is discussed extensively in ~Ref.\cite{Damour:2007yf}). 
The left panel shows the phasing obtained  with the general, {\tt TEOBResumSGeneral} EOB model that 
is discussed here, while the right panel is obtained using \TEOBResumS{}.
The top panel shows the (relative) amplitude and phase differences, the middle panel the real part of the EOB (red) and
NR (black) waveforms and the bottom  panel the  gravitational frequencies. In the bottom panel is also shown 
twice the EOB orbital frequency 2$\Omega$. The dash-dotted vertical lines indicate the alignment frequency region, 
that adopts the standard procedure discussed in Ref.~\cite{Damour:2007yf}.
The NR dataset we chose to compare with is SXS:BBH:1221.
This is a 27-orbits long simulations that was {\it not} used at all 
to inform any of the two models. The phasing plots illustrate that the eccentric EOB model 
accumulates a secular dephasing with respect to the NR waveform that is of the order of 0.1~rad 
at merger (vertical dashed line), slightly larger than, but compatible with 
the  \TEOBResumS{} one.
As usual, the figure of merit of the quality of the EOB waveform is given by the
EOB/NR unfaithfulness weighted by the Advanced LIGO noise. Considering two 
waveforms $(h_1,h_2)$, the unfaithfulness is a function of the total mass $M$ 
of the binary and is defined as
\be
\label{eq:barF}
\bar{F}(M) \equiv 1-F=1 -\max_{t_0,\phi_0}\dfrac{\langle h_1,h_2\rangle}{||h_1||||h_2||},
\ee
where $(t_0,\phi_0)$ are the initial time and phase. We used $||h||\equiv \sqrt{\langle h,h\rangle}$,
and the inner product between two waveforms is defined as 
$\langle h_1,h_2\rangle\equiv 4\Re \int_{f_{\rm min}^{\rm NR}(M)}^\infty \tilde{h}_1(f)\tilde{h}_2^*(f)/S_n(f)\, df$,
where $\tilde{h}(f)$ denotes the Fourier transform of $h(t)$, $S_n(f)$ is the zero-detuned,
high-power noise spectral density of Advanced LIGO~\cite{aLIGODesign_PSD} and
$f_{\rm min}^{\rm NR}(M)=\hat{f}^{\rm NR}_{\rm min}/M$ is the initial frequency of the
NR waveform at highest resolution, i.e. the frequency measured after the junk-radiation
initial transient. Waveforms are tapered in the time-domain so as to reduce high-frequency 
oscillations in the corresponding Fourier transforms. The EOB/NR unfaithfulness is addressed as 
$\bar{F}_{\rm EOB/NR}$.
We will also consider $\bar{F}_{\rm NR/NR}$, as Eq.~\eqref{eq:barF} computed between the
highest and second highest resolution waveforms available, that we will quote as an indication
of the NR uncertainty (see also Sec.III~B of Ref.\cite{Nagar:2020pcj}).
When computing this $\bar{F}_{\rm EOB/NR}$ for SXS:BBH:1221, we obtain
$\bar{F}^{\rm max}_{\rm EOB/NR}\equiv \max_M\left[\bar{F}_{\rm EOB/NR}(M)\right]=0.14\%$, 
that, though a satisfactory value, is one order of magnitude larger than the value 
$\bar{F}^{\rm max}_{\rm EOB/NR}=0.0157\%$ obtained with \TEOBResumS{}
(see Table~XIX of~\cite{Nagar:2020pcj}) corresponding to the time-domain phasing comparison 
in the right panel of Fig.~\ref{fig:q3}. This is the typical case for the nonspinning NR datasets 
we have considered (see Table~\ref{tab:equal_SXS}). For example, for
SXS:BBH:0180 we obtain $\bar{F}^{\rm max}_{\rm EOB/NR}=0.65\%$ to be contrasted
with $\bar{F}^{\rm max}_{\rm EOB/NR}=0.0873\%$ obtained with \TEOBResumS{}
(see Table XVIII of Ref.~\cite{Nagar:2020pcj}). The main technical motivation behind this 
difference is the fact that the eccentric model uses $\hat{\F}_r\neq 0$, while \TEOBResumS{}
was imposing by construction $\hat{\F}_r=0$. Even though, as pointed out in Appendix~\ref{sec:Fr_derivation},
the quasi-circular expression of $\hat{\F}_r$ that we are using has a relatively mild behavior towards 
merger that allows for a reasonably accurate NR-tuning of $a_6^c(\nu)$, the global model turns 
out to perform worse than when imposing $\hat{\F}_r=0$ by construction\footnote{This can be considered 
as a sort of gauge condition for the circularized dynamic~\cite{Bini:2012ji}.}. One has to remember, however, 
that all these statements are relative to an EOB conservative dynamics that employs a Pad\'e $(1,5)$ resummation 
of the $A$ potential, with the (resummed) $B$ (or $D$) function taken at 3PN accuracy. 
Given the effective character of the resummation procedures 
used (and in particular of the NR-tuning of the parameters) a priori one  cannot  exclude that either 
(i) a different resummation of the potentials or (ii) different PN truncations may eventually 
result in an increased flexibility of the model that could  yield
a closer EOB/NR agreement. These issues deserve detail and systematic 
investigations that will hopefully  be carried out in future work. 

The new functional form of $a_6^c(\nu)$ given by Eq.~\eqref{eq:a6c_fit} calls for a similarly
new determination of the effective spin-orbit parameter $c_3$. We do so using a set of NR data
that is slightly different from the one used in Ref.~\cite{Nagar:2020pcj} so to improve the
robustness of the model in certain corners of the parameter space. The NR datasets used 
are listed in Table~\ref{tab:c3}. Following Ref.~\cite{Nagar:2020pcj}, for each dataset we 
report  the value of $c_3^{\rm first\;guess}$ obtained by comparing
EOB and NR phasing so that the accumulated phase difference is of the order of the NR 
uncertainty (and/or so that consistency between NR and EOB frequencies around merger 
is achieved as much as possible). Similarly to the case of $a_6^c$, the robustness of the model
allows us to efficiently do this by hand without any automatized procedure. 
The $c_3^{\rm first\;guess}$ values of Table~\ref{tab:c3} 
are fitted with a global function of the spin variables $\tilde{a}_i\equiv S_i/(m_i M)$ with $i=1,2$
of the form
\begin{align}
  \label{eq:c3fit}
%& c_3(\tilde{a}_1,\tilde{a}_2,\nu)=
%p_0\dfrac{1+n_1\tilde{a}_0+n_2\tilde{a}_0^2+n_3\tilde{a}_0^3+n_4\tilde{a}_0^4}{1+d_1\tilde{a}_0}\nonumber\\
%&+ p_1 \tilde{a}_0\nu\sqrt{1-4\nu}+ p_2\left(\tilde{a}_1-\tilde{a}_2\right)\nu^2 + p_3 \tilde{a}_0^2\nu\sqrt{1-4\nu},
& c_3(\tilde{a}_1,\tilde{a}_2,\nu)=
p_0\dfrac{1+n_1\tilde{a}_0+n_2\tilde{a}_0^2+n_3\tilde{a}_0^3+n_4\tilde{a}_0^4}{1+d_1\tilde{a}_0}\nonumber\\
&+ p_1 \tilde{a}_0\nu\sqrt{1-4\nu}+ p_2\left(\tilde{a}_1-\tilde{a}_2\right)\nu^2 + p_3 \tilde{a}_0\nu^2\sqrt{1-4\nu},
\end{align}
where $\tilde{a}_0\equiv \tilde{a}_1+\tilde{a}_2$ and the functional form is the same of previous works\footnote{Note that this function is not 
symmetric for exchange of $1\leftrightarrow2$. This can create an ambiguity for $q=1$, so that the value 
of $c_3$ for $(1,0.6,0.4)$ is in fact different from the one for $(1,0.4,0.6)$. In fact, our convention and 
implementations are such that for $q=1$, $\chi_1$ is {\it always} the largest spin.}. 
This term helps in improving the fit flexibility as the mass ratio increases. The fitting coefficients read
\begin{align}
p_0 &=  35.482253,\\ 
 n_1 &=  -1.730483,\\ 
 n_2 &=   1.144438,\\
 n_3 &=   0.098420,\\ 
 n_4 &=  -0.329288,\\ 
 d_1 &=  -0.345207,\\
p_1  &=  244.505,\\
p_2 &= 148.184,\\
p_3  &= -1085.35.
%p_0 &=  42.720039, \\
%n_1 &=  -0.710937, \\
%n_2 &=  -0.142503, \\
%n_3 &=   0.183764, \\ 
%n_4 &=   0.025030, \\ 
%d_1 &=   0.331651, \\
%p_1 & = 100.743,\\
%p_2 & =  -5.01934,\\
%p_3 & =  -61.6306.
\end{align}
We finally test the performance of Eq.~\eqref{eq:c3fit} by computing  $\bar{F}_{\rm EOB/NR}$
over a set of 43 spinning configurations that is only partially overlapping with the one used
to determine it. The configurations are chosen so to efficiently cover the parameter space,
in particular including its difficult corners (i.e. large mass ratio and large spins). This choice 
is motivated by the fact that previous work showed that most of the 595, spin-aligned, 
SXS simulations available are in fact  redundant among themselves and do not bring 
additional information from the 
EOB/NR testing point of view~\cite{Nagar:2020pcj}. The behavior of $\bar{F}_{\rm EOB/NR}(M)$  
is shown in Fig.~\ref{fig:barF_eobnr_circ},  that also includes 10 nonspinning datasets.
All values of  $\bar{F}^{\rm max}_{\rm EOB/NR}$ 
are listed in Tables~\ref{tab:equal_SXS}-\ref{tab:unequal_SXS} and are also plotted
in the right panel of Fig.~\ref{fig:barF_eobnr_circ} versus $\tilde{a}_0$, so to have an 
immediate perception of the model performance all over the parameter space. 
The outliers, slightly above $1\%$, occur for $0.4\lesssim \tilde{a}_0\lesssim 0.8$ for $q=1$ 
as well as for large values of $\tilde{a}_0$ when $q=8$. Although it is probably 
further improvable, we believe that this level of EOB/NR agreement 
is sufficient for our current purposes.

%==============
% q=6 - nonspinning
%==============
\begin{figure*}[t]
\center
\includegraphics[width=0.4\textwidth]{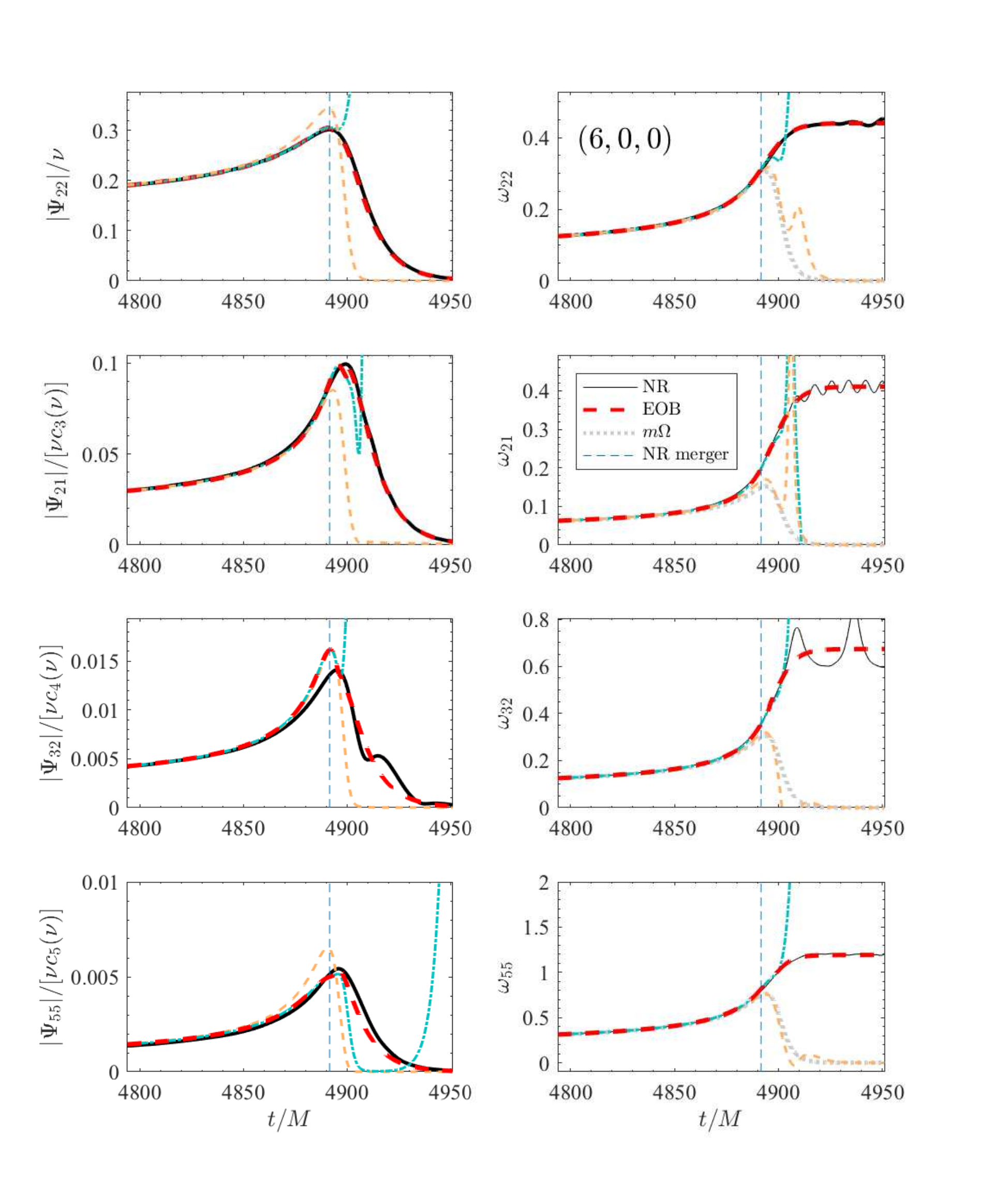}
\includegraphics[width=0.4\textwidth]{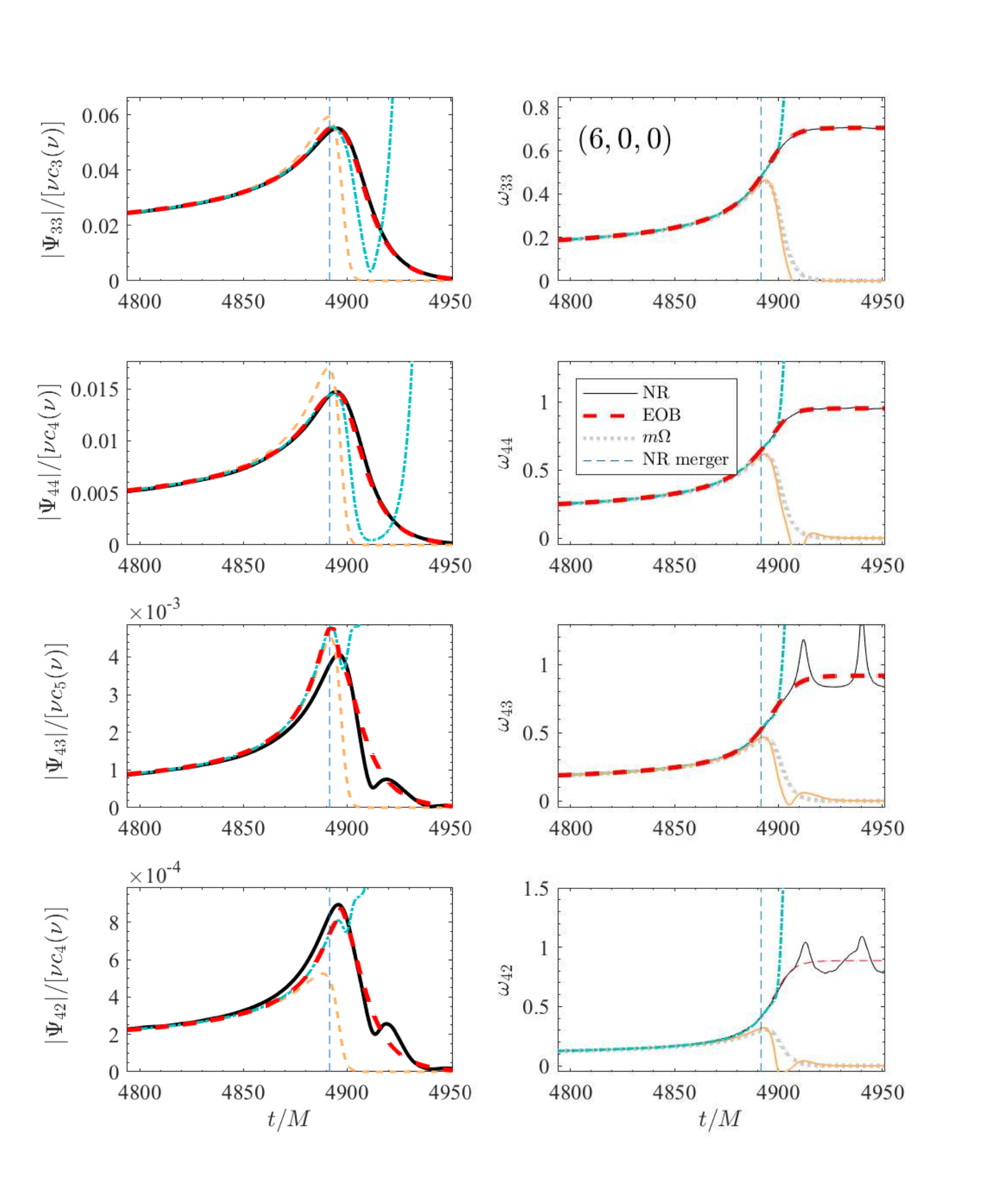}
\caption{\label{fig:merger_q6s0} EOB/NR multipolar comparison of amplitude and frequency for $(q,\chi_1,\chi_2)=(6,0,0)$,
referring to SXS:BBH:0166 dataset.
The grey, dotted, line reports $m\Omega$, where $\Omega$ is the EOB orbital frequency.
The picture also displays the EOB analytical waveform (orange online) and the NQC completed one (light-blue online). 
The vertical line indicates the NR merger time, i.e. the peak of the $(2,2)$ amplitude.}
\end{figure*}
%==============
% (3,0.3,0.3)
%==============
\begin{figure*}[t]
\center
\includegraphics[width=0.35\textwidth]{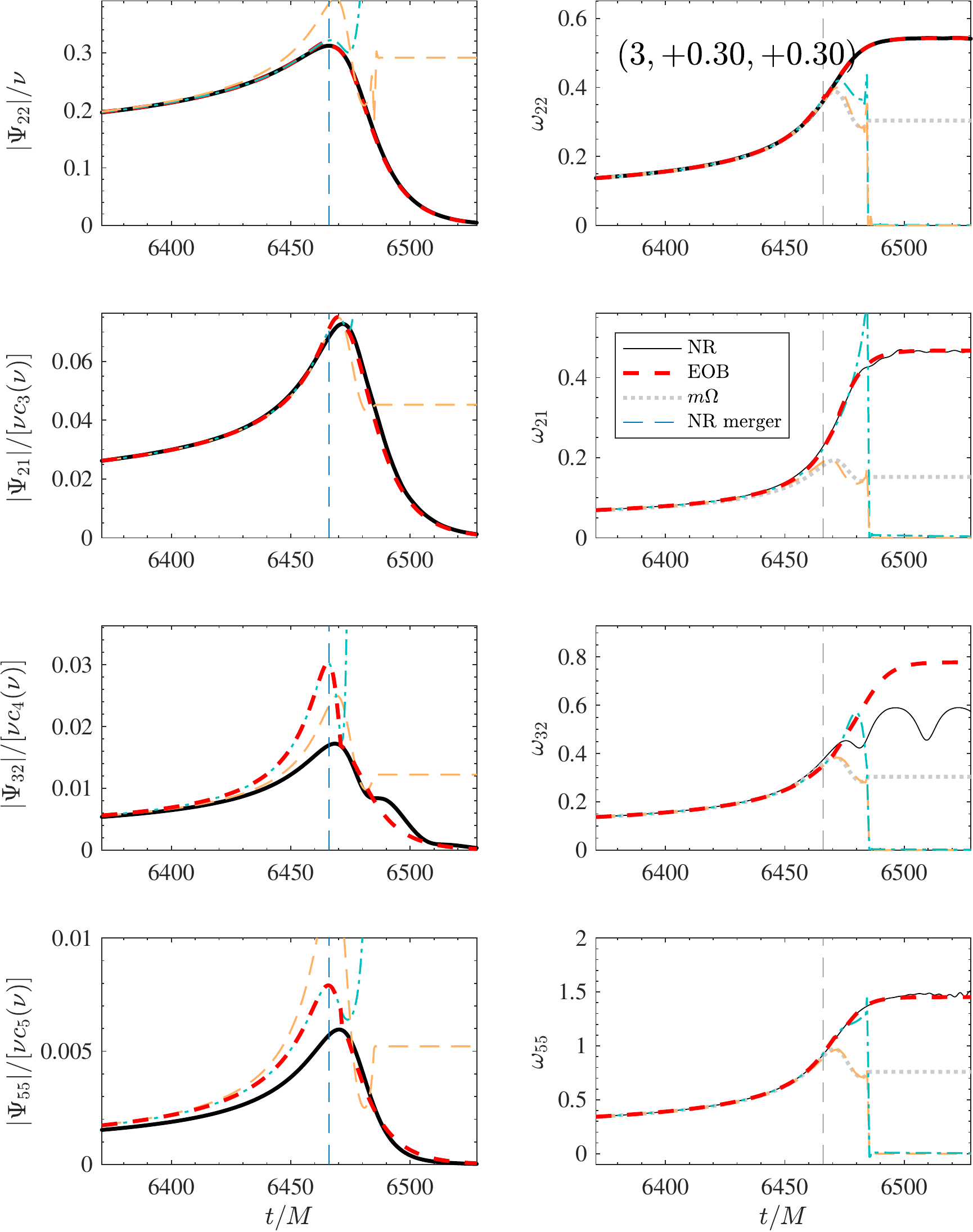}
\hspace{10mm}
\includegraphics[width=0.35\textwidth]{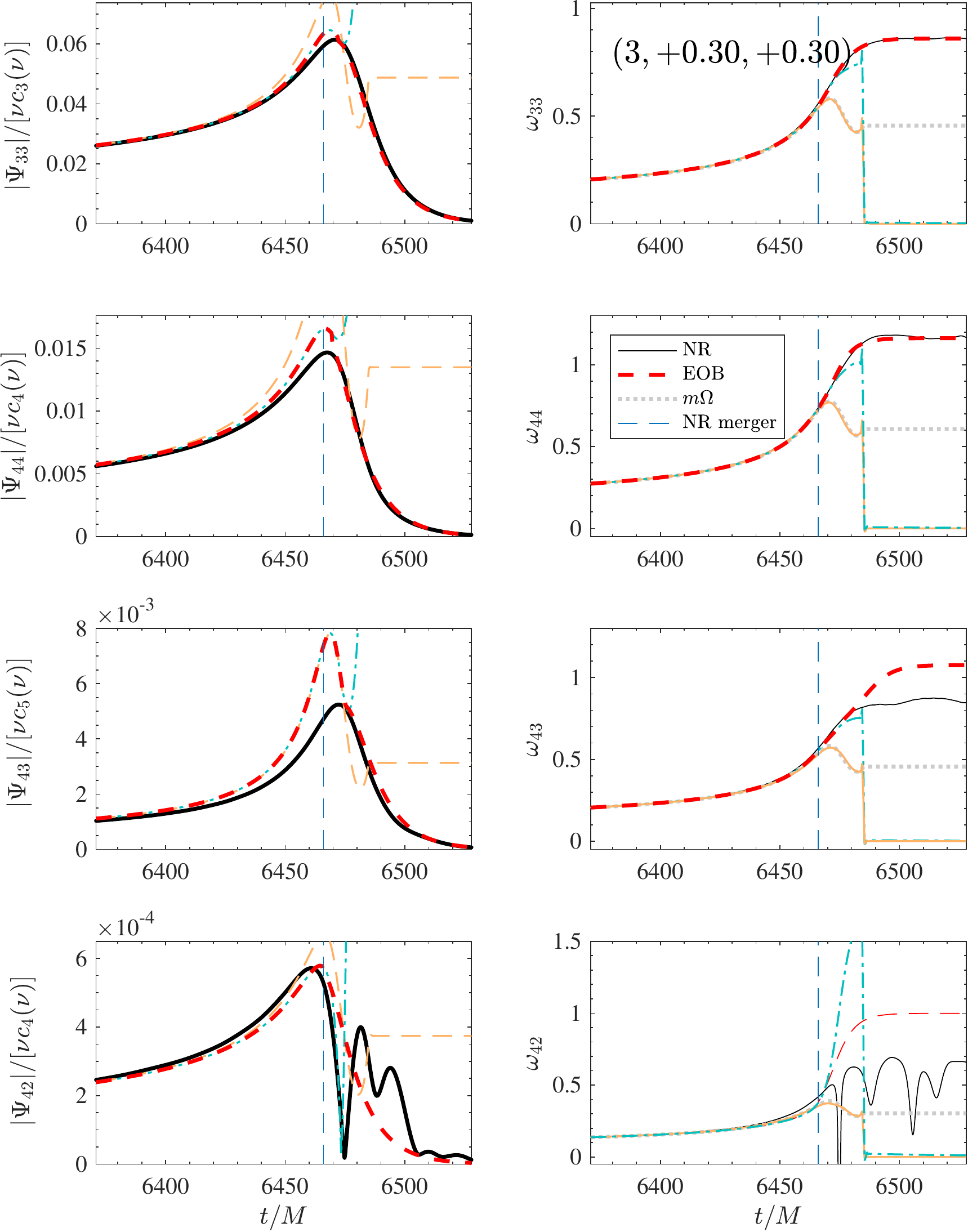}
\caption{\label{fig:merger_spin} EOB/NR multipolar comparison of amplitude and frequency for $(q,\chi_1,\chi_2)=(3,+0.30,+0.30)$,
referring to SXS:BBH:2155 dataset. The grey, dotted, line reports $m\Omega$, where $\Omega$ is the EOB orbital frequency.
The picture also displays the EOB inspiral analytical waveform (orange online) and the NQC completed one (light-blue online). 
The vertical line indicates the NR merger time, i.e. the peak of the $(2,2)$ amplitude. Note the EOB amplitude of some 
subdominant modes is overestimated towards merger.}
\end{figure*}

\subsection{Higher modes}
\label{sec:HM_circ}
The performance of the higher modes is similar to the one of {\tt TEOBResumS}, although we have 
seen that the robustness of the waveform completion through the NQC correction factor tends
to decrease as the mass ratio or the individual spins are increased. This problem was already
pointed out in Ref.~\cite{Nagar:2019wds}, and it is known to appear when the peak of the considered 
subdominant  multipole is significantly delayed with respect to the $(2,2)$ one. This is typically the
case for modes with $m\neq \ell$ when spins are large and anti-aligned with the angular momentum.
Therefore, the modes with $\ell=m$ are, on average, the most robust through merger and ringdown all 
over the parameter space. 
However, when spins are large  and anti-aligned with the angular momentum, the dynamics is such to 
prevent the NQC factor to work correctly already for modes like $(3,3)$ and $(4,4)$. 
This is for example the case of $(3,-0.85,-0.85)$, while for $(3,-0.50,-0.50)$ the application of the 
NQC factor yields a perfectly sound multipolar waveform.
The other, most relevant, subdominant modes $(2,1)$, $(3,2)$, $(4,3)$, $(4,2)$ are certainly always 
reliable through merger and ringdown when the spin magnitudes are small (i.e. up to approximately $|\chi_i|\simeq 0.5$) 
and $q\lesssim 6$, but it might not be the case for larger, negative, spins.
To highlight a few cases where everything falls correctly into place, 
Fig.~\ref{fig:merger_q6s0} refers to EOB/NR comparisons of amplitudes and frequencies 
for configuration $(q,\chi_1,\chi_2)=(6,0,0)$. For completeness of information, the figure also displays the
EOB quantities without the NQC factor (orange online) and without the ringdown attachment (light-blue online).
The vertical dashed line indicates the NR merger location, i.e. the peak of the $(2,2)$ NR waveform. The
frequency panel also displays, as a gray dotted line, the time evolution of the orbital frequency.
Note that the behavior of both the orange and light-blue lines after the merger point is unphysical, though
we prefer to keep it as additional information. As usual, the behavior of the bare EOB waveform, both
amplitude and frequency, is already highly consistent with the NR waveform, without need of additional
tuning, up to about $50M$ (or less) before merger.  The differences in the ringdown part, especially
large for $(3,2)$, $(4,3)$ and $(4,2)$ come from the absence of mode-mixing~\cite{London:2014cma,Taracchini:2014zpa,London:2018nxs}
in the ringdown on the EOB side. 

As a spinning example, Fig.~\ref{fig:merger_spin} illustrates the same multipoles as Fig.~\ref{fig:merger_q6s0} 
but for $(q,\chi_1,\chi_2)=(3,+0.30,+0.30)$. Qualitatively, the most relevant visual difference between
Fig.~\ref{fig:merger_q6s0} and Fig.~\ref{fig:merger_spin} on the EOB side concerns the amplitude of
modes $(3,2)$ and $(4,3)$ that is larger than the corresponding NR one. There are other configurations 
where  the NQC-corrected EOB waveform performs similarly or worse, possibly affecting also the frequency 
that  becomes very large and unphysical before the ringdown attachment. This effect is more and more
marked (until it eventually gets unphysical) the more the peak of the considered NR multipole is displaced 
on the right with respect to the $(2,2)$ one.  A precise quantification of the limits of reliability and robustness 
of the current EOB multipolar model is beyond the scope of this work. Qualitatively, however, the considerations 
already driven in Ref.~\cite{Nagar:2019wds} with special focus on the $(2,1)$ mode for large, anti-aligned spins
still hold here, with no new conceptual findings. A quick analysis that we performed for several mass ratios 
and spin values points, again, to the fact that the Achilles' heel of the EOB models based on the {\tt TEOBResumS} 
paradigm resides in the  NQC determination procedure. We think that the procedure of informing the multipolar  
NQC parameters with NR information extracted {\it after} the multipole peak as it is done now should be revised, 
since it necessarily relies  on a part of the EOB dynamics (i.e. beyond the location of the usual merger) that, 
at the moment, is not fully under control for certain configurations~\cite{Nagar:2018zoe}.  

\section{Eccentric configurations}
\label{sec:ecc}
%--------------------------
% SXS eccentric simulations
%--------------------------
\begin{table*}[t]
   \caption{\label{tab:SXS} SXS simulations with eccentricity analyzed in this work. From left to right: the 
   ID of the simulation; the mass ratio $q\equiv m_1/m_2\geq 1$ and the individual dimensionless 
   spins $(\chi_1,\chi_2)$; the time-domain NR phasing uncertainty at merger $\delta\phi^{\rm NR}_{\rm mrg}$; 
   the estimated NR eccentricity at first apastron $e_{\omega_a}^{\rm NR}$; 
   the NR frequency of first apastron $\omega_{a}^{\rm NR}$; 
   the initial EOB eccentricity $e^{\rm EOB}_{\omega_a}$ and apastron frequency $\omega_{a}^{\rm EOB}$ used to start the EOB evolution; 
   the maximal NR unfaithfulness uncertainty, $\bar{F}^{\rm max}_{\rm NR/NR}$ and the maximal EOB/NR unfaithfulness, 
   $\bar{F}_{\rm EOB/NR}^{\rm max}$. }
   \begin{center}
     \begin{ruledtabular}
\begin{tabular}{c| c  c c c c |l l| c c} 
  $\#$ & id & $(q,\chi_1,\chi_2)$ & $\delta\phi^{\rm NR}_{\rm mrg}$[rad]& $e^{\rm NR}_{\omega_a}$ & $\omega_a^{\rm NR}$ &$e^{\rm EOB}_{\omega_a}$ & $\omega_{a}^{\rm EOB}$ & $\bar{F}_{\rm NR/NR}^{\rm max}[\%]$  &$\bar{F}_{\rm EOB/NR}^{\rm max}[\%]$ \\
  \hline
  \hline
1 & SXS:BBH:1355 & $(1,0, 0)$ & $+0.92$ & 0.0620  & 0.03278728 & 0.0890    & 0.02805750  & 0.012 & 0.96\\
2 & SXS:BBH:1356 & $(1,0, 0)$& $+0.95$ & 0.1000 &  0.02482006  & 0.15038  & 0.019077  & 0.0077 &0.91\\
3 & SXS:BBH:1358 & $(1,0, 0)$& $+0.25$ & 0.1023 & 0.03108936 & 0.18078   & 0.021238 & 0.016 &1.07\\
4 & SXS:BBH:1359 & $(1,0, 0)$& $+0.25$  & 0.1125 & 0.03708305 & 0.18240   & 0.02139 & 0.0024&0.88 \\
5 & SXS:BBH:1357 & $(1,0, 0)$& $-0.44$  & 0.1096 & 0.03990101 & 0.19201   & 0.01960 & 0.028&0.88\\
6 & SXS:BBH:1361 & $(1,0, 0)$& +0.39    & 0.1634 & 0.03269520  & 0.23557   & 0.0210   & 0.057&1.090\\
7 & SXS:BBH:1360 & $(1,0, 0)$& $-0.22$ & 0.1604 & 0.03138220 & 0.2429  & 0.01959   &0.0094  &1.04\\
8 & SXS:BBH:1362 & $(1,0, 0)$& $-0.09$ & 0.1999 & 0.05624375 & 0.3019     & 0.01914 & 0.0098 &0.84\\
9 & SXS:BBH:1363 & $(1,0, 0)$& $+0.58$ & 0.2048 & 0.05778104 &  0.30479    & 0.01908 & 0.07 &1.04\\
10 & SXS:BBH:1364 & $(2,0, 0)$& $-0.91$ & 0.0518 &  0.03265995   & 0.08464    & 0.025231   & 0.049  &0.42\\
11 & SXS:BBH:1365 & $(2,0, 0)$& $-0.90$ & 0.0650 &  0.03305974   & 0.11015     & 0.023987 & 0.027&0.50\\
12 & SXS:BBH:1366 & $(2,0, 0)$& $-6\times 10^{-4}$ & 0.1109 & 0.03089493 & 0.1496   & 0.02580 &  0.017  &0.84 \\
13 & SXS:BBH:1367 & $(2,0, 0)$& $+0.60$ & 0.1102 & 0.02975257 & 0.15065    & 0.026025  & 0.0076  &0.50\\
14 & SXS:BBH:1368 & $(2,0, 0)$& $-0.71$ & 0.1043 & 0.02930360 &  0.14951  & 0.02527    & 0.026 &0.41\\
15 & SXS:BBH:1369 & $(2,0, 0)$& $-0.06$ & 0.2053 & 0.04263738 & 0.3134     & 0.01735  & 0.011&0.58\\
16 & SXS:BBH:1370 & $(2,0, 0)$& $+0.12$ & 0.1854 &  0.02422231 &  0.31445  & 0.016915  & 0.07&0.88\\
17 & SXS:BBH:1371 & $(3,0, 0)$& $+0.92$ & 0.0628 & 0.03263026  & 0.0912     & 0.029058   & 0.12  &0.39\\
18 & SXS:BBH:1372 & $(3,0, 0)$& $+0.01$& 0.1035 & 0.03273944 & 0.14915      & 0.026070 & 0.06  &0.32\\
19 & SXS:BBH:1373 & $(3,0, 0)$& $-0.41$ & 0.1028 & 0.03666911 & 0.15035    & 0.0253 & 0.0034 &0.23\\
20 & SXS:BBH:1374 & $(3,0, 0)$& $+0.98$ & 0.1956  & 0.02702594 & 0.31388   & 0.016946   & 0.067 &0.23\\
\hline
21 & SXS:BBH:89   & $(1,-0.50, 0)$         &  $\dots$  & 0.0469  & 0.02516870 & 0.07201    & 0.01779  & $\dots$ &0.60  \\
22 & SXS:BBH:1136 & $(1,-0.75,-0.75)$   &  $-1.90$ & 0.0777  &0.04288969 &0.12105      & 0.02728 & 0.074 &0.41  \\
23 & SXS:BBH:321  & $(1.22,+0.33,-0.44)$& $+1.47$ & 0.0527  & 0.03239001 &0.07621     & 0.02694 & 0.015  &0.71  \\
24 & SXS:BBH:322  & $(1.22,+0.33,-0.44)$& $-2.02$  & 0.0658  &  0.03396319 &0.0984       & 0.026895 & 0.016 &0.93  \\
25 & SXS:BBH:323  & $(1.22,+0.33,-0.44)$& $-1.41$ & 0.1033  & 0.03498377 &0.1438      & 0.02584 & 0.019 &0.77 \\
26 & SXS:BBH:324  & $(1.22,+0.33,-0.44)$& $-0.04$ & 0.2018  & 0.02464165 &0.29414        & 0.01894 & 0.098 &1.06\\
27 & SXS:BBH:1149 & $(3,+0.70,+0.60)$  &  $+3.00$  & 0.0371  &0.03535964 &$0.0623$   & $0.02664$ &0.025  &0.33\\
28 & SXS:BBH:1169 & $(3,-0.70,-0.60)$    &  $+3.01$ & 0.0364  &0.02759632 &$0.04895$     & $0.024285$ & 0.033 & 0.096     % 

 \end{tabular}
 \end{ruledtabular}
 \end{center}
 \end{table*}

\subsection{Numerical Relativity waveforms}
\label{sec:nr_error}
\subsubsection{NR eccentricity}
\begin{figure}[t]
\center
\includegraphics[width=0.45\textwidth]{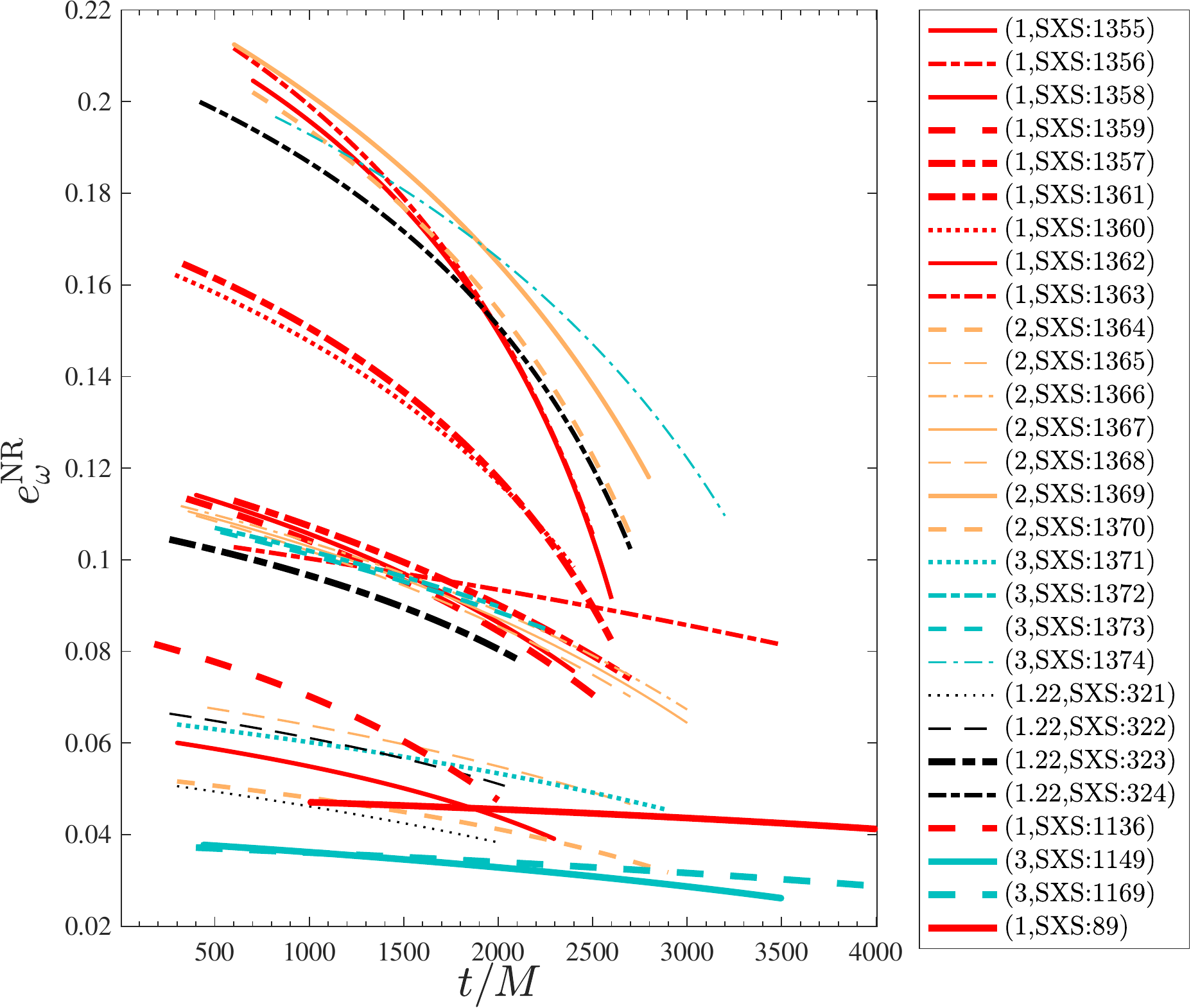}
\caption{\label{fig:NRecc} Time evolution of the NR eccentricity estimated using Eq.~\ref{eq:e_Omg} applied to the
GW frequency at apastron and periastron, Eq.~\eqref{eq:omg_ap}. The intial point of each curve corresponds to
the values at the first apastron, $e^{\rm NR}_{\omega_a}$, listed in Table~\ref{tab:SXS}.}
\end{figure}

Here we consider all the 28 eccentric SXS datasets currently
public, the nonspinning ones being published in Ref.~\cite{Hinder:2017sxy}.
As done in Ref.~\cite{Chiaramello:2020ehz}, it is instructive to evaluate the initial eccentricity of each 
NR simulation using some eccentricity estimator, so to obtain a simple
intuition about the waveform properties\footnote{Note however that this eccentricity 
is not going to play any role within the EOB model.}.
Inspired by previous work, Ref.~\cite{Ramos-Buades:2019uvh} introduced an eccentricity estimator 
\begin{align}
\label{eq:e_Omg}
    e_{\Omega}(t)=\frac{\sqrt{\Omega^{\rm NR}_p}-\sqrt{\Omega_a^{\rm NR}}}{\sqrt{\Omega_p}+\sqrt{\Omega_a}} \ ,
\end{align}
that uses the NR {\it orbital} frequencies $\Omega^{\rm NR}_a$ and $\Omega^{\rm NR}_p$ 
at apastron and periastron extracted from the puncture trajectories. 
Reference~\cite{Ramos-Buades:2019uvh} made this choice because
the gravitational wave frequency computed from the NR simulations 
(obtained using the Einstein Toolkit~\cite{Zilhao:2013hia}  and not the {\tt SpEC} code) 
considered there was too noisy to be used in Eq.~\eqref{eq:e_Omg}.
By contrast, here (as it was previously done in~\cite{Chiaramello:2020ehz}) we find 
that it is actually possible to apply such eccentricity estimator directly on the GW frequency  
$\omega_{22}^{\rm NR}$ of the SXS simulations. 
To do so efficiently and more reliably, however, we  need to remove 
some high frequency noise present at the very beginning of the simulations just after the 
junk radiation, especially in those with larger eccentricities. The noise removal 
is simply done using a Savitzy-Golay filter implemented in the commercial 
software {\tt Matlab}. Such straightforward  procedure allows us to cleanly identify the maxima 
(periastron, $\omega_{p}^{\rm NR}$) and the minima (apastron, $\omega_a^{\rm NR}$) 
of $\omega_{22}^{\rm NR}$. These quantities are then fitted 
versus time using a rational function of the form
\begin{align}
\label{eq:omg_ap}
    \omega_{a,p}^{\rm NR} = c_{a,p}\, \frac{1+n_{a,p} t}{1+d_{a,p} t} \ .
\end{align}
Our estimate of the NR eccentricity, $e_\omega^{\rm NR}$, is then obtained
from Eq.~\eqref{eq:e_Omg} by replacing $\Omega^{\rm NR}_{a,p}$ with $\omega^{\rm NR}_{a,p}$.
This procedure is applied to all NR simulations of Table~\ref{tab:SXS} and Fig.~\ref{fig:NRecc} shows
the time evolution of the corresponding $e^{\rm NR}_\omega$. The first point of each curve corresponds
to the frequency of the first apastron, $\omega_a^{\rm NR}$. The third column of Table~\ref{tab:SXS} precisely 
lists these values $e^{\rm NR}_{\omega_a}$, while the fourth column displays the corresponding frequency $\omega_a^{\rm NR}$.

%=======================
% Fig.6: eccentricity computation
%=======================

%======
% Fig.7
%======
\begin{figure}[t]
\center
\includegraphics[width=0.4\textwidth]{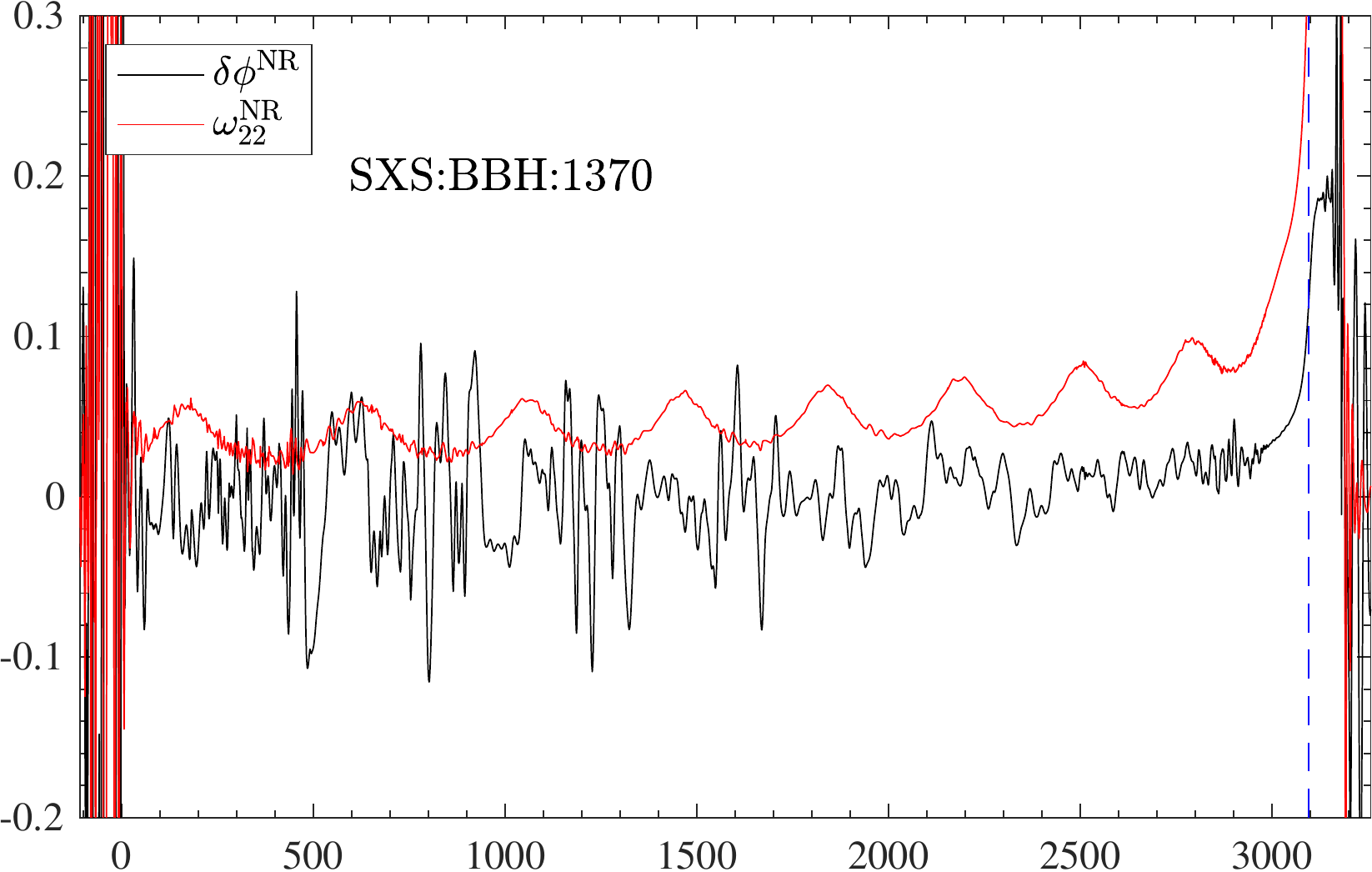}\\
\vspace{5mm}
\includegraphics[width=0.4\textwidth]{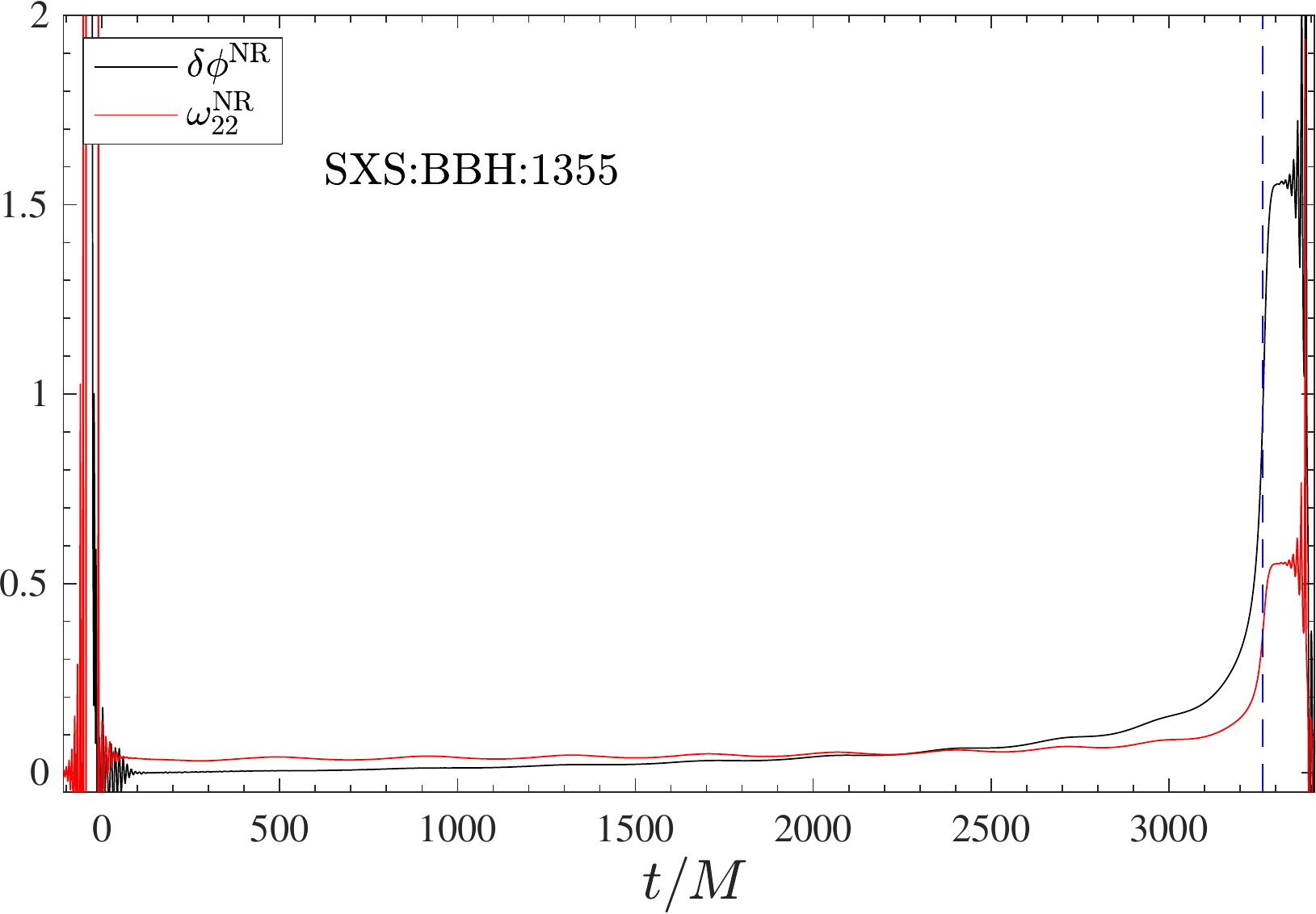}
\caption{\label{fig:delta_phi}Time domain estimate of the nominal NR phase uncertainty $\delta\phi^{\rm NR}$ obtained
taking the phase difference between the two highest resolutions available. 
We also report the NR frequency of the $\ell=m=2$ mode $\omega_{22}^{\rm NR}$.
The vertical dashed line corresponds to the
merger of the dataset with the highest resolution. For SXS:BBH:1370 the large oscillations
present during the inspiral suggest that the quality of the waveform is lower than for the SXS:BBH:1355 case.
Note however the clean evolution of $\delta\phi^{\rm NR}$ towards merger and the corresponding rather small
value of $\delta\phi^{\rm NR}_{\rm mrg}$. }
\end{figure}

%=================
% NR-NR unfaithfulness
%================= 
\begin{figure}[t]
\center
\includegraphics[width=0.42\textwidth]{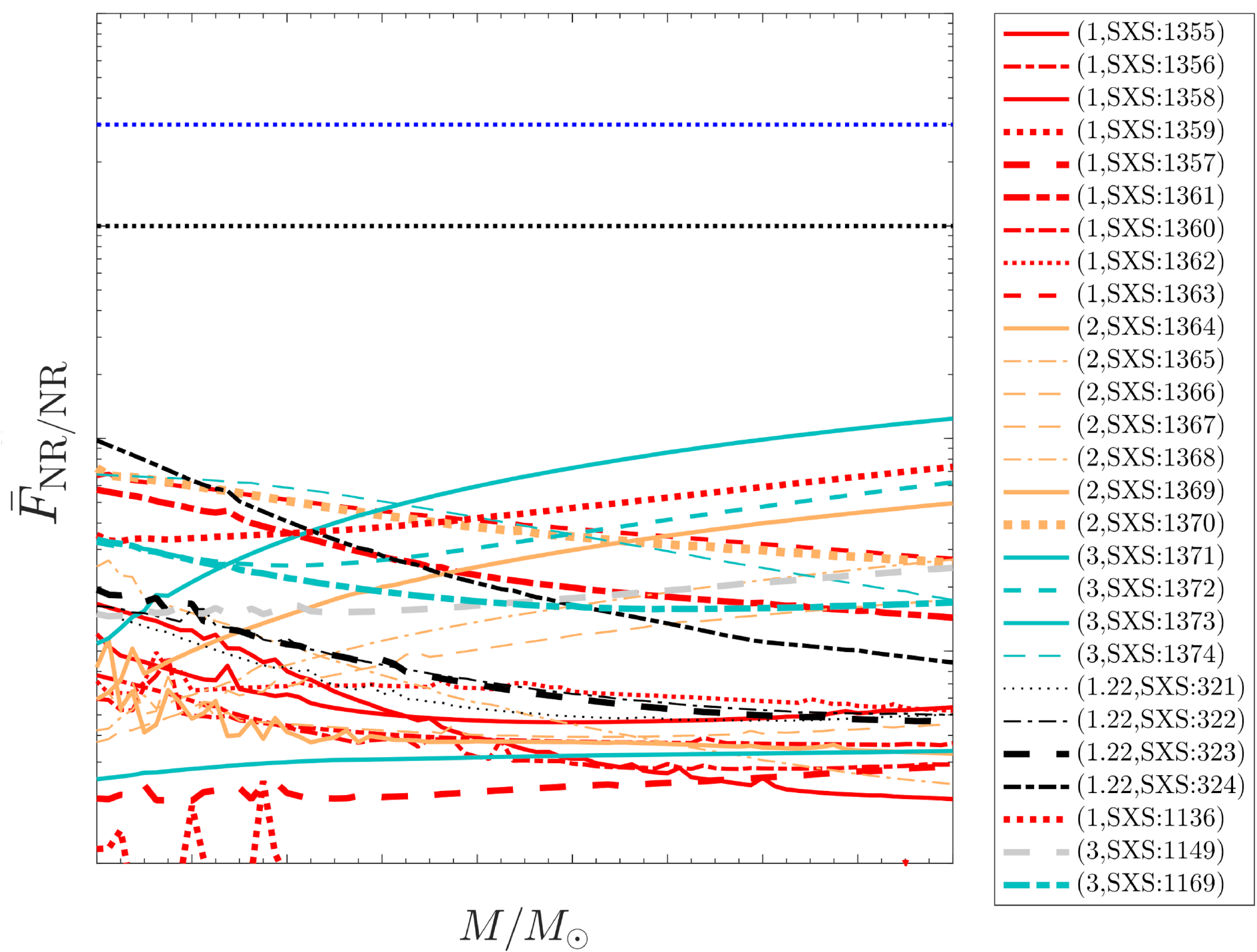}
\caption{\label{fig:barF_nrnr}Estimate of the NR uncertainty $\bar{F}_{\rm NR/NR}$ 
computing $\bar{F}$ from Eq.~\eqref{eq:barF}  between the highest and second highest resolution level for 
each NR simulation.The horizontal lines mark the $0.03$ and $0.01$ values.
Datasets SXS:BBH:324, SXS:BBH:1361 and SXS:BBH:1369 have the 
largest uncertainties during the inspiral.}
\end{figure}

\begin{figure}[t]
\center
\includegraphics[width=0.4\textwidth]{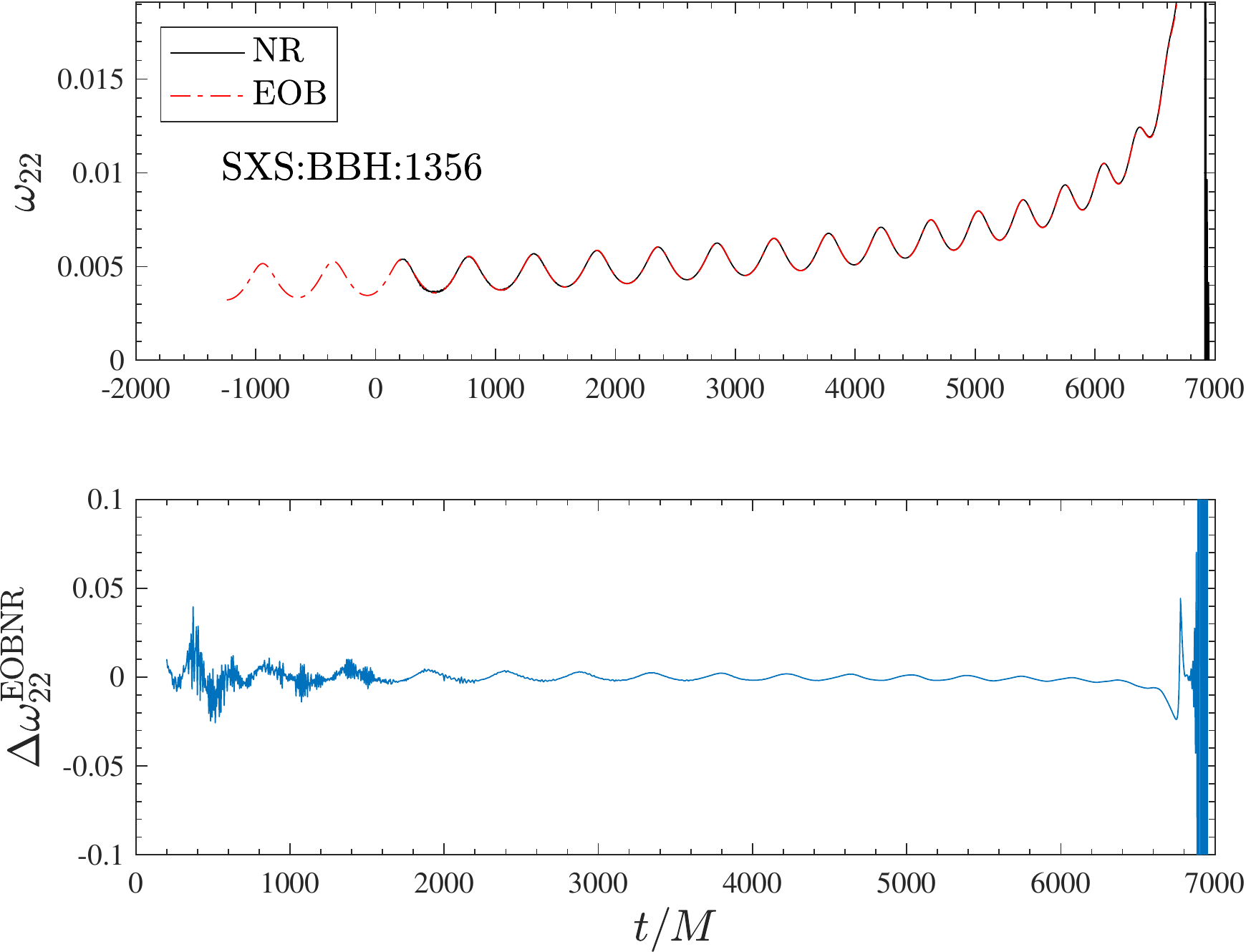}\\
\vspace{5mm}
\includegraphics[width=0.4\textwidth]{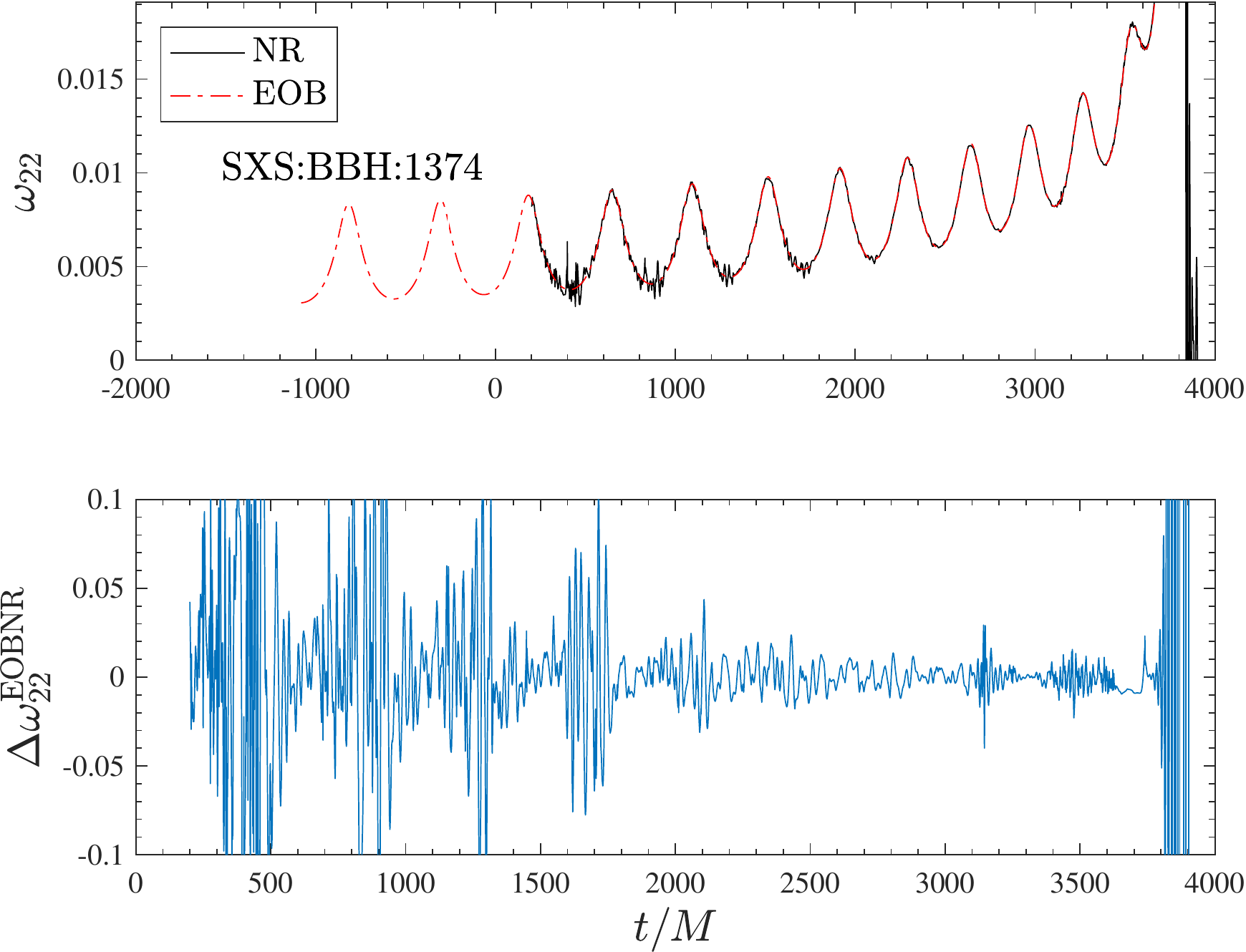}
\caption{\label{fig:sxs_Domg}EOB/NR gravitational wave frequency, $\omega_{22}^{\rm EOB/NR}$, agreement for 
two illustrative SXS datasets with smaller eccentricity (top panels) and larger eccentricity (bottom panels) 
obtained suitably varying the initial parameters $(\omega_a^{\rm EOB},e^{\rm EOB})$. The parameters are chosen
so that the fractional phase difference $\Delta\omega_{22}^{\rm EOBNR}\equiv \omega_{22}^{\rm EOB}-\omega_{22}^{\rm NR}$ 
averages zero for the longest time interval (that excludes merger and ringdown) and with small amplitude, secular, oscillations. }
\end{figure}

%====================================
% Amplitude, phasing and frequency comparison
%====================================
\begin{figure*}[t]
\center
\includegraphics[width=0.4\textwidth]{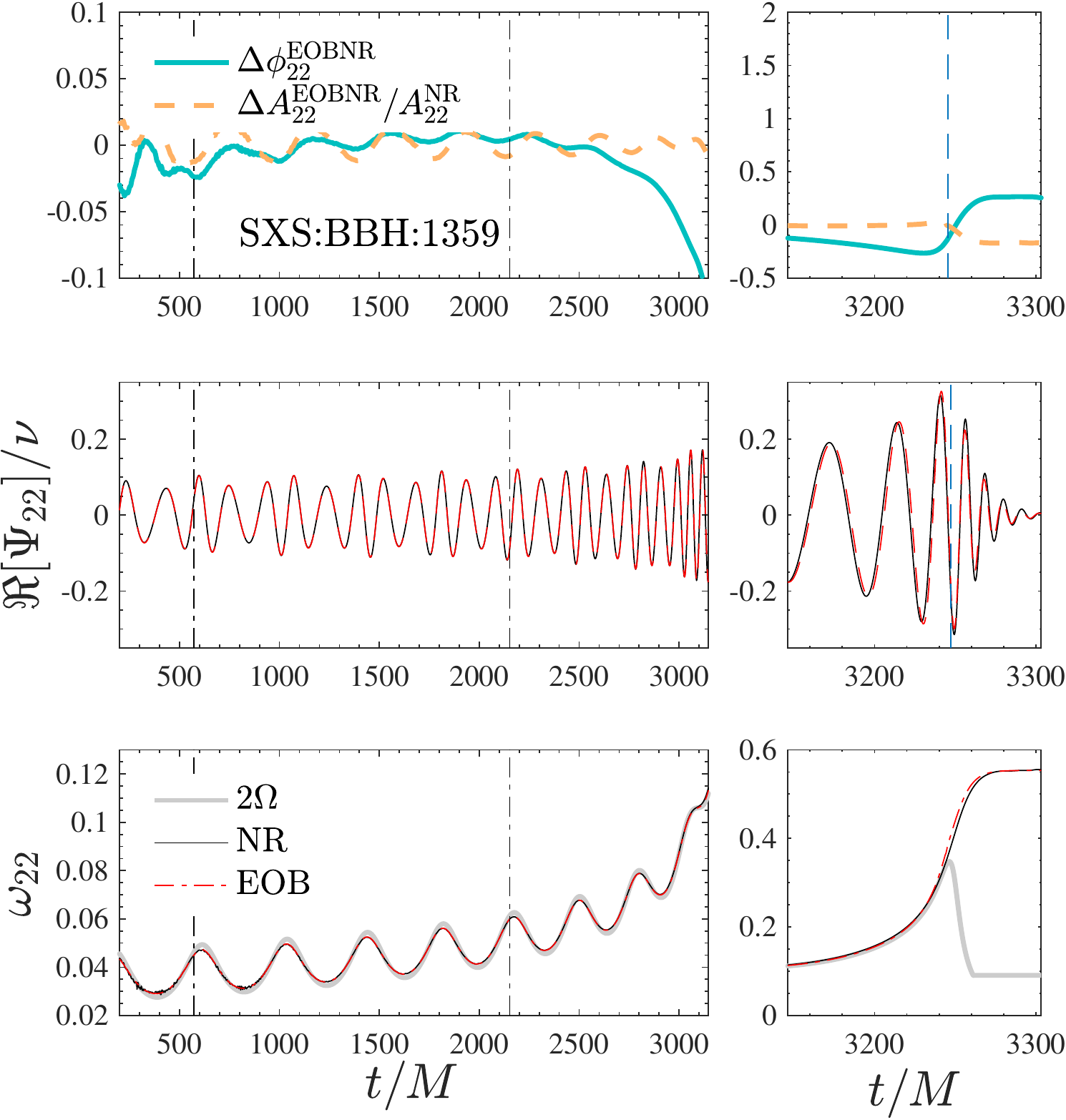}
\qquad
\includegraphics[width=0.4\textwidth]{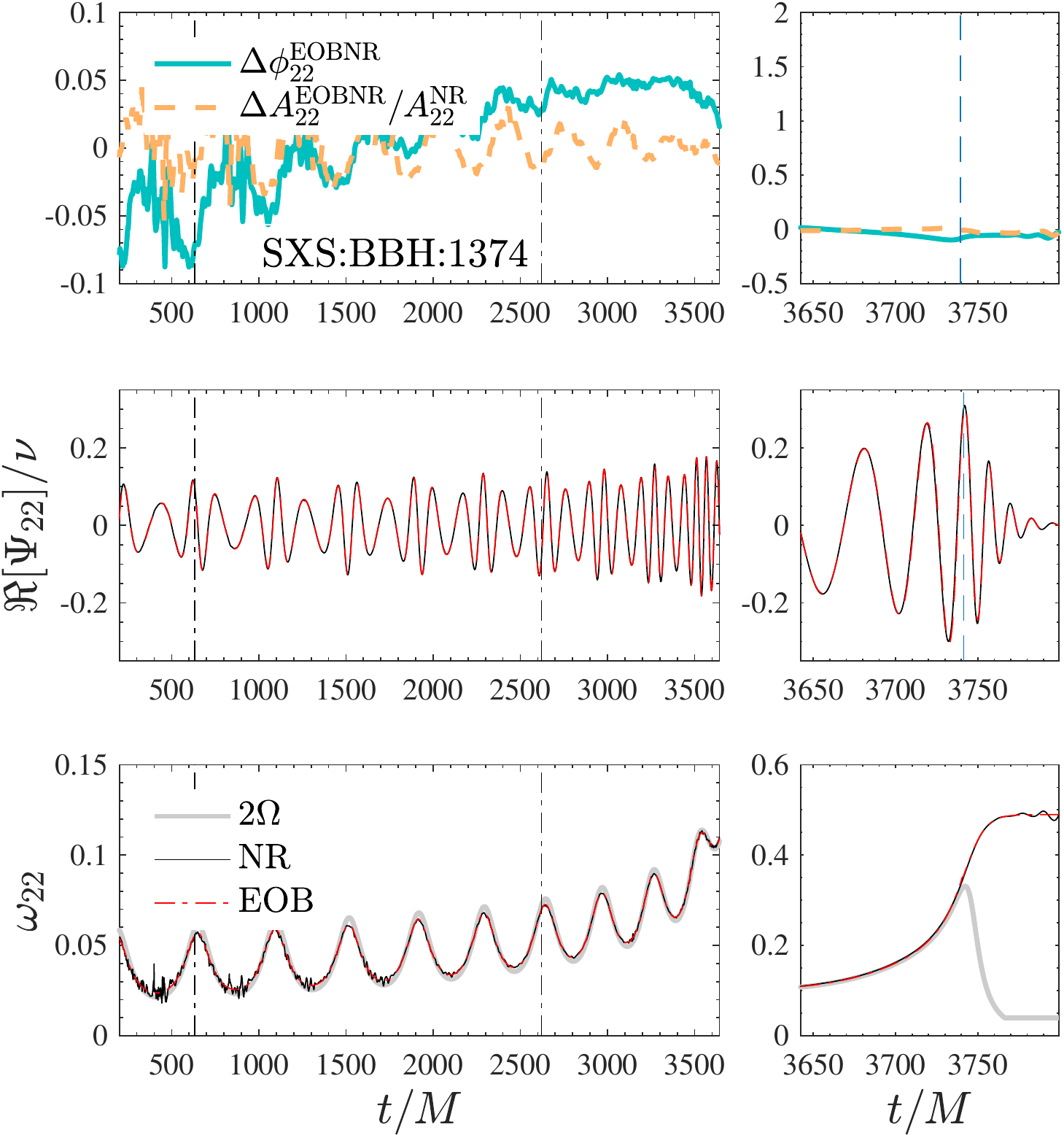}
\caption{\label{fig:sxs_1359} EOB/NR time-domain phasing comparison for nonspinning configurations.
Left panels: SXS:BBH:1359, $q=1$, $e_{\omega_a}^{\rm NR}\simeq 0.11$. Right panels: SXS:BBH:1374, 
$q=3$, $e_{\omega_a}^{\rm NR}\simeq 0.2$. The vertical dash-dotted lines indicate the alignment interval,
while the dashed vertical line the merger time. With $\Omega$ we address the EOB orbital frequency (gray online).
Note the excellent EOB/NR phasing agreement during the eccentric inspiral.}
\end{figure*}
%===============
% EOB/NR Unfaithfulness
%===============
\begin{figure}[t]
\center
\includegraphics[width=0.45\textwidth]{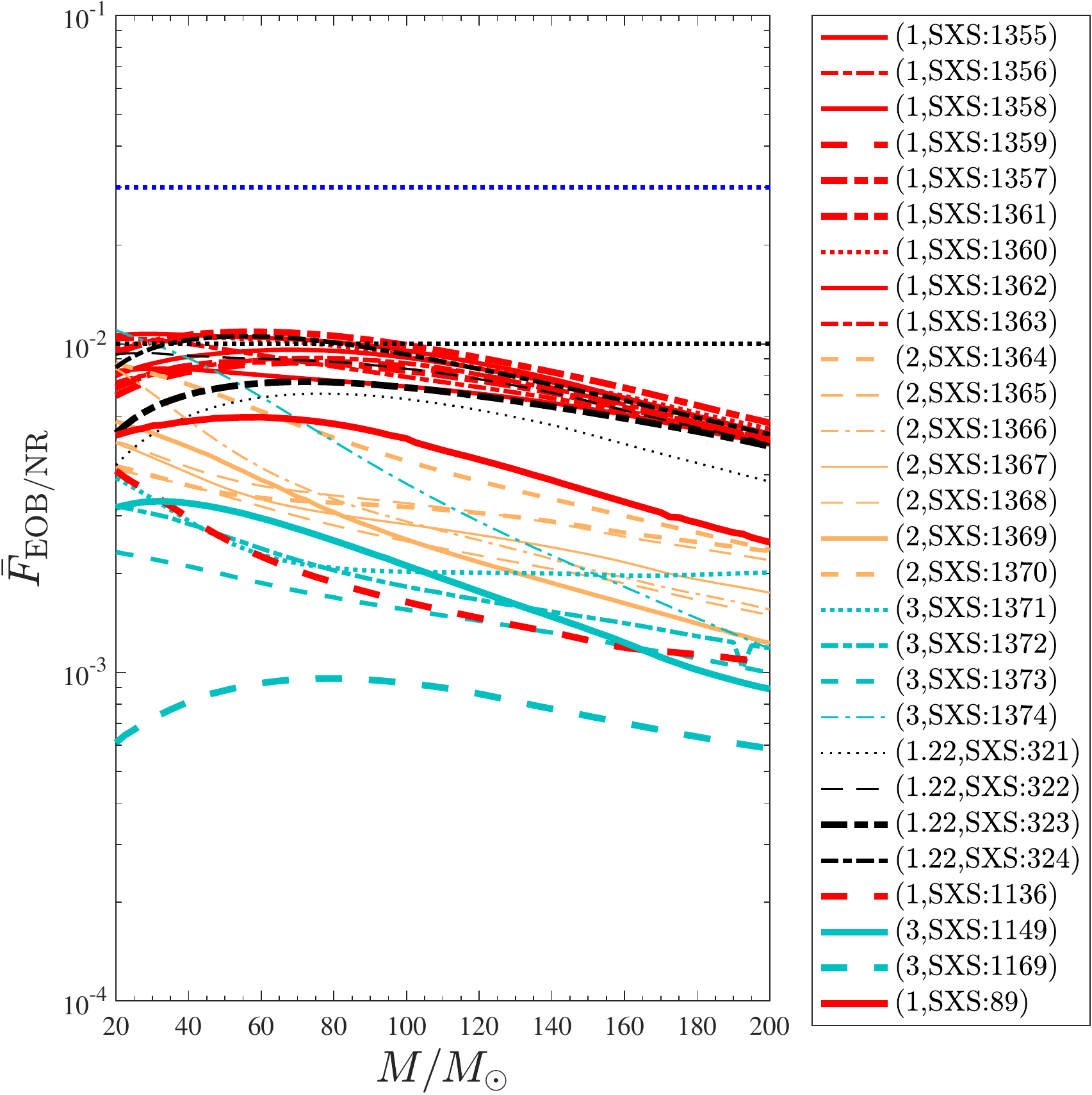}
\caption{\label{fig:barF}EOB/NR unfaithfulness computed over the eccentric SXS
  simulations publicly available. The horizontal lines mark
  the $0.03$ and $0.01$ values.}
\end{figure}
%------------------------------------
%================= 
%==============
%  phasing for 1371
%==============
\begin{figure*}[t]
\center
\includegraphics[width=0.4\textwidth]{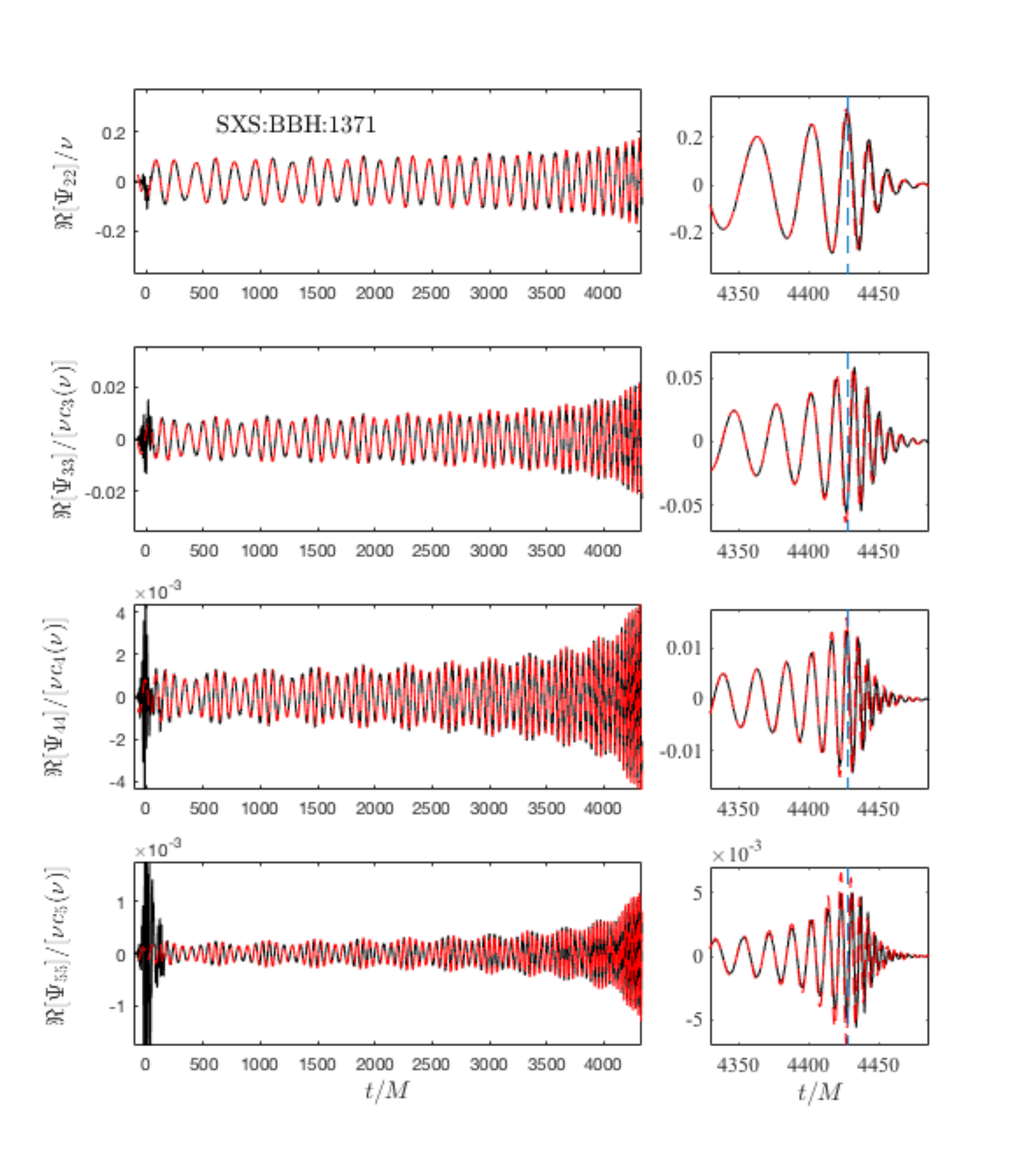}
\includegraphics[width=0.4\textwidth]{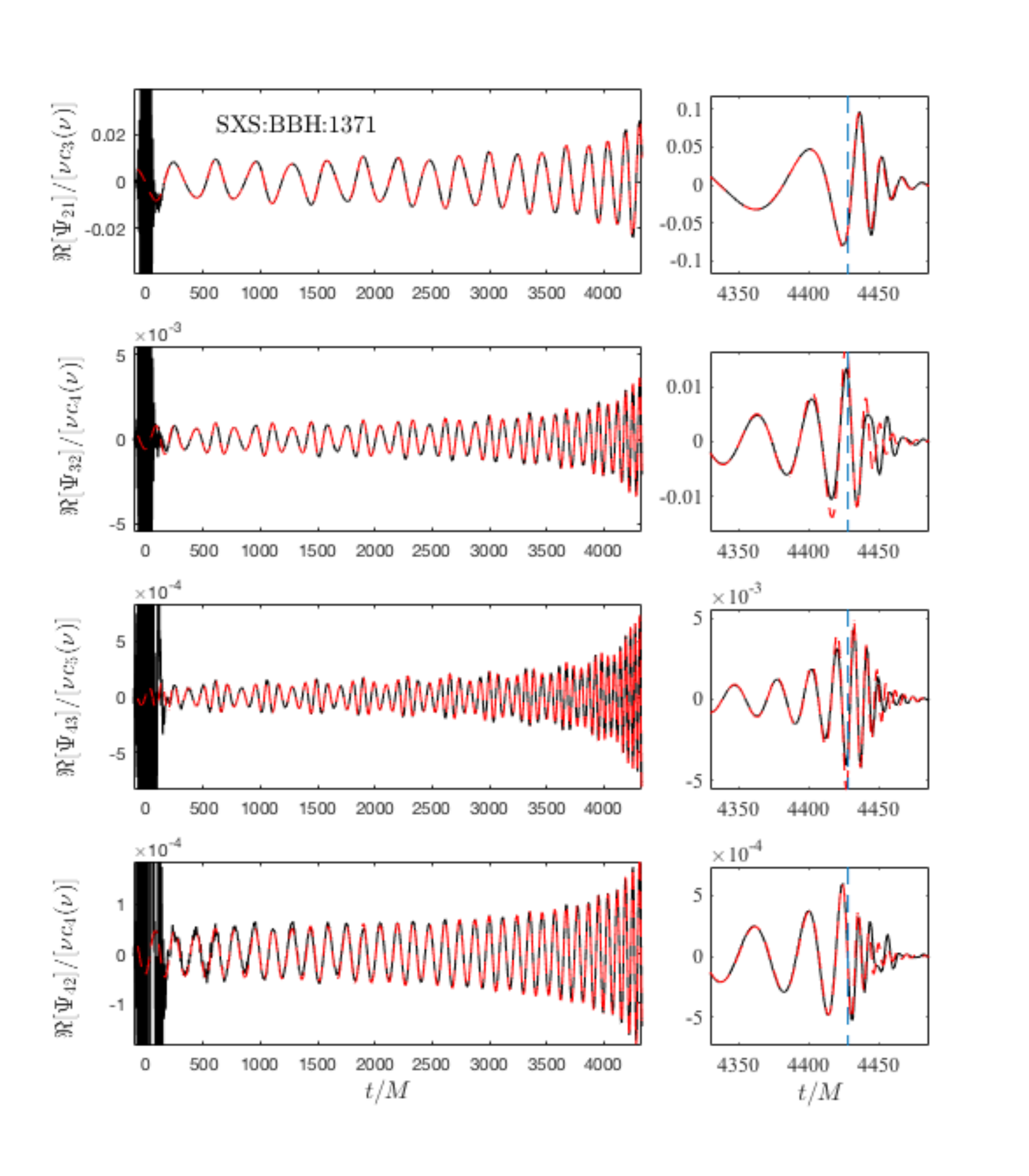}
\caption{\label{fig:1371} EOB/NR multipolar phasing agreement for SXS:BBH:1371, with $(3,0,0)$ and initial $\e_{\omega_a}^{\rm NR}\simeq 0.06$.
Waves are aligned during the early inspiral. The vertical dashed line indicate the
NR merger location. The EOB waveform inspiral is robust and reliable {\it also} for modes like $(5,5)$ and $(4,2)$, where the
corresponding NR data are typically plagued by high-frequency numerical noise.}
\end{figure*}
%==============
%  phasing for 1374
%==============
\begin{figure*}[t]
\center
\includegraphics[width=0.4\textwidth]{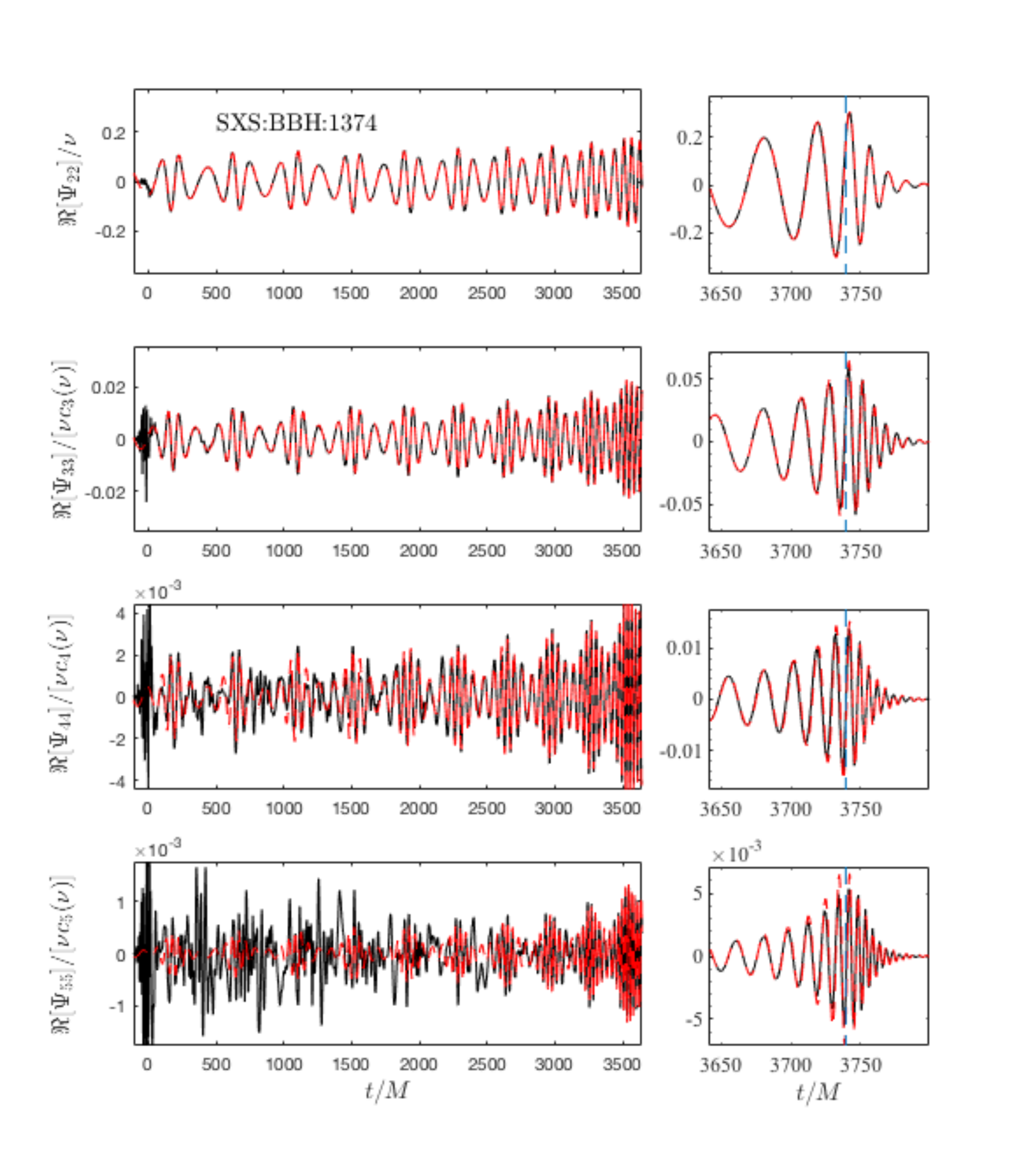}
\includegraphics[width=0.4\textwidth]{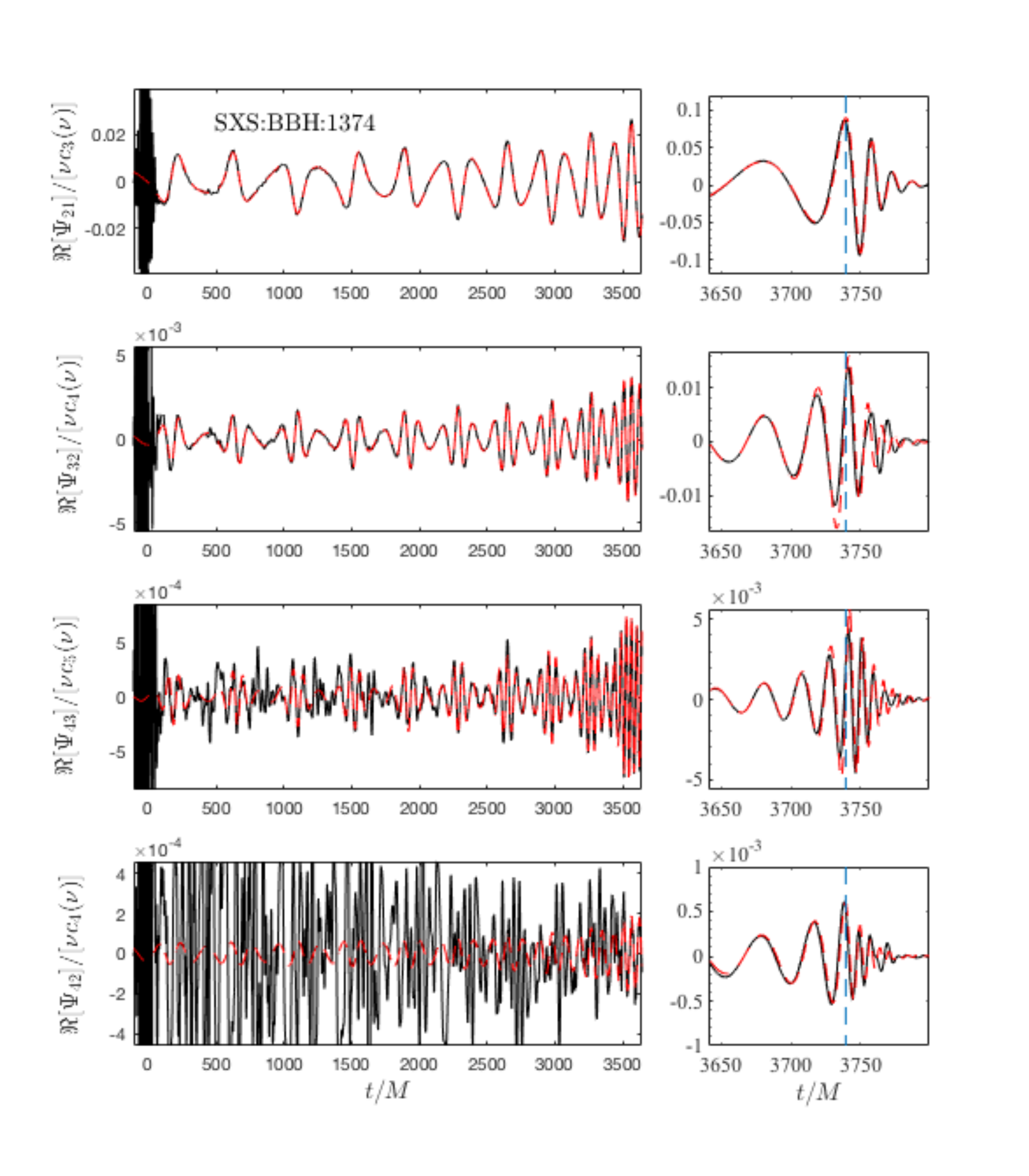}
\caption{\label{fig:1374} EOB/NR multipolar phasing agreement for SXS:BBH:1374, with $(3,0,0)$ and initial $\e_{\omega_a}^{\rm NR}\simeq 0.2$.
Waves are aligned during the early inspiral. The vertical dashed lines indicate the
NR merger location. The EOB/NR phasing and amplitude agreement is excellent all over
the inspiral up to merger and ringdown, modulo mode-mixing effects during ringdown for 
$(3,2)$, $(4,3)$ and $(4,2)$ multipoles.}
\end{figure*}
%==============
%==============
%  phasing for 324
%==============
\begin{figure*}[t]
\center
\includegraphics[width=0.4\textwidth]{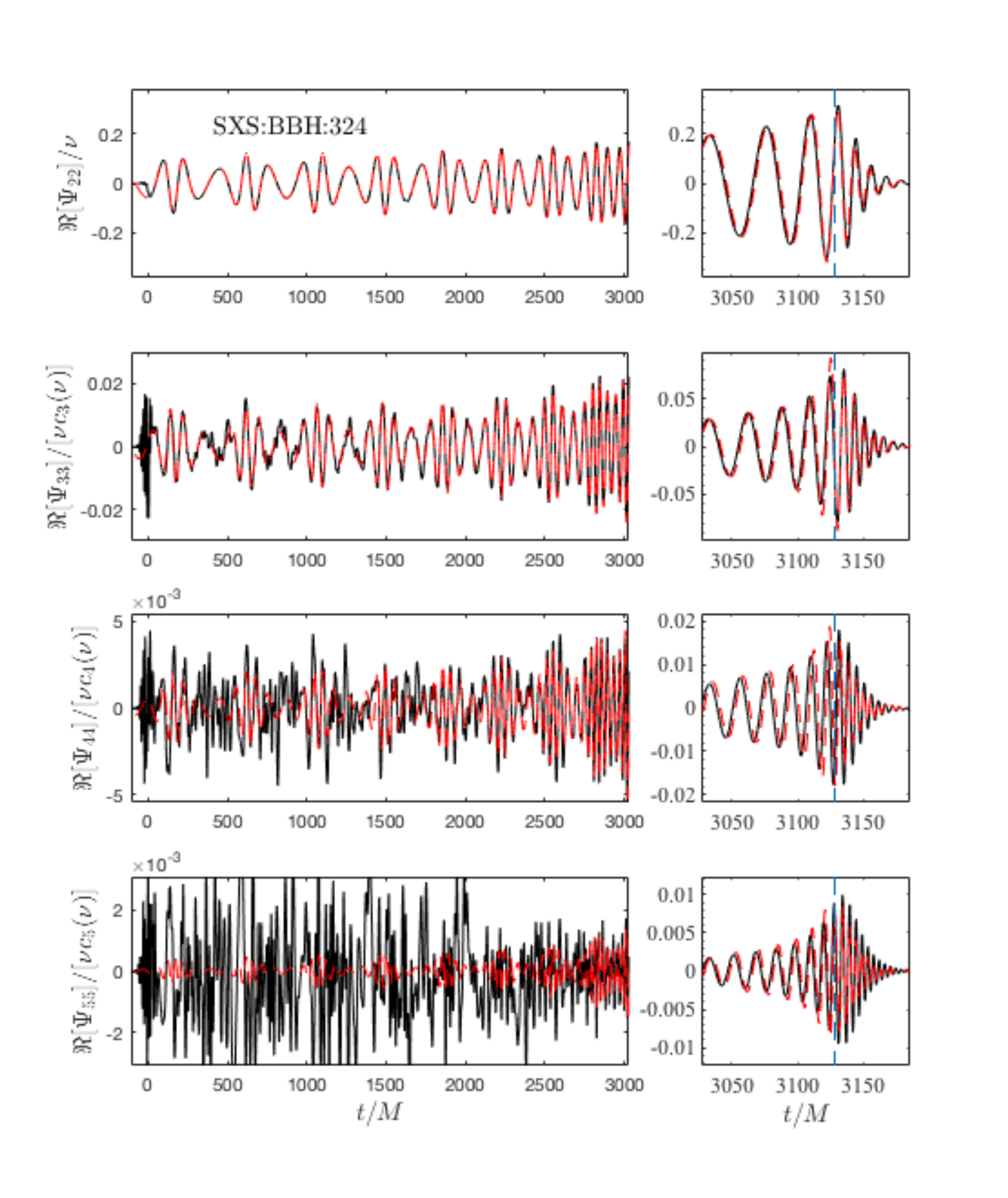}
\includegraphics[width=0.4\textwidth]{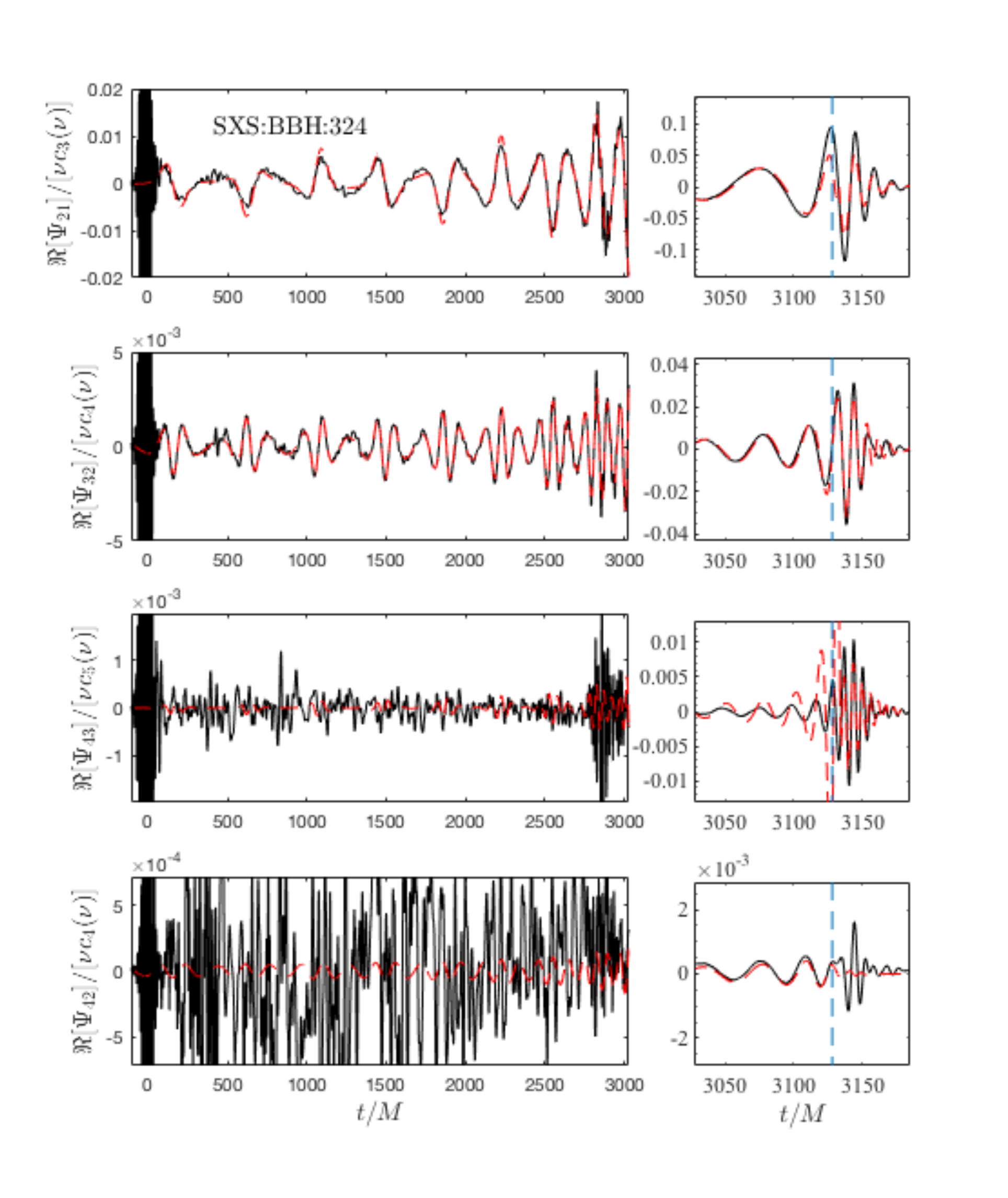}
\caption{\label{fig:324} Multipolar phasing agreement between EOB (red online) and NR (black online) waveform for SXS:BBH:324, 
with $(1.22,+0.33,-0.44)$ and $\e_{\omega_a}^{\rm NR}\simeq 0.2$. This is currently the most eccentric and general spin-aligned configuration 
available in the SXS catalog. Waves are aligned during inspiral. The vertical dashed lines indicate the NR merger location. 
Note the rather large numerical noise in many of the NR subdominant multipoles, especially the $(4,2)$ that is completely
unreliable.   For mode $(4,3)$, the NQC correction factor introduces well known unphysical effects in the amplitude during 
late inspiral.}
\end{figure*}
%==============
\begin{figure}[t]
\center
\includegraphics[width=0.45\textwidth]{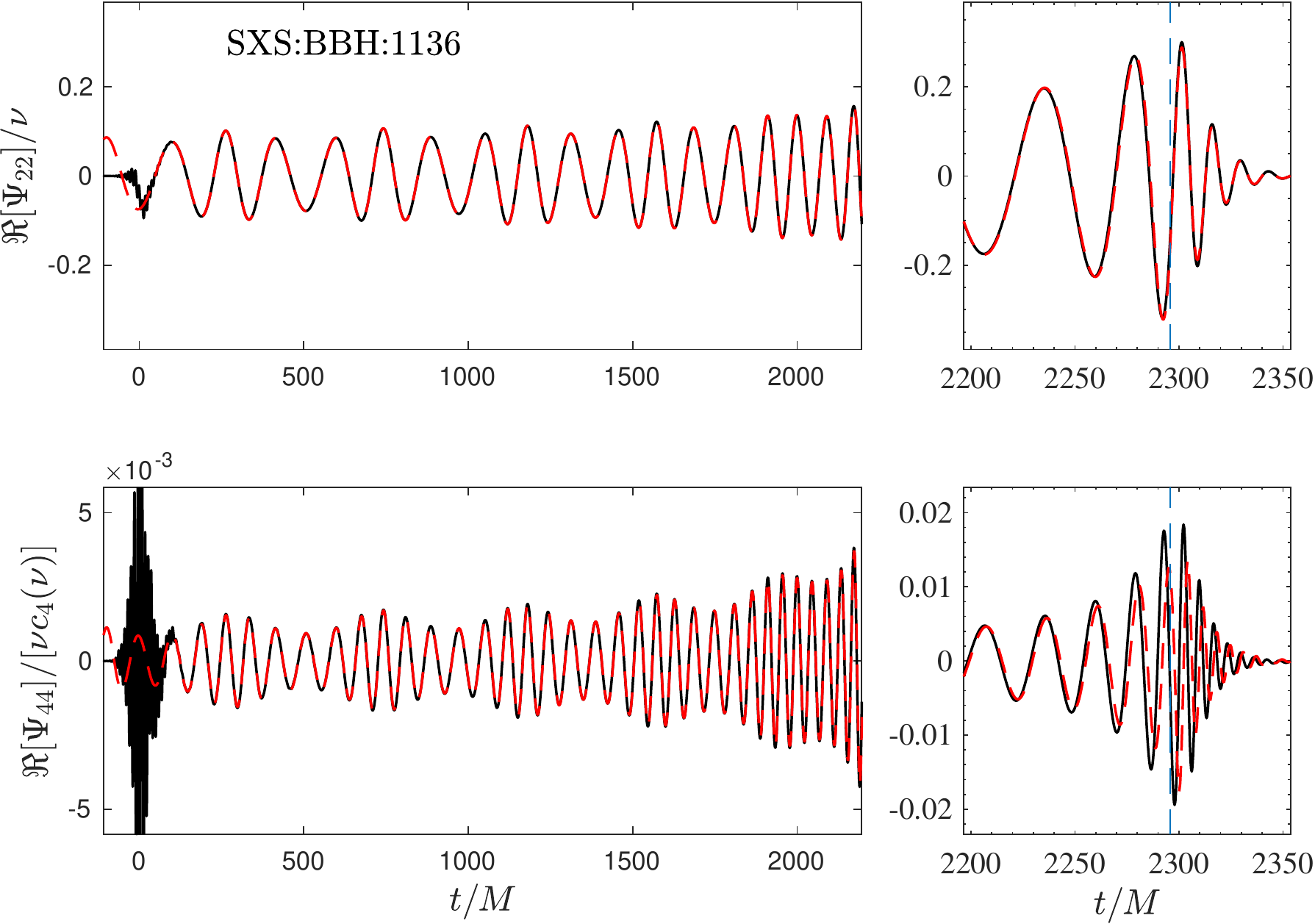}
\caption{\label{fig:1136}EOB/NR phasing comparison for the nonzero even-parity modes for SXS:BBH:1136 dataset, with $(1,-0.75,-0.75)$
and $e_{\omega_a}^{\rm NR}\simeq 0.08$. Waveforms are aligned during the inspiral, while the vertical dashed line indicates 
the merger position. The corresponding $(2,2)$ EOB/NR phase difference oscillates between $-0.05$ and $0.05$ rad during 
the inspiral, to eventually accumulate $-0.5$~rad at merger point.}
\end{figure}

\subsubsection{NR uncertainty}
To complement what we did already in Ref.~\cite{Chiaramello:2020ehz}, 
and to better evaluate the performance of the EOB model in the next section, 
we provide here an explicit error estimate on the eccentric NR simulations 
that we are considering, repeating and extending the analysis 
of Ref.~\cite{Hinder:2017sxy}, that was limited to nonspinning configurations. 
All SXS datasets we use are listed in Table~\ref{tab:SXS}.
We use the highest and second highest resolutions available in the SXS catalog to 
give two, complementary, error estimates. On the one hand, we compute the 
time-domain phase difference $\delta\phi^{\rm NR}$ for  the $\ell=m=2$ waveform mode 
between the highest and second highest resolution and retain its value at the high-resolution 
merger point $\delta\phi_{\rm mrg}^{\rm NR}$. This value appears in the third column of Table~\ref{tab:SXS}.
A few comments are in order. First, for all nonspinning dataset, this quantity is {\it at most} 
$\sim 1$~rad, often less. Although this number is a useful indicator of a, probably overestimated,
accumulated NR uncertainty, in itself it might hide details that require the inspection of the 
full time evolution of  $\delta\phi^{\rm NR}(t)$. In fact one easily realizes that not all NR simulations
seem to have the same quality. For some dataset, $\delta\phi^{\rm NR}(t)$ is smooth and clean,
with a quality comparable to the one of standard quasi-circular simulations. 
In other cases, $\delta\phi^{\rm NR}(t)$ is very noisy, with large oscillations within 
the $\pm 0.1$ band. This is clearly illustrated in
Fig.~\ref{fig:delta_phi} for the two situations mentioned above: the highly eccentric configuration 
SXS:BBH:1370, that presents large amplitude oscillations, and a mildly eccentric one, SXS:BBH:1355, 
where $\delta\phi^{\rm NR}(t)$ looks much better behaved. These details can influence the quality
of EOB/NR phasing comparisons, as we will mention below. 
The $\delta\phi^{\rm NR}_{\rm mrg}$ for spinning datasets, last eight rows of Table~\ref{tab:SXS}, 
are typically rather large, and do not really give a useful, stringent, measure of the error bar. 
We note that $\delta \phi^{\rm NR}$  for SXS:BBH:324 shows a similar behavior to the one of SXS:BBH:1370,
with large oscillations during the inspiral, so that the value $\delta\phi^{\rm NR}_{\rm mrg}=-0.04$
may not faithfully reflect the actual quality of the data.

As a complementary accuracy estimate, we also evaluate the unfaithfulness  
$ \bar{F}_{\rm NR/NR}\equiv 1-F_{\rm NR/NR}$, using Eq.~\eqref{eq:barF} above,
between the two highest NR resolutions using the {\tt zero\_det\_highP}~\cite{dcc:2974}  
Advanced-LIGO power spectral density.   Figure~\ref{fig:barF_nrnr} displays $\bar{F}_{\rm NR/NR}$ 
for all SXS datasets available except for SXS:BBH:0089, that is given in the catalog 
with a single resolution. The corresponding values of $\bar{F}_{\rm NR/NR}^{\rm max}\equiv \max(\bar{F}_{\rm NR/NR})$  
are listed in the sixth column of Table~\ref{tab:SXS}. The picture highlights that there are three 
simulations that have larger uncertainties during the inspiral, that  corresponds to the 
small $M$ region. As we will see below, especially by inspecting higher 
modes of SXS:BBH:324, these are datasets whose quality should possibly be improved 
to better exploit them in the future for EOB/NR comparison purposes.

\subsection{EOB waveforms: choosing initial conditions}
\label{sec:eccID}
To provide meaningful EOB/NR comparison, the EOB dynamics should be started in such
a way that the eccentricity induced modulations in the EOB waveform are consistent with
the corresponding ones present in the NR simulations. We follow previous work 
(including the generalization of ID setup of Ref.~\cite{Hinderer:2017jcs} to spinning binaries)
so to setup initial data using a nominal EOB eccentricity parameter $e^{\rm EOB}$. This is not 
strictly necessary\footnote{The EOB dynamics is ruled by initial energy and angular momentum.
So, one could simply setup the system at apastron in the same way as it is done for
hyperbolic configurations~\cite{Nagar:2020xsk}.} (especially because, differently from Ref.~\cite{Hinderer:2017jcs} we
do not express the EOB dynamics using it), but it is just intuitively
convenient. So, as it was done in Ref.~\cite{Chiaramello:2020ehz}, the eccentricity-related
modulation of the EOB dynamics and waveform are controlled via the initial gravitational wave
quadrupole frequency at apastron, $\omega_a^{\rm EOB}$ and initial EOB eccentricity at the same frequency
$e^{\rm EOB}_{\omega_a}$. 
From these parameters, one then determines, via quasi-Newtonian formulas following
Ref.~\cite{Hinderer:2017jcs}, an initial semilatus rectum and from  this an initial 
separation and angular momentum. The radial momentum is always set to zero since 
the EOB dynamics is started, by construction, at apastron. The procedure for correctly 
finding the EOB values that  allow for a best match between waveforms is slightly 
tricky and cannot be fully  automatized. Let us discuss it in some detail. For an initial 
guess of $(\omega_a^{\rm EOB},e^{\rm EOB}_{\omega_a})$, motivated by the values of the corresponding
NR quantities, the EOB and NR waveforms are aligned in the time-domain with our usual 
procedure~\cite{Damour:2012ky} during the inspiral. Then, we compute the fractional EOB/NR 
frequency difference  $\Delta\omega_{22}^{\rm EOBNR}\equiv \omega_{22}^{\rm EOB}-\omega_{22}^{\rm NR}$
and inspect it versus time. We then vary, recursively, first  $\omega_a^{\rm EOB}$  and 
then $e^{\rm EOB}_{\omega_a}$, until we make $\Delta\omega_{22}^{\rm EOBNR}$ as small as possible 
($\simeq 10^{-2}$ or less) and as nonoscillatory as possible over the longest inspiral time interval.
In practice, we monitor $\Delta\omega_{22}^{\rm EOBNR}$ and allow it to eventually grow only during 
the plunge up to merger and ringdown. For doing so efficiently one has to remember that increasing $e^{\rm EOB}$ 
translates into a stronger GW emission at the periastron and thus a shorter waveform,
i.e. the inspiral gets accelerated. If $e^{\rm EOB}_{\omega_a}$ is decreased, it is true the opposite
and the inspiral gets longer. The procedure is very sensitive to minimal variations of the 
parameters $(\omega_a^{\rm EOB},e^{\rm EOB}_{\omega_a})$. Typically, one has to vary 
$\omega_a^{\rm EOB}$ in steps of order $10^{-4}$ and $e^{\rm EOB}$ in steps of 
order $10^{-5}$ to achieve an acceptable phasing agreement. 
Figure~\ref{fig:sxs_Domg} displays two examples of the outcome of this procedure,
for SXS:BBH:1356 (smaller eccentricity, $e_\omega^{\rm NR}\sim 0.1$) 
and SXS:BBH:1374 (larger eccentricity, $e_\omega^{\rm NR}\sim 0.2$). 
For each simulations we show, superposed, the EOB and NR frequencies, together
with $\Delta\omega_{22}^{\rm EOBNR}$. The figure corresponds to the values 
of $(\omega_a^{\rm EOB},e^{\rm EOB})$ in Table~\ref{tab:SXS}. For SXS:BBH:1356
$\Delta\omega_{22}^{\rm EOBNR}$ is found to oscillate around zero 
within $\pm 5\times 10^{-3}$ for most of the time, but it grows during plunge up to merger time.
Interestingly, despite being marred by high-frequency noise,
$\Delta\omega_{22}^{\rm EOBNR}$ essentially averages zero also for SXS:BBH:1374, 
thus confirming the quality of our parameter choice. The picture remains essentially the same
for all simulations we considered, except for SXS:BBH:1370. For this dataset, it
doesn't seem possible to optimize $(\omega_a^{\rm EOB},e^{\rm EOB})$ to flatten
the secular oscillation in  $\Delta\omega_{22}^{\rm EOBNR}$ during the early inspiral.
As mentioned above, this is probably due to the slightly lower quality  of the 
SXS:BBH:1370 simulation with respect to the others.
Modulo this case, we found that the procedure is robust and reliable all over the dataset
at our disposal, although it is sensitive to small variations and a careful fine tuning is
needed. Our current best choices of $(\omega_a^{\rm EOB},e^{\rm EOB})$ are 
listed in Table~\ref{tab:SXS}.

\subsection{EOB/NR phasing and unfaithfulness: the quadrupole mode}
\label{sec:barF_ecc}
We start the discussion of the EOB/NR phasing comparisons focusing on Fig.~\ref{fig:sxs_1359}.
The figure shows together EOB/NR phasings for SXS:BBH:1359, a mildly eccentric dataset with $e_{\omega_a}^{\rm NR}=0.11$ and
mass ratio $q=1$, as well as for SXS:BBH:1374 that has larger initial eccentricity, $e^{\rm NR}_{\omega_a}\sim 0.2$ and $q=3$.
For each dataset, the figure shows: the EOB/NR phase difference $\Delta\phi_{22}^{\rm EOBNR}$ (top panel); the EOB and
NR real part of the waveform (middle panel); the EOB and NR frequencies (bottom panel). 
The figure highlights that our choice of  $(e^{\rm EOB},\omega_a^{\rm EOB})$ is such to also yield a rather 
flat phase difference during the inspiral, with just small amplitude residual oscillations, that are always confined
between  $\pm 0.05$. These differences are of the same order (actually a bit larger) than the corresponding 
NR error uncertainties obtained by taking the phase difference between the two highest resolutions available.
The behavior of the phase difference illustrated in Fig.~\ref{fig:sxs_1359}  remains approximately 
the same for all other configurations. As mentioned above, an exception is represented by the,
apparently less accurate, SXS:BBH:1370 dataset, where it does not  seem possible to reduce 
$\Delta\phi^{\rm EOBNR}_{22}$ below $\pm 0.2$ rad. Similarly, despite varying and fine tuning
methodically the initial parameters, it doesn't seem possible to obtain phase differences that look
perfectly flat on the $[-0.1,+0.1]$~rad scale, likewise the quasi-circular case,
see Fig.~\ref{fig:q3}. This does not look surprising in view of the fact that
noncircular effects in both waveform and radiation reaction are incorporated only through the leading 
order Newtonian factors. Additional improvements are expected to occur by  including, probably
in some resummed form, up to 3PN noncircular terms, see e.g.~\cite{Mishra:2015bqa,Ebersold:2019kdc}

The quality of the model  is finally assessed by computing the EOB/NR unfaithfulness 
$\bar{F}_{\rm EOB /NR}$.
As discussed in Ref.~\cite{Hinder:2017sxy}, the clean computation of the Fourier transform 
is trickier than the quasi-circular case and a more aggressive tapering is needed to 
avoid effects due to Gibbs phenomenon. Figure~\ref{fig:barF} shows the result of 
$\bar{F}_{\rm EOB/NR}$ computation using, as in the circular case, the Advanced 
LIGO sensitivity curve. The improvement with respect to Fig.~4  of~\cite{Chiaramello:2020ehz} 
is dramatic, with $\bar{F}^{\rm max}_{\rm EOB/NR}$ at most of order $1\%$, visually
rather similar to the performance of the circularized case, Fig.~\ref{fig:barF_eobnr_circ}.
The figure is complemented by the values of $\bar{F}^{\rm max}_{\rm EOB/NR}$ listed
in the last column of Table~\ref{tab:SXS}. The most interesting thing to note is that there
are little differences between the low-eccentricity and high-eccentricity cases, and in all
cases $\bar{F}^{\rm max}_{\rm EOB/NR}\simeq 1\%$. This suggests, somehow surprisingly, 
that the treatment of noncircular effects via the general Newtonian prefactor
could be acceptably accurate (i.e. $\bar{F}^{\rm max}_{\rm EOB/NR}\lesssim 3\%$) 
also for {\it larger eccentricities}. A thorough assessment of this statement would require 
additional SXS eccentric simulations at least as accurate as the ones currently available.

\subsection{Subdominant modes}
\label{sec:hm}
Reference~\cite{Chiaramello:2020ehz} already pointed out that it is also possible
to obtain a good agreement between EOB and NR higher multipolar modes by
simply replacing each circularized Newtonian factor with the corresponding noncircular
counterpart. The purpose of this section is to show this explicitly for a sample of
significative NR datasets. We focus on 4 specific configurations: SXS:BBH:1371 (large mass ratio, 
low eccentricity), Fig.~\ref{fig:1371};  SXS:BBH:1374 (large mass ratio, large initial eccentricity, $e_{\omega_a}^{\rm NR}\simeq 0.2$), 
Fig.~\ref{fig:1374}; SXS:BBH:324, a configuration with large eccentricity $e_{\omega_a}^{\rm NR}\simeq 0.2$, 
unequal mass and unequal spins, Fig.~\ref{fig:324};
and SXS:BBH:1136, equal-mass, small eccentricity $e_{\omega_a}^{\rm NR}\simeq 0.08$, 
but large spins, anti-aligned with the orbital angular momentum, Fig.~\ref{fig:1136}.
Figures~\ref{fig:1371}-\ref{fig:324} display all meaningful multipoles, that is $(2,2)$, $(2,1)$, $(3,3)$, $(3,2)$, $(4,3)$, $(4,4)$ 
and $(5,5)$, while Fig.~\ref{fig:1136} only shows $(2,2)$ and $(4,4)$.
In all figures, the vertical dashed line indicates the NR merger location. Waveforms are always
aligned during the inspiral. The phase and amplitude agreement is satisfactory in all cases. 
In particular, it is interesting to note that the waveforms remain well in phase,
including the higher modes, {\it also} when the eccentricity is large, i.e. for SXS:BBH:1374 and SXS:BBH:324,
Figs.~\ref{fig:1374} and~\ref{fig:324}, even if in these two cases some of the NR higher modes 
(see e.g. the $(5,5)$) are marred by high-frequency numerical noise. This is an interesting fact 
that allows the EOB waveform to be used as benchmark to further improve the quality of the NR simulations.
This is especially interesting for a small-amplitude mode like the $(4,2)$: once the NR 
high-frequency noise clears up (see e.g. the bottom right panels of Figs.~\ref{fig:1374} and~\ref{fig:324}), 
the late-inspiral waveform is found to be well consistent with the analytical prediction.
By contrast, focusing on Fig.~\ref{fig:324}, let us also note that the amplitude of the analytical $(4,3)$ mode
gets progressively too large towards merger due to the unphysical action of the NQC correction
factor that we discussed above. Similarly, differences during ringdown in modes like $(3,2)$, $(4,3)$ 
and $(4,2)$ are due to the absence of mode-mixing effects~\cite{London:2014cma,Taracchini:2014zpa,London:2018nxs}.

\section{Hyperbolic encounters and scattering angle}
\label{sec:scattering}
\begin{figure}[t]
\center
\includegraphics[width=0.45\textwidth]{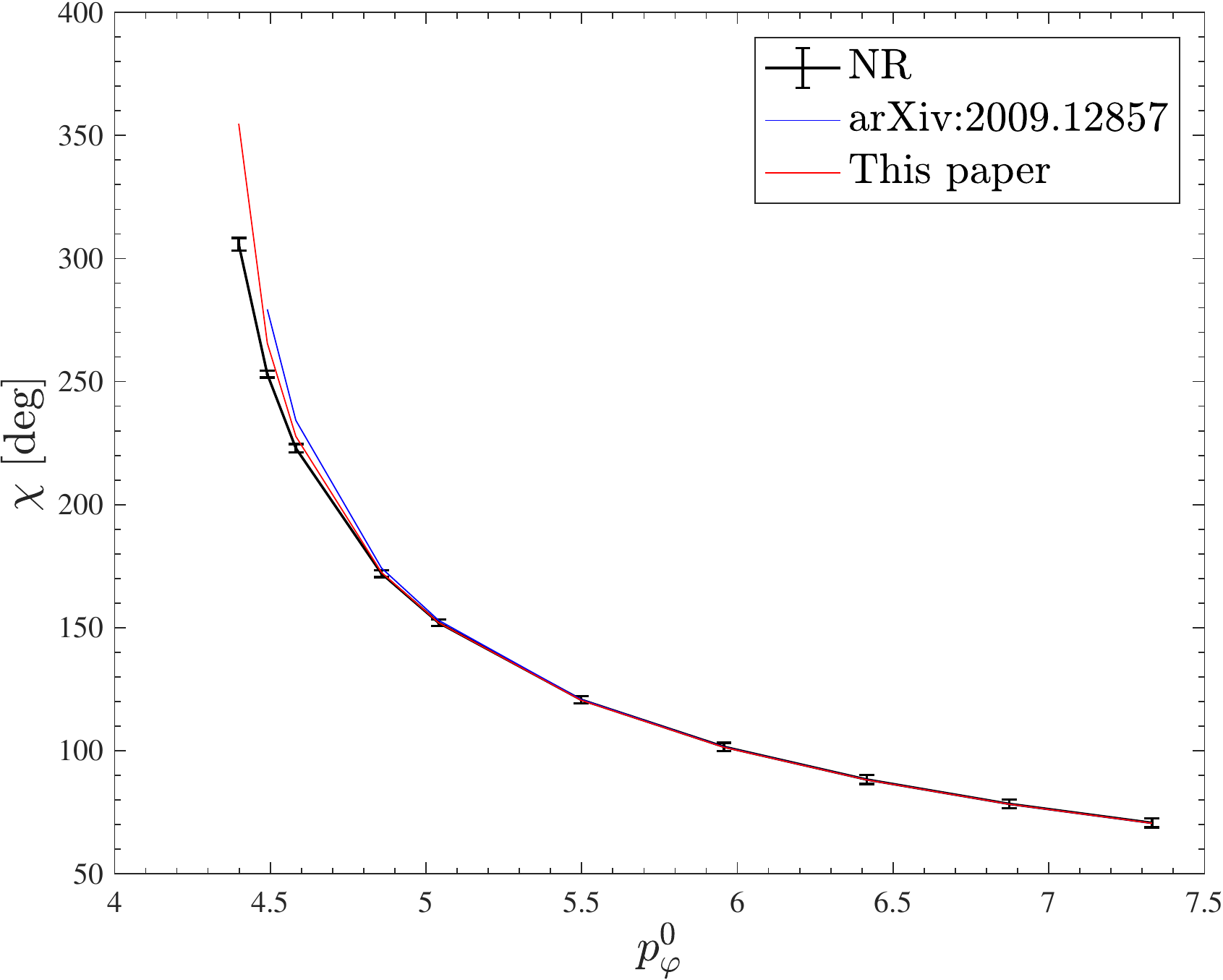}
\caption{\label{fig:chi}Visual EOB/NR comparison of the scattering angles of Table~\ref{tab:chi_scattering}. 
To appreciated the improvement reached here we also list the EOB points computed in 
Table~I of~\cite{Nagar:2020xsk}.}
\end{figure}

Recently, Ref.~\cite{Nagar:2020xsk} showed how the EOB model can be used to compute
dynamics and waveforms from hyperbolic encounters. See Ref.~\cite{Loutrel:2020kmm,Loutrel:2020jfx} 
for a recent overview. Recently, Ref.~\cite{Mukherjee:2020hnm} also pointed out that these events 
may be detectable by the present and next-generation ground-based observatories. Accurate modelization 
of dynamics and waveform is then needed. 
Both the (i) new expressions of $(\hat{\F}_\varphi,\hat{\F}_r)$ and (ii) the new determination 
of $(a_6^c,c_3)$ will have a quantitative impact on the results of~\cite{Nagar:2020xsk},
although the basic phenomenology of hyperbolic encounters and dynamical 
captures remains the same discussed there. It is however informative to repeat here the EOB calculation 
of the scattering angle $\chi$ for the 10 configurations simulated in NR~\cite{Damour:2014afa}
and that are discussed in Table~I of~\cite{Nagar:2020xsk}. The EOB outcome,
together with the original NR values, $(\chi^{\rm EOB},\chi^{\rm NR})$ is listed in Table~\ref{tab:chi_scattering},
that is visually complemented by Fig.~\ref{fig:chi}. The Table also reports the GW energy, $\Delta E$, and
angular momentum, $\Delta J$, losses for both the NR simulations and the EOB 
dynamics~\footnote{Let us specify that while the NR losses are computed from the waveform, the EOB
losses are computed subtracting the initial and final energy and angular momentum, i.e. effectively
accounting for the action of the radiation reaction on the dynamics.}.
The figure also plots for convenience the results of Ref.~\cite{Nagar:2020xsk}.
The EOB/NR agreement is now rather good, with a marked improvement with respect 
to~\cite{Nagar:2020xsk} for the first 4 configurations, that correspond to the smallest values of the 
EOB impact parameter $r_{\rm min}$. Notably, configuration $\# 1$, that was directly plunging 
in Ref.~\cite{Nagar:2020xsk}, is now found to have the correct qualitative scattering behavior, 
with a quantitative EOB/NR fractional difference that is of $12\%$. 
This fact is a reliable cross check of the consistency and robustness of our procedure 
to obtain $a_6^c(\nu)$: although the function was determined using quasi-circular 
configurations, its impact looks to be essentially correct {\it also} for scattering configuration.
For what concerns the comparisons between the energy and angular momentum losses, 
the agreement between $(\Delta E^{\rm NR},\Delta E^{\rm EOB})$ is pretty stable, with 
absolute fractional differences of few percents, ranging from $8.9\%$ for $\#10$ to 
$6.75\%$ for $\#1$. By contrast, the  $(\Delta J^{\rm NR},\Delta J^{\rm EOB})$ difference
is rather large, $\sim 44\%$ for $\#10$, to get progressively better and better as the impact
parameter diminishes, up to only $\sim 0.6\%$ for $\#1$. This looks like a promising starting
point for future investigations, that aim at improving both the conservative dynamics, e.g.
including higher order terms in the EOB potentials (see e.g. Ref.~\cite{Nagar:2020xsk}),
and the radiation reaction beyond the quasi-circular limit.
%=====================
% Scattering angle
%=====================
\begin{table*}[t]
 \caption{\label{tab:chi_scattering} Comparison between EOB and NR scattering angle. From left to right the columns report:
 the ordering number; the EOB impact parameter $r_{\rm min}$; the NR and EOB radiated energies, 
 $(\Delta E^{\rm NR}/M,\Delta E^{\rm EOB}/M)$; the NR and EOB radiated angular momentum, 
 $(\Delta J^{\rm NR}/M^2,\Delta J^{\rm EOB}/M^2)$; the NR and EOB scattering angles $(\chi^{\rm NR},\chi^{\rm EOB})$ and
 their fractional difference  $\hat{\Delta}\chi^{\rm NREOB}\equiv |\chi^{\rm NR}-\chi^{\rm EOB}|/\chi^{\rm NR}$. 
 The EOB/NR agreement is improved with to Ref.~\cite{Nagar:2020xsk}, see Table~I there and Fig.~\ref{fig:chi}.}
   \begin{center}
     \begin{ruledtabular}
\begin{tabular}{c c c c c c c c c } 
$\#$  & $r_{\rm min}$ & $\Delta E^{\rm NR}/M$ & $\Delta E^{\rm EOB}/M$ & $\Delta J^{\rm NR}/M^2$ & $\Delta J^{\rm EOB}/M^2$ & $\chi^{\rm NR}$ [deg] & $\chi^{\rm EOB}$[deg] & $\hat{\Delta}\chi^{\rm NREOB}[\%]$ \\
\hline
%1 & 3.47    & 0.01946(17)       & 0.020773    &   0.17007(89)  & 0.169076    &   305.8(2.6)  & 343.022077   &  12.1720\\
%2 & 3.82    & 0.01407(10)       & 0.013549    &  0.1380(14)     &  0.120819   & 253.0(1.4)   &  262.926831  &   3.9236\\
%3 & 4.12    & 0.010734(75)     & 0.009858    &   0.1164(14)    &  0.094955  &  222.9(1.7)   &226.490660    &   1.6109 \\
%4 & 4.91    & 0.005644(38)     & 0.004900    & 0.076920(80)  &  0.057046  &  172.0(1.4)   & 172.180675   &   0.1050 \\
%5 & 5.39    & 0.003995(27)     &  0.003424   & 0.06163(53)    &  0.044310  &152.0(1.3)   &  151.909122  &   0.0598\\
%6 & 6.53     & 0.001980(13)    & 0.001687    & 0.04022(53)    &   0.027309  &  120.7(1.5)  & 120.465819   &   0.1940\\
%7 & 7.62     & 0.0011337(90)  & 0.000975    &  0.029533(53) &   0.018985  &  101.6(1.7)  &101.393669    &   0.2031\\
%8 & 8.68    & 0.007108(77)    & 0.000622   &  0.02325(47)    &  0.014170   & 88.3(1.8)    &  88.157623    &   0.1612\\
 %9 &  9.74   &  0.0004753(75) & 0.000424   &  0.01914(76)    &  0.011092   &78.4(1.8)    & 78.274386      &  0.1602\\
%10 & 10.79 & 0.0003338(77)  & 0.000304   &   0.0162(11)     &   0.008987   &  70.7(1.9)    &   70.542117    &  0.2233\\
1 & 3.43    & 0.01946(17)       & 0.021348    &   0.17007(89)  & 0.176789    &   305.8(2.6)  & 354.782050   &  16.0177\\
2 & 3.79    & 0.01407(10)       & 0.013453    &  0.1380(14)     &  0.123928   & 253.0(1.4)   &  265.604256  &   4.9819\\
3 & 4.09    & 0.010734(75)     & 0.009610    &   0.1164(14)    &  0.096685  &  222.9(1.7)   &227.722810    &   2.1637 \\
4 & 4.89    & 0.005644(38)     & 0.004623    & 0.076920(80)  &  0.057576  &  172.0(1.4)   & 172.446982    &   0.2599 \\
5 & 5.37    & 0.003995(27)     &  0.003183  & 0.06163(53)     &  0.044608  &152.0(1.3)   &  152.028075   &   0.0181\\
6 & 6.52     & 0.001980(13)    & 0.001529    & 0.04022(53)    &   0.027405  &  120.7(1.5)  & 120.481058   &   0.1814\\
7 & 7.61     & 0.0011337(90)  & 0.000870    &  0.029533(53) &   0.019026  &  101.6(1.7)  &101.389990    &   0.2067\\
 8 & 8.68    & 0.007108(77)    & 0.000549   &  0.02325(47)    &  0.014190   & 88.3(1.8)    &   88.150825    &   0.1689\\
 9 &  9.74   &  0.0004753(75) & 0.000372   &  0.01914(76)    &  0.011103   &78.4(1.8)    & 78.268033       &  0.1683\\
10 & 10.79 & 0.0003338(77)  & 0.000266   &   0.0162(11)     &   0.008993  &  70.7(1.9)    &   70.536902    &  0.2307\\
  \end{tabular}
 \end{ruledtabular}
 \end{center}
 \end{table*}
 %-------------------------------------------------------------------
 %======================
% Fr comparison: PN vs resum
%=======================
%=============

\section{Conclusions}
\label{sec:conclusions}
We have presented an improvement of the EOB model of Ref.~\cite{Chiaramello:2020ehz}
for generic, spin-aligned, coalescing black hole binaries, {\tt TEOBResumSGeneral}.
The main, new, technical aspect of the model  concerns the fact that its quasi-circular limit 
is now correctly informed by NR waveform
data. This allows us to have a single, NR-faithful, waveform model for spin-aligned binaries 
that can deal with any kind of configuration, from quasi-circular to eccentric and hyperbolic. 
Our main findings are as follows:
\begin{enumerate}
\item[(i)] In the quasi-circular limit, the model is compared with a significative fraction of the SXS
waveform catalog (including mass ratios up to $q=10$ and the largest spins available) 
by providing the usual EOB/NR unfaithfulness $\bar{F}_{\rm EOB/NR}$ computation 
using the Advanced LIGO power spectral density. We find that  $\bar{F}_{\rm EOB/NR}\simeq 1\%$ 
except for a single outlier, $(8,+0.85,+0.85)$ (obtained with the BAM code) that still is $<2\%$.
Although the performance is not  as good as the one of the standard quasi-circular 
\TEOBResumS{} model~\cite{Nagar:2020pcj},  it is below the usual $3\%$ threshold used 
as figure of merit. We postpone to future studies a precise investigation of how this performance 
translates on parameter estimation. 
We would also like to stress that the availability of two different, though rather similar, EOB 
models based on the same analytical structure, and with comparable EOB/NR performances, 
will allow one to put on a very solid ground any statement about {\it analytic systematics}. 
In particular, it will be interesting to understand to which extent a minimal degradation of the 
$\bar{F}_{\rm EOB/NR}$ function, determined by well defined modifications in the model 
(e.g., the presence or absence of $\hat{\F}_r$) impacts  the inferred parameters. 

\item[(ii)] For eccentric binaries, in stable configurations, we have obtained a notable improvement 
with respect to the results of Ref.~\cite{Chiaramello:2020ehz}, with $\bar{F}_{\rm EOB/NR}^{\rm max}\lesssim 1\%$
for all available eccentric SXS configurations. We stress that eccentric NR data are used {\it only} 
to test the model and not to inform it. On the one hand, this indicates 
that our eccentric model is mature enough for being used in parameter estimation on all GW 
BBH sources already detected. On the other hand, the rather easy, though successful, approach that 
we followed already calls for improvements, either to accurately deal with even larger eccentricities or 
to see whether $\bar{F}^{\rm max}_{\rm EOB/NR}$ can be lowered further, possibly to reach the same level
of the standard quasi-circular model, $\bar{F}^{\rm max}_{\rm EOB/NR}\simeq 10^{-3}$. 
Concerning larger initial eccentricities, it will be useful  to have additional, public, SXS simulations 
of the same quality (or possibly higher) of the current ones. NR simulations with higher initial 
eccentricities, up to $\simeq 0.4$, do exist~\cite{Ramos-Buades:2019uvh}, 
but this data are private\footnote{Note however that this data were obtained using the 
Einstein Toolkit~\cite{Zilhao:2013hia} and how their quality compares to the one of the SXS 
ones should be carefully studied.}  and  not available for our purposes.
\item[(iii)] We have shown that higher modes in presence of eccentricity are also largely reliable,
and often more accurate than the corresponding NR ones, especially during the inspiral. 
\item[(iv)] For hyperbolic scattering, the model provides values of the scattering angle 
(and of the GW energy losses) that are highly consistent with few NR computations currently 
available, especially for small values of the impact parameter, improving quantitatively
the results of Ref.~\cite{Nagar:2020xsk}.
\item[(v)] The waveform model discussed here is publicly available 
as stand-alone $C$-implementation, that is released via a {\tt bitbucket} git repository~\cite{teobresums}, 
within the eccentric branch. See also Appendix~\ref{sec:Ccode} for additional technical details.
Although a precise assessment of the performance of the model for parameter estimation purposes
is beyond the scope of this work, let us mention a few interesting features. The generation of each 
of the 28 datasets in Table~\ref{tab:SXS} typically requires computational times $\simeq 0.1$~s, 
though for some configurations like $\#21$, $\#27$ and $\#28$ it can reach up to $\sim 0.25$~s
because of the rather low starting frequency. The initial EOB frequencies 
$\omega_a^{\rm EOB}$ of  Table~\ref{tab:SXS} correspond, via $GM_\odot/c^3=4.925490947\times 10^{-6}$~s, 
to physical gravitational wave frequencies between $\sim 11$~Hz and $\sim 19$~Hz for a $M=50M_\odot$  
binary. When attempting a preliminary  parameter estimation  run on GW150914, this 
yielded a total running time of approximately four days on a single Intel Xeon at 2.2GHz CPU 
with 32 cores. This paves the way to new, independent, estimates of eccentricity on LIGO-Virgo
events~\cite{Liu:2019jpg,Lower:2018seu,Romero-Shaw:2019itr,Romero-Shaw:2020thy,Romero-Shaw:2020thy,CalderonBustillo:2020odh,CalderonBustillo:2020srq,Gayathri:2020coq} via analytical waveform models, possibly not limited by the hypothesis of mild eccentricities.

\end{enumerate}

\begin{acknowledgments}
We are grateful to R.~Gamba and M.~Breschi preliminary tests of the performance of the 
$C$ implementation on GW150914 data and for comments on the manuscript.
A.~N. is grateful to S.~Albanesi and T.~Damour for discussions, and especially to S.~Bernuzzi
for the daily exchange of ideas and for the music. We are grateful to D.~Chiaramello 
for collaboration in the early stages of this project.
We appreciate technical comments from I.~Romero-Shaw and M.~Zevin.
\end{acknowledgments}

\appendix
\section{Derivation of $\hat{\F}_r$ as used in Refs.~\cite{Chiaramello:2020ehz,Nagar:2020xsk}}
\label{sec:Fr_derivation}
%=================================
% Phasing comparison Fr_circ vs Fr_noncirc
%=================================
\begin{figure*}[t]
\center
\includegraphics[width=0.31\textwidth]{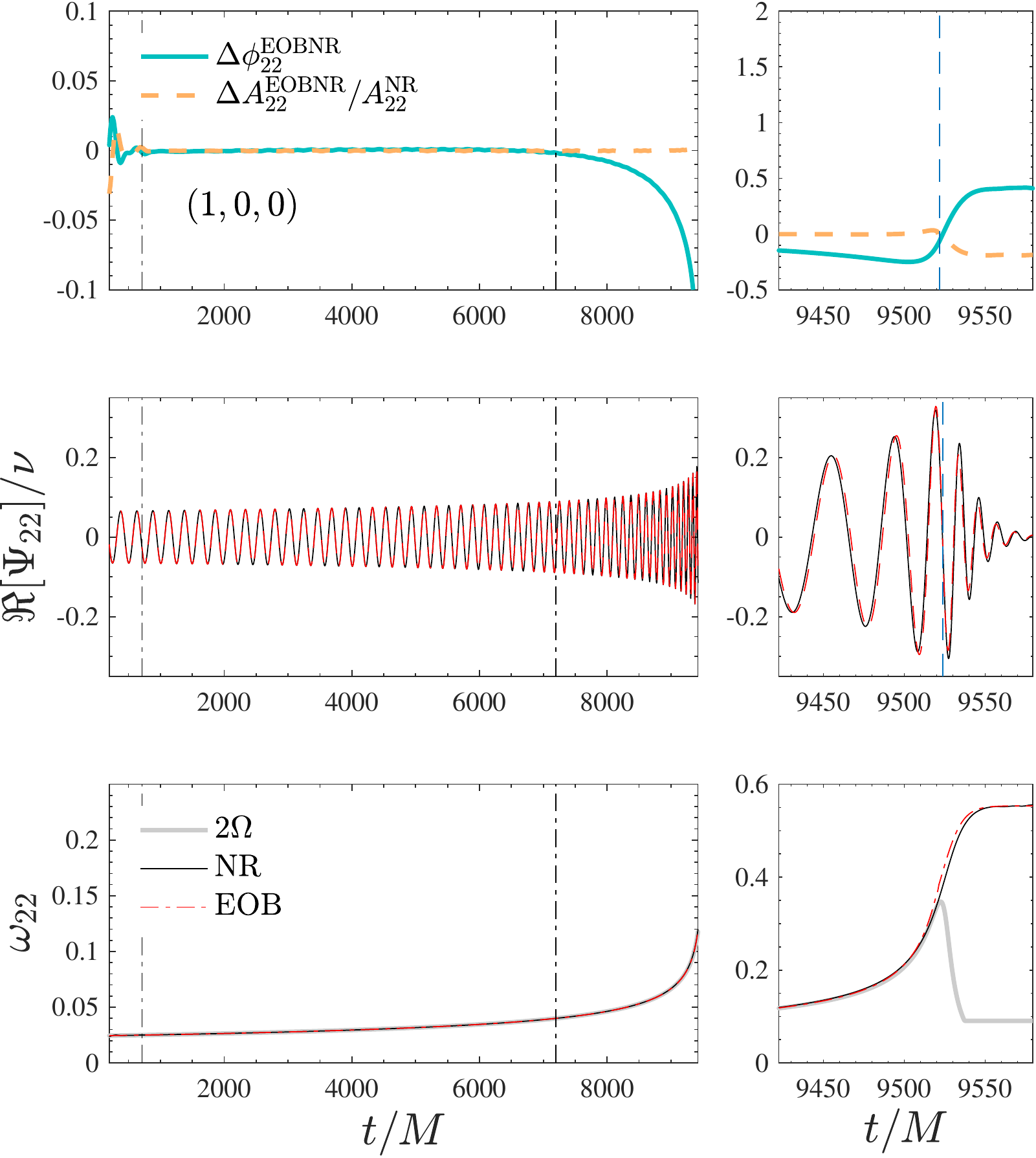}\;
\includegraphics[width=0.31\textwidth]{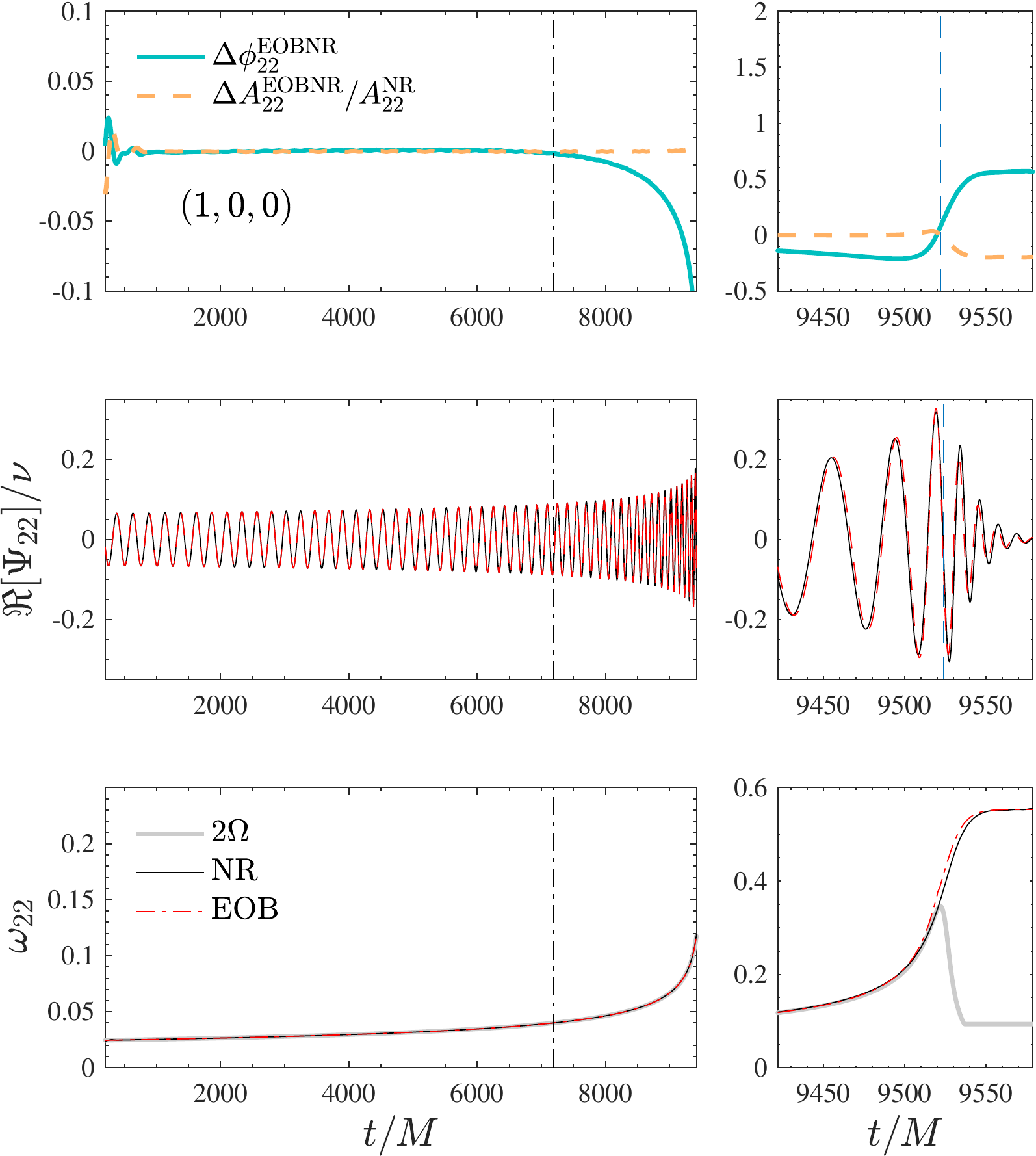}\;
\includegraphics[width=0.31\textwidth]{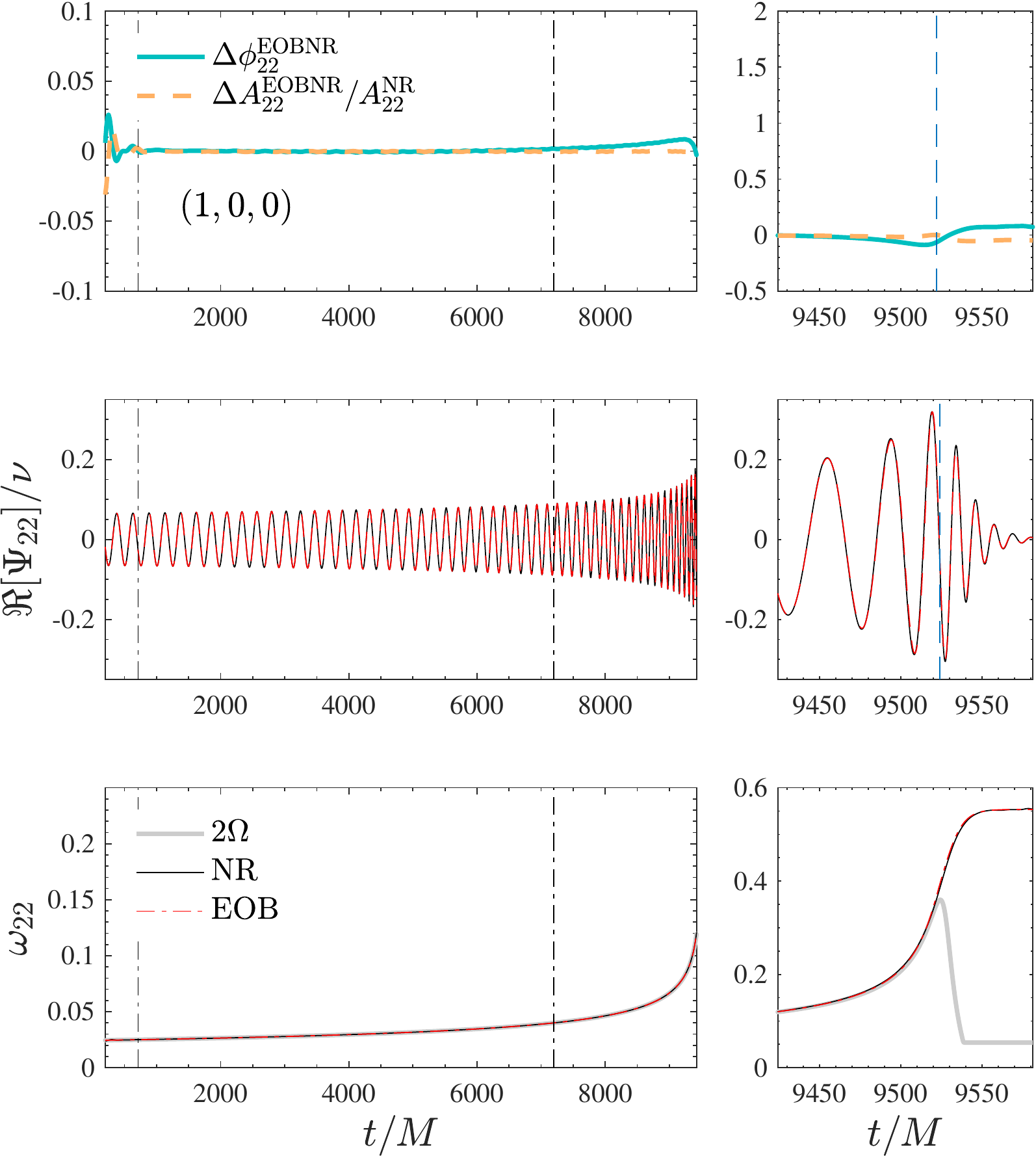}
\caption{\label{fig:phasing_cnc} Left panel: EOB/NR phasing comparison for $q=1$, with the quasi-circular $\hat{\F}_r$ and $a_6^c=281.62$,
Eq.~\eqref{eq:Fr_circ_resum}. Middle panel: same comparison, with the same value of $a_6^c=281.62$, but using the general expression 
of $\hat{\F}_r$ given by the resummed version of Eq.~\eqref{eq:Fr_noncirc}. In this case case one sees a slightly larger phase difference accumulated 
during plunge up to merger and ringdown. Right panel: phasing comparison with the standard, quasi-circular, {\tt TEOBResumS} model~\cite{Nagar:2020pcj}.}
\end{figure*}
%===================================
In Refs.~\cite{Chiaramello:2020ehz,Nagar:2020xsk} it was used the general, 
non circularized,  2PN-accurate resummed form of the radial force $\hat{\F}_r$,
whose expression was not written explicitly. The purpose of this Appendix
is two fold: on the one hand, fill the information missing in previous literature;
on the other hand, highlight that the action of $\hat{\F}_r$ becomes unacceptably
large in strong field and prevents us from exploiting in full the natural flexibility
of the EOB $A$ function through $a_6^c(\nu)$ that has been key in previous
EOB/NR works, e.g.~\cite{Nagar:2017jdw,Nagar:2019wds,Rettegno:2019tzh,Nagar:2020pcj}.
To overcome this difficulty, that we will discuss below, we decided to adopt the 
quasi-circular expression for $\hat{\F}_r$ of Eq.~\eqref{eq:FRcirc} used in the main text.
Before entering in the details of this issue, let us review the analytical steps that
brought us to the expression of $\hat{\F}_r$ used in Refs.~\cite{Chiaramello:2020ehz,Nagar:2020xsk}
Let us start by recalling the expression of $\hat{\F}_r$ given in Eq.~(3.70) of  Ref.~\cite{Bini:2012ji},
that reads
\begin{multline}
  \hat{\F}_r=\frac{p_r}{r^3} \left(\sum_{i=1}^{4} T_{i} X^{i}+\epsilon^2 \sum_{i,j=1}^{4} T_{ijk} X^{ij}+\right.\\\left.\epsilon^4 \sum_{i,j,k=1}^{4} T_{ijk} X^{ijk} \right)
\end{multline} 
where $\epsilon\equiv c^{-1}$ as a reminder of the explicit PN expansion at 2PN.
The EOB scalars $X_i$, as introduced in Eqs.~(3.17) and (3.41) of Ref.~\cite{Bini:2012ji},
read 
\begin{align}
\label{eq:scalars}
    X_1 &=p^2 , \\
    X_2  & =p_{r}^2 , \\
    X_3 &=u \\
    \label{eq:X4}
    X_4 &= r \partial_r\hat{H}_{\rm EOB} \ ,
\end{align}
where $p\equiv \left(p_{\varphi}^2 u^2 + B^{-1} p_{r}^2 \right)^{1/2}$
and $A$ e $B$ are the EOB potentials and in the equation above
we introduced the shorthands $X_{ij}\equiv X_i X_j$ and $X_{ijk}\equiv  X_i X_j X_k$.
The $(T_iT_{ij},T_{ijk})$ are expressed in Eqs.~(D9-D11) of Ref.~\cite{Bini:2012ji}
and we report them here explicitly for convenience. The $T_i$ read
\begin{align}
    T_2&=0 , \qquad T_3=\frac{32 \nu }{3} , \qquad T_4=-\frac{56 \nu }{5} \ .
\end{align}
The $T_{ij}$ read
\begin{align}  
    T_{22}    &=0  \ , \\
    T_{23}    &=\nu  \left(\frac{4 \nu }{7}+\frac{100}{21}\right) , \\
    T_{24}    &=\nu  \left(-\frac{76 \nu }{105}-\frac{232}{105}\right) , \\
    T_{33}    & =\nu  \left(-\frac{3776 \nu }{105}-\frac{4532}{105}\right),\\
    T_{34}    & =\nu  \left(\frac{1172 \nu }{35}+\frac{998}{105}\right) , \\
    T_{44}    &=\nu  \left(\frac{368}{21}-\frac{400 \nu }{21}\right) \ .
\end{align}
\begin{align}
   T_{222} &=0, \\
   T_{223} &=\nu  \left(-\frac{206 \nu ^2}{315}-\frac{94 \nu }{35}-\frac{14}{15}\right),\\
    T_{224}&=\nu  \left(\frac{88 \nu ^2}{189}+\frac{1382 \nu }{945}+\frac{1088}{945}\right),\\
   T_{233}&=\nu  \left(-\frac{1312 \nu ^2}{2835}-\frac{17678 \nu }{567}-\frac{1024}{135}\right),\\
   T_{234}&=\nu  \left(-\frac{263 \nu ^2}{945}+\frac{550 \nu }{27}+\frac{209}{21}\right),\\
    T_{244}&=\nu  \left(\frac{104 \nu ^2}{315}-\frac{1786 \nu }{315}+\frac{562}{315}\right),\\
    T_{333}&=\nu  \left(\frac{1748 \nu ^2}{35}+\frac{1138 \nu }{5}-\frac{351098}{2835}\right),\\ 
   T_{334}&=\nu  \left(-\frac{73384 \nu ^2}{945}-\frac{63173 \nu }{945}+\frac{16148}{189}\right),\\ 
   T_{344} &=\nu  \left(\frac{7544 \nu ^2}{135}-\frac{5584 \nu }{315}-\frac{33976}{945}\right),\\ 
   T_{444} &=\nu  \left(-\frac{836 \nu ^2}{105}+\frac{5393 \nu }{315}-\frac{968}{315}\right).
\end{align}
Then, the $X_4$ we use in practice is given by Eq.~\eqref{eq:X4} expanded at 2PN
accuracy, and similarly $X_1$ at 2PN reads (see Eq.~(3.47) of Ref.~\cite{Bini:2012ji})
\begin{align}
    &X_1=X_2+X_3-X_4+\epsilon^2 \left(2 X_{23}-\frac{1}{2} (\nu +1) X_{24}+3 X_{33} \right. \nonumber\\
    &\left. +\frac{1}{2} (\nu -5) X_{34}+\frac{1}{2} (\nu +1) X_{44}\right)+\epsilon^4 \left(\frac{1}{8} \left(\nu ^2-\nu +1\right) X_{224} \right. \nonumber\\
    &\left.+(2-6 \nu ) X_{233}-\frac{1}{4} \left(\nu ^2+\nu +3\right) X_{234}+\frac{3 \nu}{4} X_{244}\right. \nonumber\\
    &\left.-3 (\nu -3) X_{333}+\frac{1}{8} \left(\nu ^2+7 \nu -63\right) X_{334}+\frac{1}{4} (5 \nu +8) X_{344}\right.\nonumber\\
    &\left.-\frac{1}{8} \left(\nu ^2+5 \nu +1\right) X_{444}\right) \ .
\end{align}

As a last step, we replace the radial momentum $p_r$ with $p_{r_*}$ using the usual relation $p_r=p_{r_*} \sqrt{B/A}$ 
at 2PN so that we finally obtain (setting $\epsilon=1$)
\be
\label{eq:Fr_noncirc}
\hat{\F}_r = \nu p_{r_*} u^4(\hat{f}_r^{\rm N} + \hat{f}_r^{\rm 1PN} + \hat{f}_r^{\rm 2PN})  \ ,
\ee
where
\begin{align}
\hat{f}^{\rm N}_r &= -\dfrac{8}{15}+\dfrac{56}{5}p_\varphi^2 u  \ ,\\
\hat{f}_r^{\rm 1PN} &= \left(\frac{556 \nu }{105}-\frac{1228}{105}\right) p_{r*}^2+\del{\frac{16 \nu
   }{21}-\frac{1984}{105}} u\nonumber\\
   &+\left(-\frac{436 \nu }{105}-\frac{124}{105}\right) p_{r*}^2 p_{\varphi}^2u\nonumber\\
   &+ \left(\frac{1252}{105}-\frac{2588 \nu }{105}\right) p_{\varphi}^4u^3\nonumber\\
   &+ \left(-\frac{1268 \nu }{105}-\frac{1696}{35}\right) p_{\varphi}^2 u^2 \ , \\
   \hat{ f}_r^{\rm 2PN} & = \left(-\frac{1273 \nu ^2}{315}+\frac{1061 \nu }{315}+\frac{323}{315}\right)p_{r*}^4\nonumber\\
&+\del{-\frac{3548 \nu ^2}{315}+\frac{9854 \nu }{105}+\frac{59554}{2835}}u^2\nonumber \\ 
   &+ \left(-\frac{8804\nu ^2}{315}+\frac{10292 \nu }{315}-\frac{1774}{21}\right) p_{r*}^2 p_{\phi}^2 u^2 \nonumber\\
   & + \left(\frac{194 \nu ^2}{7}-\frac{1052 \nu}{105}-\frac{628}{105}\right) p_{r*}^2 p_{\phi}^4 u^3\nonumber \\ 	
   & + \left(-\frac{1752 \nu
   ^2}{35}+\frac{9568 \nu }{315}-\frac{29438}{315}\right) p_{\phi}^2 u^3\nonumber\\
   &+ \left(\frac{131 \nu ^2}{63}-\frac{983 \nu}{315}-\frac{461}{315}\right) p_{r*}^4 p_{\phi}^2 u\nonumber \\ 
   &+ \left(-\frac{218 \nu
   ^2}{189}+\frac{17590 \nu }{189}+\frac{20666}{315}\right)p_{r*}^2 u\nonumber\\
   &+\left(\frac{3277 \nu ^2}{105}-\frac{718 \nu}{63}-\frac{3229}{315}\right) p_{\varphi}^6 u^5+ \\ \nonumber
   &+ \left(\frac{25217 \nu
   ^2}{315}+\frac{1606 \nu }{15}-\frac{35209}{315}\right) p_{\varphi}^4 u^4 \ .
\end{align}
%====================================
% TABLE EOB/NR values
%====================================
\begin{table}
  \caption{\label{tab:equal_SXS} 
  Maximal values of unfaithfulness for spinning datasets with $q=1$. 
  From left to right, the columns report: the number of dataset; the SXS simulation number; 
  mass ratio and dimensionless spins $(q,\chi_1,\chi_2)$; the maximum value of the unfaithfulness
  $\bar{F}^{\rm max}_{\rm NR/NR}$ taken between the two highest NR resolutions, see Ref.~\cite{Nagar:2020pcj},
  and and between EOB and NR waveforms, $\bar{F}^{\rm max}_{\rm EOB/NR}$, see Fig.~\ref{fig:barF_eobnr_circ}.}
\begin{center}
\begin{ruledtabular}
\begin{tabular}{r | l l l l}
\# & {\rm id} & $(q,\chi_1,\chi_2)$ & $\bar{F}^{\rm max}_{\rm NR/NR}[\%]$ & $\bar{F}^{\rm max}_{\rm EOB/NR}[\%]$\\\hline
$1$ & BBH:0180 & $(1,0,0)$ & $0.0035$ & 0.65$$\\
$2$ & BBH:0007 & $(1.5,0,0)$ & $0.0020$ & $0.40$\\
$3$ & BBH:0169 & $(2,0,0)$ & $0.0032$ & $0.19$\\
$4$ & BBH:1221 & $(3,0,0)$ & $ 0.0016$ & $0.14$\\
$5$ & BBH:0294 & $(3.5,0,0)$ & $0.0102$& $ 0.066 $\\
$6$ & BBH:0167 & $(4,0,0)$ & $0.0057$ & $0.082$\\
$7$ & BBH:0056 & $(5,0,0)$ & $0.0158$ & $0.058$\\
$8$ & BBH:0166 & $(6,0,0)$ & .. & $0.057$\\
$9$ & BBH:0063 & $(8,0,0)$ & $0.0754$ & $0.056$\\
$10$ & BBH:0303 & $(10,0,0)$ & $0.0045$ & $0.059$\\
\hline
11 & BBH:1124    & $(1,+0.998,+0.998)$ & $\dots $ & 0.18 \\
12 & BBH:0178 & $(1,+0.9942,+0.9942)$ & $0.0066$ &   0.23\\
13 & BBH:0177 & $(1,+0.9893,+0.9893)$ & $0.0021$ & 0.15 \\
14 & BBH:0172 & $(1,+0.9794,+0.98)$ & $0.0022$ & 0.21 \\
15  &BBH:0158   & $(1,+0.97,+0.97)$ &   0.31        & 0.25  \\
16 & BBH:0157 & $(1,+0.95,+0.95)$ & $0.0027$ & 0.20 \\
17 & BBH:0160 & $(1,+0.9,+0.9)$ & $0.0118$ & 0.48\\
18 & BBH:0153 & $(1,+0.85,+0.85)$ & .. & 0.61\\
19 & BBH:0230 & $(1,+0.8,+0.8)$ & $0.0016$ & 0.63 \\
20 & BBH:0228 & $(1,+0.6,+0.6)$ & $0.0080$ & 0.62 \\
21 & BBH:1122  & $(1,+0.44,+0.44)$ & 0.0031  & $0.62$  \\
22 & BBH:0150 & $(1,+0.2,+0.2)$ & $0.0027$ & 0.86\\
23 & BBH:0149 & $(1,-0.2,-0.2)$ & $0.0037$ & 0.63\\
24 & BBH:0148 & $(1,-0.44,-0.44)$ & $0.0013$ & 0.38\\
25 & BBH:0215 & $(1,-0.6,-0.6)$ & $0.0040$ & 0.22\\
26 & BBH:0154 & $(1,-0.8,-0.8)$ & $0.0036$ & 0.19\\
27 & BBH:0159 & $(1,-0.9,-0.9)$ & $0.0069$ & 0.20\\
28 & BBH:0156 & $(1,-0.95,-0.95)$ & $0.0055$ &0.23 \\
29 & BBH:0156 & $(1,-0.97,-0.97)$ & $0.0055$ & 0.24 \\
30 & BBH:0231 & $(1,+0.9,0)$ & $0.0046$ & 1.00\\
31 & BBH:0232 & $(1,+0.9,+0.5)$ & $0.0073$ & 1.19\\
32 & BBH:0224  &$(1,+0.40,-0.80)$ & $0.002$ & 0.29

\end{tabular}
\end{ruledtabular}
\end{center}
\end{table}

\begin{table}
  \caption{\label{tab:unequal_SXS} Maximal values of unfaithfulness for spinning datasets with $q\neq 1$. 
  From left to right, the columns report: the number of dataset; the SXS simulation number; 
  mass ratio and dimensionless spins $(q,\chi_1,\chi_2)$; the maximum value of the unfaithfulness
  $\bar{F}^{\rm max}_{\rm NR/NR}$ taken between the two highest NR resolutions, see Ref.~\cite{Nagar:2020pcj}, 
  and between EOB and NR waveforms, $\bar{F}^{\rm max}_{\rm EOB/NR}$, see Fig.~\ref{fig:barF_eobnr_circ}.}
\begin{center}
\begin{ruledtabular}
\begin{tabular}{r | l l l l}
\# & {\rm id} & $(q,\chi_1,\chi_2)$ & $\bar{F}^{\rm max}_{\rm NR/NR}[\%]$ & $\bar{F}^{\rm max}_{\rm EOB/NR}[\%]$\\
\hline
33 & BBH: 1146 & $(1.5,+0.95,+0.95)$ & \dots & 0.23 \\
34 & BBH:0234& $(2,-0.85,-0.85)$ & 0.0049 & 0.16 \\
35 & BBH:0239 & $(2,-0.37,+0.85)$ & 0.0005 & 0.13\\
36 & BBH:0252 & $(2,+0.37,-0.85)$ & 0.0029 & 0.44 \\
37 & BBH:0257 & $(2,+0.85,+0.85)$ & 0.0024 & 0.070 \\
38 & BBH:0260& $(3,-0.85,-0.85)$   & 0.0004 & 0.12 \\
39 & BBH:0268& $(3,-0.40,-0.60)$  & 0.0016& 0.13\\
40 & BBH:0285& $(3,+0.40,+0.60)$  & 0.0013 & 0.36\\
41 & BBH:0293& $(3,+0.85,+0.85)$  & 0.0046& 0.013\\
42 & BBH:1434 & $(4.368,+0.80,+0.80)$  & $\dots$ & 0.096\\
43 & BBH:0208 & $(5,-0.90,0)$  & 0.0385 & 0.091\\
44 & BBH:1463 & $(5,+0.61,0.24)$  & 0.0032 & 0.49\\
45 & BBH:1432 & $(5.841,+0.66,0.80)$  & 0.0192 & 0.38 \\
46 & BBH:0207& $(7,-0.60,0)$  & 0.011& 0.059 \\
47 & BBH:0205& $(7,-0.40,0)$  & 0.0040& 0.062 \\
48 & BBH:0203& $(7,+0.40,0)$  & 0.0095& 0.23 \\
49 & BBH:0202& $(7,+0.60,0)$  &0.048 & 0.75 \\
50 & BBH:1375& $(8,-0.90,0)$  & $\dots$ & 0.22 \\
51 & BBH:1419& $(8,-0.80,-0.80)$  & $\dots$ & 0.16 \\
52 & BBH:1426& $(8,+0.48,+0.75)$  & 0.0378 & 0.091\\
53 & BAM & $(8,+0.85,+0.85)$  & $\dots$ & 1.108
\end{tabular}
\end{ruledtabular}
\end{center}
\end{table}
%=====================
%==============
% Initial frequency
%==============
\begin{figure}[t]
\includegraphics[width=0.45\textwidth]{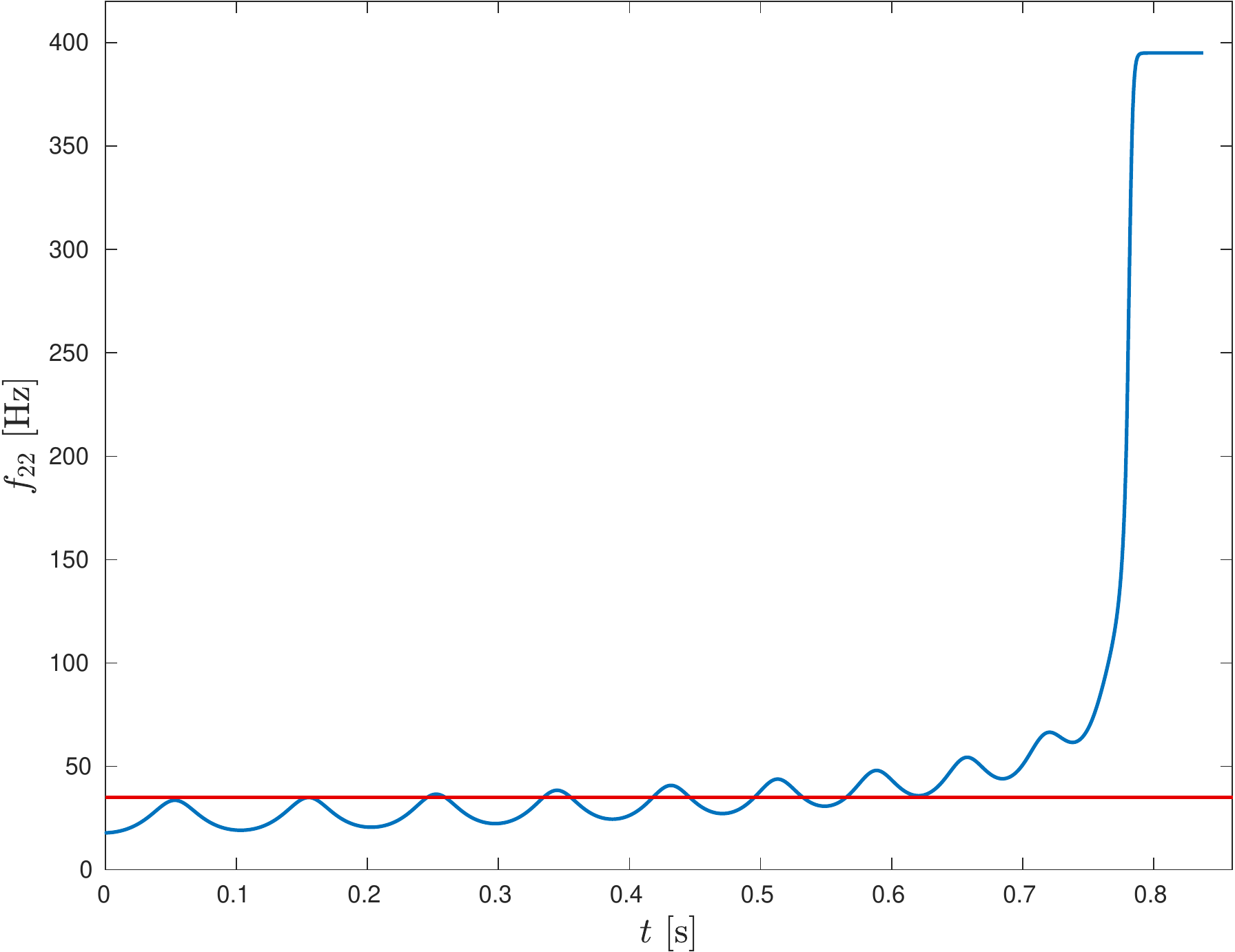}
\caption{\label{initialfreq} 
Gravitational wave frequency $f_{22}$ of the $\ell=m=2$ mode of signal generated with {\tt TEOBResumSGeneral} 
using $M = 45M_\odot$, $q = 1.24138$, $\chi_{1,2} = 0$, $D_L = 410$, $\iota = 1.309$, $e^{\rm EOB}_{\omega_{\rm 0}} = 0.2$ and initial frequency at periastron equal to $35$~Hz (red line).
Changing the prescription for the initial frequency of 35~Hz would approximately mean 
starting the system at the apastron around $t = 0.62$~s when using $\omega_a^{\rm EOB}$ or around $t = 0.47$~s for $\omega_{\rm secular}^{\rm EOB}$.}
\end{figure}

The function $\tilde{\F}_r^{\rm 2PN}\equiv \hat{f}_r^{\rm N}+\hat{f}_r^{\rm 1PN}+\hat{f}_r^{\rm 2PN}$ is
then written as $\tilde{\F}_r^{\rm 2PN}=\hat{f}_r^{\rm N}\tilde{f}_r$, where $\tilde{f}_r\equiv 1 + c^{\rm 1PN} u + c_{\rm 2PN} u^2$,
and $c^{\rm 1PN,2PN}\equiv \hat{f}_r^{\rm 1PN,2PN}/\hat{f}_r^{\rm N}$, and it is then resummed
using a $(0,2)$ Pad\'e approximant. This is the prescription of $\hat{\F}_r$ used in previous\footnote{Note that due
to a calculation error, the coefficients that now correctly read $9854\nu/105$ and $9568\nu/315$
were respectively replaced by $9686\nu/105$ and  $58424\nu/315$ in Refs.~\cite{Chiaramello:2020ehz,Nagar:2020xsk}.
This does not have any meaningful impact on the results discussed therein.}
works~\cite{Chiaramello:2020ehz,Nagar:2020xsk}. As mentioned in the main text, we realized that
such expression becomes too large in strong field and prevents us to efficiently inform $a_6^c(\nu)$
so to obtain an acceptably small EOB/NR phase difference up to merger.
In particular, one realizes that the dynamics becomes practically insensitive to $a_6^c$, so that it is
necessary to increase it a lot to gain minimal improvements around merger. In particular this implies 
large differences between the best value for $q=1$ and the best value for $q=2$. This effect, though 
still there, is less dramatic with the quasi-circular expression, that allows one to obtain more easily
a good EOB/NR phasing agreement through plunge, merger and ringdown. To get a more precise
understanding of the effect we report in Fig.~\ref{fig:phasing_cnc} three EOB/NR phasing comparisons.
The left and middle panel of the figure are obtained with the eccentric model and both
share the same value of  $a_6^c=281.62$ but: (i) the leftmost panel uses the quasi-circular 
expression for $\hat{\F}_r$~Eq.~\eqref{eq:Fr_circ_resum} instead of the general one 
discussed here and shown in the middle panel. In this second case, the figure highlights that a (slightly) 
larger phase difference accumulated during the late plunge up to merger and ringdown. Note that  this 
difference cannot be reabsorbed by changing further (i.e., by increasing) $a_6^c$; it is due 
to a $\hat{\F}_r$ that, despite the resummation, keeps growing in strong field and dominates 
the dynamics in the very late plunge phase. For the sake of comparison, the right panel of 
Fig.~\ref{fig:phasing_cnc} also shows the phasing agreement yielded by the standard, quasi-circular
model {\tt TEOBResumS} of Ref.~\cite{Nagar:2020pcj} that has, by construction, $\hat{\F}_r=0$.
Although the phase disagreement shown in the figure has little impact, in terms of $\bar{F}_{\rm EOB/NR}$, 
for such an  equal-mass, nonspinning binary, it becomes unacceptably larger when spin is switched on. 
For example, we have verified that a model constructed with the general (resummed) $\hat{\F}_r$ 
leads to $\bar{F}^{\rm max}_{\rm EOB/NR}\simeq 4\%$ for a quasi-circular configuration 
with $(q,\chi_1,\chi_2)=(1,+0.95,+0.95)$, and this difference cannot be cured neither changing
$a_6^c$, nor $c_3$, because it is fundamentally due to $\hat{\F}_r$. At the moment, the choice of the 
2PN-resummed  quasi-circular expression of $\hat{\F}_r$ seems to give an acceptable compromise
to get the model consistent in terms of losses as well as accurate versus quasi-circular NR simulations.

\section{Explicit EOB/NR unfaithfulness results}
\label{sec:barF}
In this Appendix we list the explicit values of the maximum EOB/NR unfaithfulness 
$\bar{F}^{\rm max}_{\rm EOB/NR}$ on the meaningful portion of the SXS catalog that
we considered to obtain Fig.~\ref{fig:barF_eobnr_circ}.
The data are listed in Table~\ref{tab:equal_SXS}, for nonspinning and equal mass,
spinning configurations, and in Table~\ref{tab:unequal_SXS} for unequal mass, 
spinning configurations.

%==============
% Scatterplot f(e)
%==============
\begin{figure}[t]
	\center
	\includegraphics[width=0.485\textwidth]{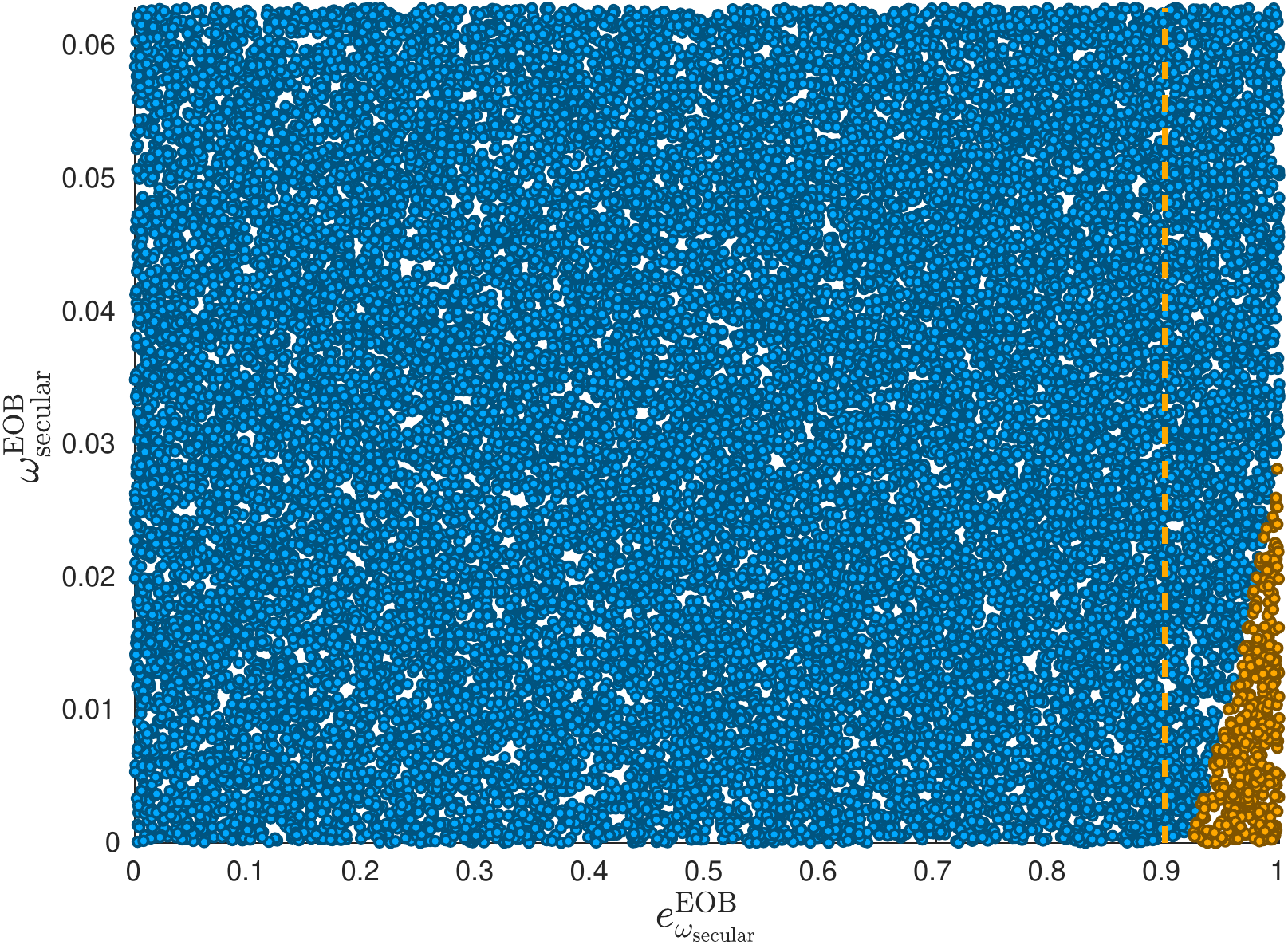}
	\vspace{5mm}
	\hspace{1.5mm}\includegraphics[width=0.48\textwidth]{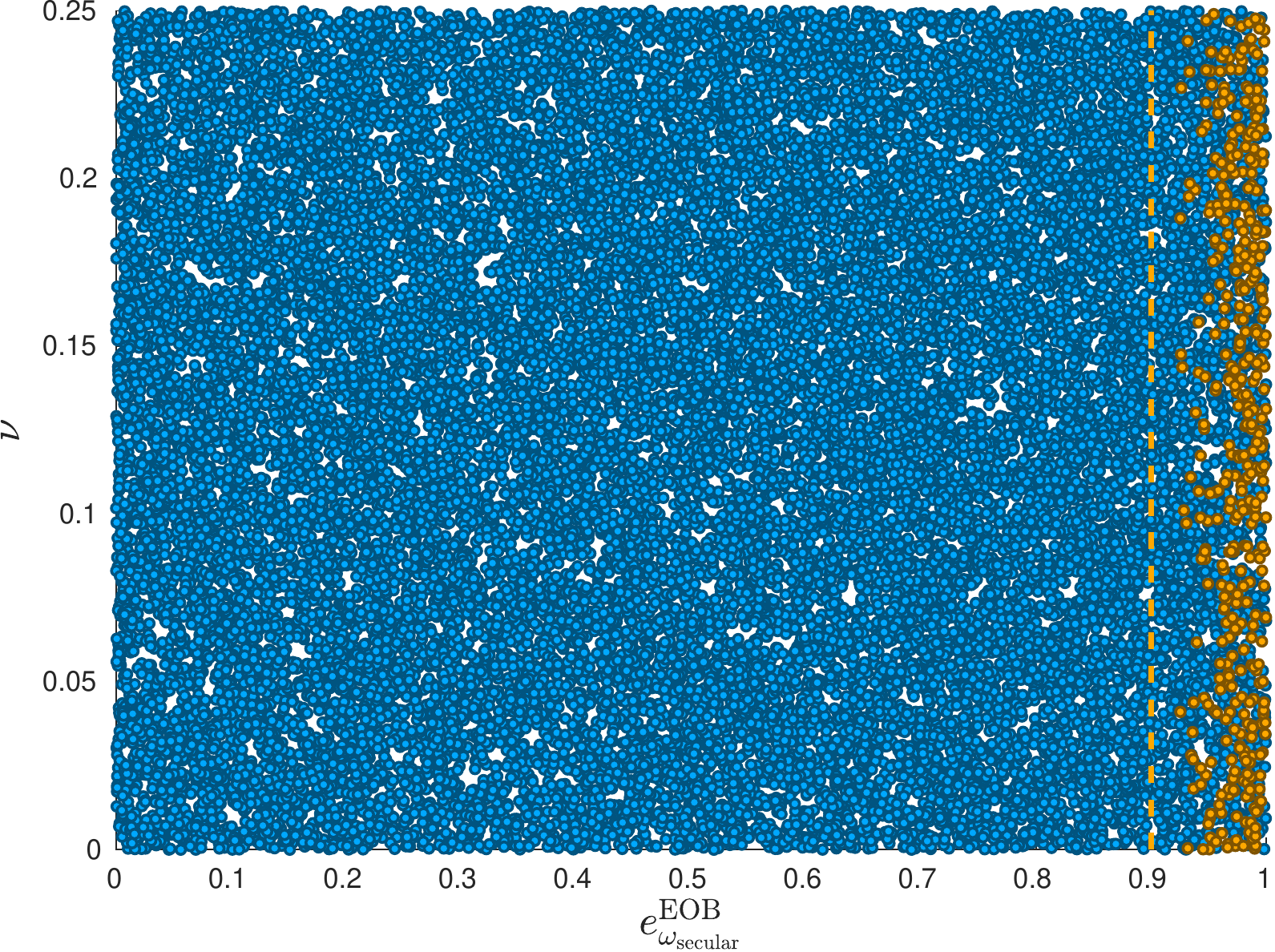}
	\caption{\label{parspace_fe} Reliability of waveform generation over the parameter space for eccentric inspirals.
	The markers in the plot correspond to waveforms generated with the {\tt TEOBResumSGeneral} code, specifying
	the initial secular frequency, $\omega_{\rm secular}^{\rm EOB}$ and the corresponding eccentricity $e_{\omega_{\rm secular}}^{\rm EOB}$. 
	We here show 25000 points, with parameters randomly generated in the ranges: $\nu \in \left(0,0.25\right)$, $(\chi_1,\chi_2) \in \left(- 1,+1\right)$; 
	$  \omega^{\rm EOB}_{\rm secular} \in \left(0,10^{-2}\right)\times 2\pi$; $e^{\rm EOB}_{\omega_{\rm secular}} \in \left(0,1\right)$.
	Blue dots mark successfully computed waveforms, while orange dots mark configurations for which the initial conditions could not be generated.
	The dashed orange line corresponds to  $e^{\rm EOB}_{\omega_{\rm secular}} =0.9$.
	}
\end{figure}

\section{{\tt TEOBResumSGeneral} $C$ code}
\label{sec:Ccode}
%==============
% Eccentric visual
%==============
\begin{figure}[t]
	\center
	\includegraphics[width=0.45\textwidth]{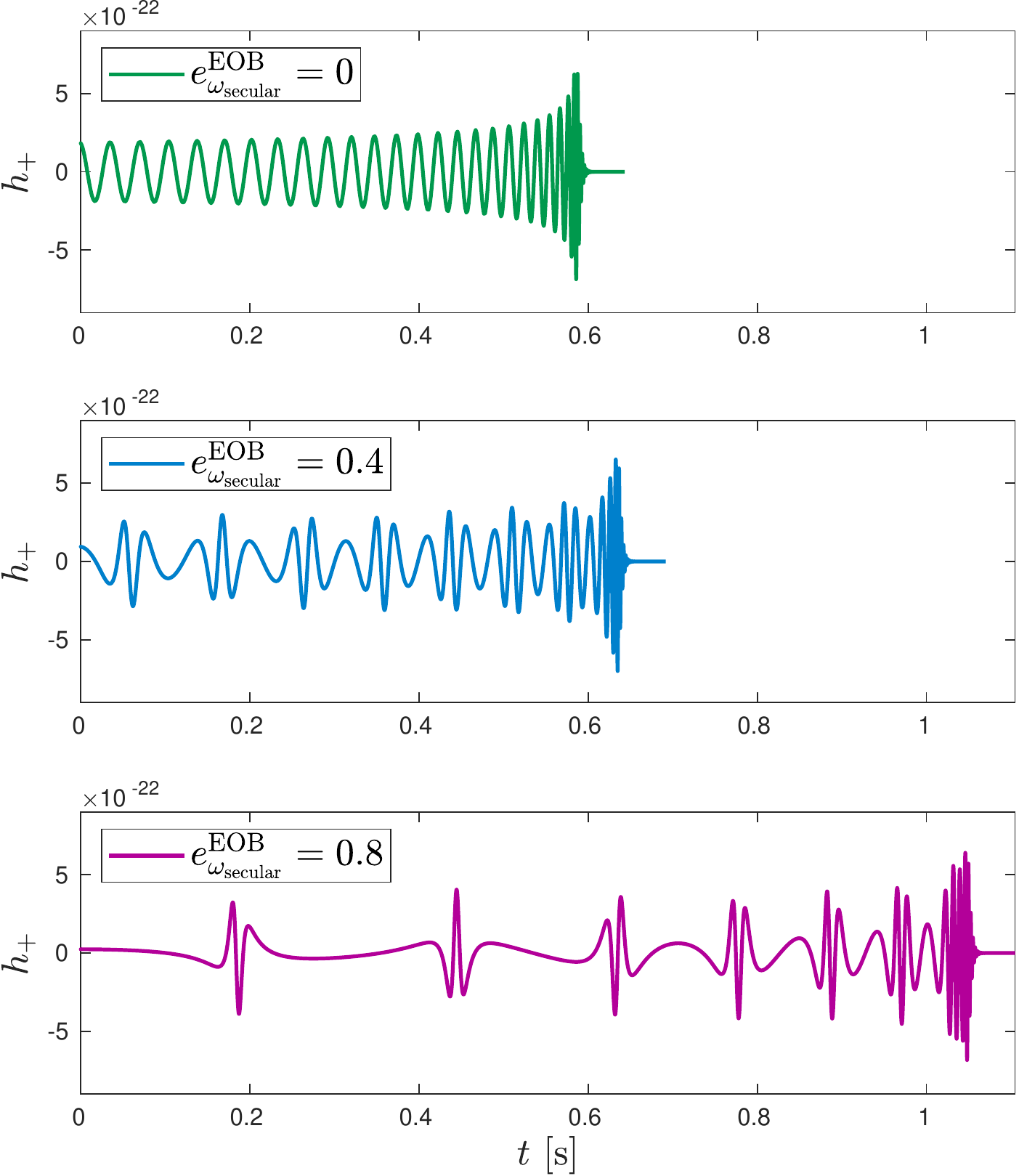}
	\caption{\label{visual_e} 
	Visual comparison of GW signals with different initial eccentricities. 
	The waveforms are generated with {\tt TEOBResumSGeneral} using $M = 45 M_\odot$, $q = 1.24138$, $\chi_i = 0$, $D_L = 410$, $\iota = 1.309$, 
	$\omega^{\rm EOB}_{\rm secular} = 35\times 2\pi$~Hz and $e^{\rm EOB}_{\omega_{\rm secular}} = (0,0.4,0.8)$.
	}
\end{figure}
%======================
% Scatterplot time(f,e)
%======================
\begin{figure}[t]
	\center
	\includegraphics[width=0.49\textwidth]{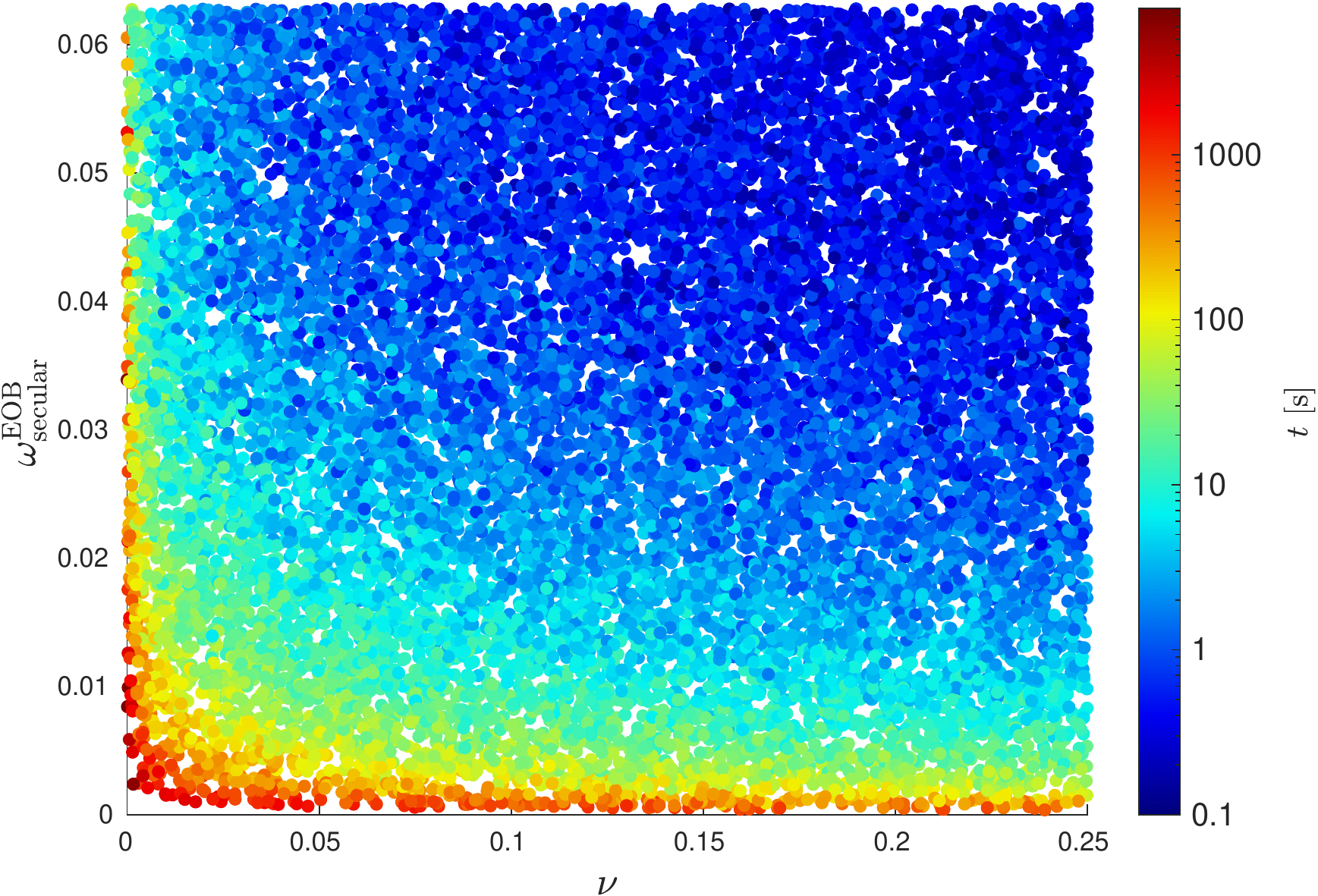} \hfill	
	\includegraphics[width=0.47\textwidth]{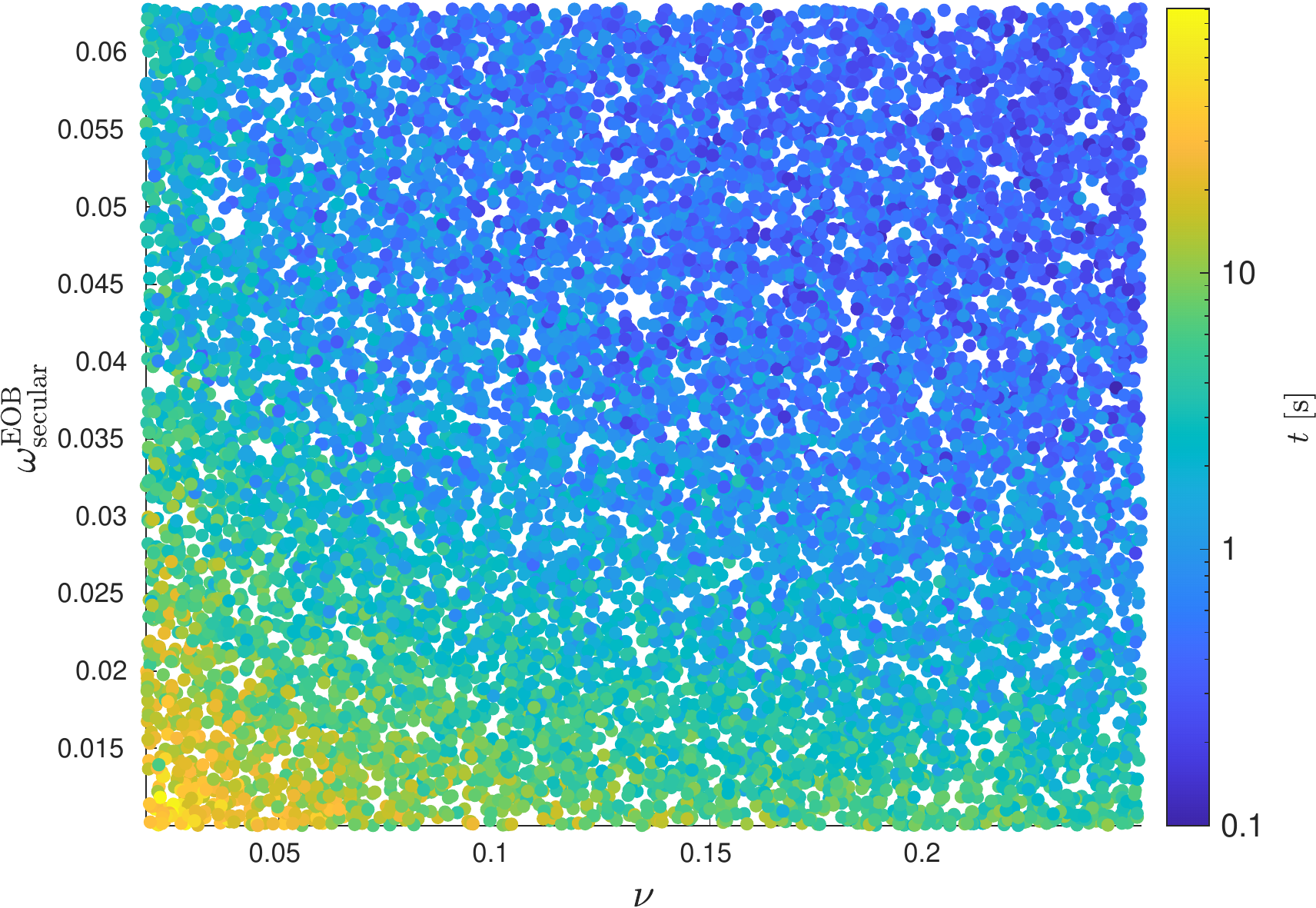}
	\caption{\label{timing} Generation time analysis for {\tt TEOBResumSGeneral}. 
		Top panel. Timing of $~25000$ configurations, with parameters randomly generated in the same ranges as in Fig.~\ref{parspace_fe}.
		The color code indicates the computation time of each waveform in the time domain without interpolation on an uniformly spaced temporal grid.
		We omitted systems for which we could not compute the computation time. This can be due to the failure of the initial data prescription 
		(equivalent to orange points in Fig.~\ref{parspace_fe}) or because the chosen configuration does not reach the merger before a time $t/M = 10^9$ 
		(very low starting frequency or very high mass ratio). Bottom panel: close-up on the region with $q < 50$ 
		and $\omega_{\rm secular}^{\rm EOB} > 0.01$, the part of parameter space that is of interest 
		for currently published LIGO-Virgo events~\cite{Abbott:2020niy}.}
\end{figure}

The EOB waveform model {\tt TEOBResumSGeneral} discussed in this paper is publicly available as a stand-alone $C$ implementation 
via a {\tt bitbucket} git repository~\cite{teobresums}. It can deal with quasi-circular configuration, eccentric inspirals or hyperbolic encounters.
As discussed in the main text, an eccentric waveform is computed by specifying an initial GW frequency $\omega^{\rm EOB}_0$ 
and a corresponding value of the EOB eccentricity at this frequency $e^{\rm EOB}_{\omega_0}$. 
It is convenient to start the system specifying the initial apastron frequency $\omega_a^{\rm EOB}$,
posing $\omega_0^{\rm EOB}=\omega_a^{\rm EOB}$. However, to maintain 
continuity with respect to the circular waveforms, the initial conditions can also be determined 
using the starting secular (i.e. average of the apastron and periastron) frequency $\omega^{\rm EOB}_{\rm secular}$. 
The difference between these two approaches is clarified in Fig.~\ref{initialfreq}.

Using frequency and eccentricity as initial conditions, extending the usual prescription for circular 
systems, {\tt TEOBResumSGeneral} can reliably generate waveforms with eccentricity 
$e^{\rm EOB}_{0} \lesssim 0.9$, see Fig.~\ref{parspace_fe}. 
For higher eccentricities, this approach fails to correctly compute initial conditions. 
Its region of validity could be extended, but it will inexorably fail somewhere near the limit $e \rightarrow 1$, i.e. 
near head-on collisions. A more robust procedure, in this case, is to initiate the dynamics specifying values
of the energy and angular momentum, coherently with what is done for hyperbolic systems~\cite{Nagar:2020xsk}, 
but selecting stable configurations. We also point out that, while the model could not be validated with NR simulations, 
it still produces sane waveforms also for extreme values of the eccentricity. An example is shown in Fig. ~\ref{visual_e},
that illustrates three cases of (nonspinning) waveforms starting at the same secular frequency but with eccentricities
$e^{\rm EOB}_{\omega_{\rm secular}}=(0,0.4,0.8)$. Note, in the third case, that, despite the sequence of short-duration
bursts, the system essentially circularizes before merger and ringdown.

A detailed study of computation times is shown in Fig.~\ref{timing}. The bottom panel of the figure selects mass ratio $q< 50$,
so to have an estimate of the timing needed for generating waveforms consistent with the currently published LIGO-Virgo 
events~\cite{Abbott:2020niy}. As mentioned in the main text, our implementation is sufficiently efficient to be used for 
inferring eccentricity measurements at least from GW150914-like events. To do so, one needs 
to consider {\it both} $(\omega_{\rm EOB}^{\rm secular},e_{\rm EOB}^{\rm secular})$ as sampling parameters.
This is necessary because the initial mean EOB anomaly is always set to zero, since the evolution of the system
begins at the apastron. Therefore, in order to cover all possible configurations and avoid biases \cite{Islam:2021mha}, 
the sampling of the mean anomaly typical of other approaches is replaced by the sampling of $\omega_{\rm EOB}^{\rm secular}$.

Finally, let us comment on extreme mass-ratio inspirals (EMRIs), given their importance for future 
space-based detectors such as LISA~\cite{Audley:2017drz}. 
The generation time for a one-year long EMRI ($M = 10^6 M_\odot$, $q = 10^5$) varies a lot depending on starting frequency and eccentricity. 
Let us consider as test case a nonspinning system with initial eccentricity $e_{\rm EOB}^{\rm secular} = 0.3$.
The code takes around 30 seconds to generate a one-year long waveform when starting at a rather low frequency ($\sim 10^{-4}$~Hz).
The time increases to about 5 minutes when considering a higher starting frequency ($\sim 10^{-3}$~Hz) and hence 
a phase in which the system is no longer adiabatic but instead it is slowly inspiralling.

\bibliography{refs20210409.bib,local.bib}

\end{document}